\documentclass[twocolumn,twocolappendix]{aa}
\usepackage[utf8]{inputenc}
\usepackage{graphicx}
\usepackage{color}
\usepackage{here}
\usepackage{ascmac}
\usepackage{fancybox}
\usepackage{bm}
\usepackage{amsmath}
\usepackage{amssymb}
\usepackage{mathrsfs, mhchem}
\usepackage[T1]{fontenc}
\usepackage{comment}
\usepackage{xcolor}

\usepackage{txfonts} 
\usepackage{hyperref}
\usepackage{natbib}
\newcommand{\Msun}{M_\odot}

\newcommand{\unit}[1]{\mathrm{\,#1}}

\newcommand{\msunyr}{\Msun {\rm yr}^{-1}}
\newcommand{\mdoti}{\dot{M}_\mathrm{infall}}
\newcommand{\mdota}{\dot{M}_\mathrm{acc}}

\title{From streamers to stars: Overcoming mass loss in protoplanetary disks}

\titlerunning{From Streamers to Stars}
\authorrunning{Ooyama et al.}

\author{
W.~Ooyama\inst{1}\thanks{ \email{ooyama.wataru.64z@st.kyoto-u.ac.jp}}
\and R.~Nakatani\inst{2}
\and T.~Hosokawa\inst{1}
\and H.~Mitani\inst{3,4}
}

\institute{
Department of Physics, Graduate School of Science, Kyoto University, Sakyo, Kyoto 606-8502, Japan
\and
Dipartimento di Fisica, Universit\`a degli Studi di Milano, Via Celoria 16, I-20133 Milano, Italy
\and
Faculty of Physics, University of Duisburg-Essen, Lotharstra{\ss}e 1, D-47057 Duisburg, Germany
\and
Department of Physics, School of Science, The University of Tokyo, 7-3-1 Hongo, Bunkyo, Tokyo 113-0033, Japan
}

\abstract{
Recent high-resolution observations have revealed filamentary accretion flows (streamers) in protoplanetary disks older than 1 Myr, suggesting that late-stage interstellar gas infall (late infall) might affect disk evolution and stellar accretion. In Lupus, observations report a positive correlation between ambient gas density and stellar accretion rate. However, it remains unclear whether infall can truly boost stellar accretion, because incoming gas might instead be lost through photoevaporation or magnetically driven disk winds; alternatively, it could remain trapped in the outer disk. In this work, we performed 1D long-term ($\sim$1--10 Myr) disk evolution simulations. We first treated late infall as a mass source and then included the effective torque arising from the angular-momentum difference between the infalling gas and Keplerian disk gas. We find that even if substantial gas reaches the outer disk ($\sim 10^{2}$ au), much of it is eventually lost through photoevaporation. Sustained stellar accretion therefore requires efficient inward gas delivery by mechanisms that locally remove angular momentum. Without an effective infall torque, strong viscosity can provide this transport, but it also ends up driving outward angular-momentum transport and excessive disk spreading, which is inconsistent with the compact disk sizes observed in Lupus. In contrast, magnetohydrodynamic (MHD) disk winds can remove angular momentum without significantly expanding the disk, allowing late infall to sustain stellar accretion, while keeping disks compact. Thus, if the Lupus accretion-density correlation is indeed caused by late infall without an effective infall torque, efficient angular-momentum removal by MHD disk winds would be required. In contrast, when the effective torque is included, the angular-momentum mismatch itself can promote inward gas transport and enhance stellar accretion, even without strong MHD disk winds.
}


\begin{document}
\maketitle
\nolinenumbers

\section{Introduction}

Planetary systems originate within protoplanetary disks and improving our understanding of the evolutionary pathways of these disks is essential for elucidating the processes of planet formation. Among the various aspects of disk evolution, the behavior of the gaseous component is particularly critical, as it provides the raw material for forming gas-giant planets.

Traditionally, class II disks ($\sim \rm Myr$) have been assumed to evolve without external mass supply, since their parental envelopes are thought to have dissipated by this stage \citep{Adamas_1987}. However, recent high-resolution observations have updated this classical picture. A number of disks show large-scale spirals and streamer-like features \citep{Garufi_2020,Huang_2021,Benisty_2023}, interpreted as evidence for continued accretion of interstellar gas (i.e., so-called late infall) even during the class II phase \citep[e.g.,][]{Pineda_2023}. These types of signatures have been reported in a number of systems including AB Aur \citep[e.g.,][]{Nakajima_1995,Grady_1999,Fukagawa_2004,Tang_2012,Speedie+2025}, HD 100546 \citep[e.g.,][]{Grady_2001,Ardila_2007}, DR Tau \citep[e.g.,][]{Mesa_2022,Huang_2023}, SU Aur \citep[e.g.,][]{Ginski_2021}, and HL Tau \citep[e.g.,][]{Hayashi_1993,Welch_2000,Yen_2019,Garufi_2022}. The overall occurrence rate of such late infall has been estimated at about 16\% \citep{Garufi_2024}.

Late infall is presumed to have a potentially profound impact on disk evolution. By replenishing gas and dust, it offers a possible solution to the long-standing discrepancy between the observed low disk masses and the amounts required by planet formation theories \citep{Greaves_2010,Williams_2012,Najita_2014,Manara_2018,Ward-Duong_2018,Tychoniec_2018}. In addition to extending disk lifetimes, late infall might also promote planetesimal formation by creating pressure bumps \citep{Zhao_2025}. Furthermore, 
the accretion of ambient material can inject energy and angular momentum into the disk, potentially driving turbulence, enhancing mass transport, and promoting accretion onto the central star \citep[e.g.,][]{Winter+2024b,Padoan_2025}. In contrast, current observational constraints on disk sizes do not reveal any clear systematic trend with environmental density, and hence offer no clear evidence for late infall. Further theoretical and observational efforts are thus needed to elucidate the role of late infall.

Numerical studies have explored late infall from complementary perspectives. Three-dimensional (3D) hydrodynamic and magnetohydrodynamic (MHD) simulations have reproduced diverse disk structures, streamer-like features, non-axisymmetric structures around class I objects, and even UX Ori stars or FU Ori-type bursts triggered by infall \citep{Vorobyov_2015,Kuffmeier_2020,Kuffmeier_2021,Kuffmeier_2023,Hanawa_2022,Hanawa_2024,Grinin_2024}. Parallel works have modeled late infall as Bondi–Hoyle–Lyttleton accretion, demonstrating its ability to reproduce streamer-like features \citep{Huhn+2025b}, spiral structures of disks \citep{Huhn+2026}, correlations between stellar mass and accretion rate \citep{Troop_2008,Klessen_2010}, stellar luminosities \citep{Padoan_2005}, disk angular momentum \citep{Padoan_2025}, and the persistence of old accretors \citep{Scicluna_2014}. However, most of these studies focus on short-term evolution or on reproducing specific observed phenomena, leaving open the question of how late infall shapes the long-term ($\sim$ Myr) evolution of class II disks.

Beyond morphological evidence, stellar accretion rates themselves provide key signatures of late infall. In the Lupus star-forming region, with an age of $\sim 2$ Myr, \citet{Winter+2024a} found that disks located in denser environments exhibit systematically higher accretion rates. This trend is difficult to reconcile with the conventional scenario of steadily declining accretion once the envelope has dispersed. Instead, a natural explanation posits that ambient interstellar gas continues to feed the disks, with the accretion rate scaling with the local gas density. This observational trend thus constitutes strong indirect evidence that late infall can penetrate to the inner disk and fuel stellar accretion. Theoretical models reinforce this view: Bondi–Hoyle–Lyttleton accretion predicts a strong dependence on ambient density and relative velocity, consistent with the observed scatter in accretion rates \citep{Padoan_2005}, while more recent models show that late infall can decisively shape the long-term accretion histories of class II stars \citep[e.g.,][]{Winter+2024b}. These findings suggest that late infall may not be a marginal process but a fundamental factor in regulating disk and stellar evolution.

Nevertheless, a critical uncertainty remains: whether late-infalling material can actually survive competition with dispersal processes such as photoevaporation \citep[e.g.,][]{Shu_1993,Hollenbach1994} and MHD disk winds \citep[e.g.,][]{Blandford_1982,Suzuki_2009,Bai+2013,Tabone+2022}. If infalling gas is dissipated before reaching the inner disk, its contribution to stellar accretion may be negligible.
At present, the physical conditions that allow late infall to reach the star remain poorly constrained. Addressing this requires models that simultaneously incorporate both infall and dispersal, while following disk evolution over megayear (Myr) timescales.

A major challenge in modeling late infall lies in its physical origin; in particular, how mass is supplied to protoplanetary disks. Insights can be drawn from studies of the early stages of disk formation, where disks accrete material from their parent protostellar envelopes. A key quantity in this process is the angular momentum of the infalling gas. If the disk is approximately Keplerian, gas with sub-Keplerian rotation must exchange angular momentum with the disk to settle. This mismatch exerts an effective torque that can substantially modify disk evolution \citep[e.g.,][]{Cassen+1981,Cassen+1983,Nakamoto+1994,Jin+2010,Liu+2017,Huhn+2025}.

In this study, we employed 1D models of protoplanetary disk evolution that include photoevaporation, MHD disk winds, and late infall. To separate the effect of the infalling gas as a mass reservoir from its effect as a source of torque, we performed calculations both without and with the effective torque exerted by the infalling gas. Focusing on the Lupus star-forming region, we quantified the extent to which late infall can sustain stellar accretion and identify the physical conditions under which this is possible without violating observational constraints on disk masses and radii.

The remainder of this paper is organized as follows. Section~\ref{sec:method} describes the numerical methods and model setup. Our results for the models without and with the effective torque exerted by the infalling gas are presented in Sections~\ref{Result} and~\ref{Result:Sub-Kep}, respectively. We discuss the implications of our findings in Section~\ref{Discussion} and summarize our conclusions in Section~\ref{Conclusion}.

\section{Method}
\label{sec:method}

This section describes the methodology of our study. We begin in Section~\ref{method:evolution} with the governing equations for the evolution of the gas surface density and proceed in Section~\ref{Method:late infall} and \ref{method:ic} to present the late-infall model and initial condition. Section~\ref{method:parameter} then outlines the parameter space explored and Section~\ref{method:disksize} provides the definition of disk size adopted for our comparisons with observations.

\subsection{Time evolution of gas surface density}
\label{method:evolution}
In this study, we assumed an axisymmetric disk and followed the time evolution of the surface density using 1D numerical simulations \citep[e.g.,][]{Lynden-Bell+1974, Clarke+2001, Suzuki+2016,Kunitomo+2020,Komaki+2023,Weder+2023}. 
Our methodology largely follows that of \citet{Ooyama+2025}, who incorporated MHD disk winds and photoevaporation. Here, we extended their model by including late infall of interstellar gas and far-ultraviolet (FUV) driven photoevaporation. We briefly outline the overall approach below and refer to \cite{Ooyama+2025} for a more detailed description. Unless otherwise specified, we focused on the disks around $0.2M_{\odot}$ central stars, 
representative of the dominant low-mass stellar population in Lupus \citep{Winter+2024a}.
In all of our models, the stellar mass was kept fixed in time. Since $\dot{M}{\rm acc}\leq 5\times10^{-8}\,\msunyr$ in most cases, this approximation is broadly justified, except for the most extreme cases considered in this study. The basic equation is
\begin{align}
\frac{\partial \Sigma}{\partial t}
&= \frac{1}{r} \frac{\partial}{\partial r}\Bigg[
\frac{2}{r\Omega} \Bigg\{
\frac{\partial}{\partial r} \left( r^{2} \Sigma \alpha_{r\phi}c_{s}^{2} \right)
+ r^{2}\left( \rho c_{s}^{2} \right)_{\rm mid} \alpha_{\phi z}
\Bigg\} \Bigg] \nonumber\\
&\quad {}
- C_\mathrm{w} \left( \rho c_{s} \right)_{\rm mid}
- \dot{\Sigma}_\mathrm{pw}
+ \dot{\Sigma}_{\rm infall},
\label{eq:basic_eq1}
\end{align}
where $\Sigma$, $\Omega$, $c_{s}$, and $\rho$ denote the gas surface density, the Keplerian angular velocity, the sound speed, and the gas density, respectively. 
The label "mid" in the subscript indicates that the quantity refers to the midplane value. The coefficients $\alpha_{r\phi}$, $\alpha_{\phi z}$\footnote{While \cite{Suzuki+2016} and \cite{Ooyama+2025} expressed $\alpha_{r\phi}$ and $\alpha_{\phi z}$ as $\overline{\alpha_{r\phi}}$ and $\overline{\alpha_{\phi z}}$, we omit the bars in this paper for simplicity.}, and $C_\mathrm{w}$ represent the strengths of the turbulent viscosity, the MHD disk wind torque, and the mass-loss rate due to the MHD disk wind, respectively. The source terms $\dot{\Sigma}_\mathrm{pw}$ and $\dot{\Sigma}_{\rm infall}$ indicate the mass loss rate of photoevaporation and the mass supply rate due to the late infall per unit area, respectively.
Eq.~\eqref{eq:basic_eq1} includes the effect of late infall only as a source of mass supply and is used for the models without the effective torque (Section~\ref{Result}). The corresponding equation that includes the effective torque is introduced later in Section~\ref{Method:Sub-Keplerian}.
We calculated the midplane temperature to get the sound speed, $c_s$, as described in Section~\ref{ssec:T_disk}.

We considered the following two cases for the turbulent viscosity parameter, $\alpha_{r\phi}$, 
to clarify how late infall influences disk evolution under different angular momentum redistribution regimes: 
$\alpha_{r\phi}=8\times10^{-3}$ for strong turbulent viscosity models and $\alpha_{r\phi}=8\times10^{-5}$ for weak turbulent viscosity models.

Regarding the magnetic braking torque, we considered both spatially varying and constant $\alpha_{\phi z}$.
In the fiducial models, we adopted the spatially varying, $\Sigma$-dependent model of \cite{Suzuki+2016} via
\begin{equation}
    \alpha_{\phi z} (r,t) 
    = \mathrm{min} \left[10^{-5}\left( \frac{\Sigma}{\Sigma_{\rm int}}\right)^{-0.66},1 \right] ,
    \label{alpha_pz_fiducial}
\end{equation}
where $\Sigma_{\rm int}$ is the initial surface density (see Eq.~\eqref{eq:sgm_ini} later). In the constant $\alpha_{\phi z}$ models, we adopted $10^{-3}$ and $10^{-4}$. These models tend to yield enhanced torques compared to Eq.~\eqref{alpha_pz_fiducial} and thus represent the efficient angular momentum removal by disk winds.

We determined the mass-loss rate due to the MHD disk wind based on an energy balance: the energy released by accretion and viscous heating was assumed to be carried away by disk winds and radiation. The coefficient, $C_\mathrm{w}$, is then written as
\begin{equation}
    C_{\mathrm{w}}=\mathrm{min}\Bigg[ C_{\mathrm{w, 0}},  (1-\epsilon_{\mathrm{rad}})\left( \frac{3\sqrt{2\pi}c_{s}^{2}}{r^{2}\Omega^{2}}\alpha_{r\phi} + \frac{2c_{s}}{r\Omega}\alpha_{\phi z} \right) \Bigg] ,
    \label{eq:Cwe_WDW}
\end{equation}
where $C_{\mathrm{w, 0}}$ and $\epsilon_{\mathrm{rad}}$ are the upper limit of $C_{\rm w}$ and the fraction of energy allocated to radiation, respectively. We adopted the "weak disk wind" model considered in \cite{Suzuki+2016}, which assumes that 10\% of the energy is carried away by the disk wind and 90\% by radiation, namely, $\epsilon_{\mathrm{rad}} = 0.9$. 
We assumed $C_{\mathrm{w, 0}} = 2 \times 10^{-5}$ and $1 \times 10^{-5}$ for models with $\alpha_{r\phi}=8\times10^{-3}$ and $8\times10^{-5}$, respectively.

\subsection{Disk midplane temperature}
\label{ssec:T_disk}

We calculated the disk midplane temperature, $T$, by considering stellar irradiation, viscous heating, and dust cooling, following \citet{Nakamoto+1994}. The temperature, $T$, is given by 
\begin{equation}
    T^4 = T_{\rm irr}^4 + T_{\rm vis}^4,
\end{equation}
where $T_{\rm irr}$ and $T_{\rm vis}$ are the equilibrium temperatures corresponding to stellar irradiation and viscous heating, respectively, both balanced by dust cooling. The explicit expressions for each term are also given in Section~2.4 of \citet{Ooyama+2025}.

\subsection{Photoevaporation}
\label{method:photoevaporation}

In this section, we focus on photoevaporation driven by the central star. In Lupus, the external FUV field is not strong enough to drive significant mass loss \citep{Winter+2018}, so the external photoevaporation is negligible. We therefore considered only the contributions from stellar FUV, extreme UV (EUV), and X-ray radiation. The total mass loss rate, $\dot{\Sigma}_{\rm pw}$, is written as the sum of the rates caused by radiation at these different wavelengths,  
\begin{equation}
    \dot{\Sigma}_{\rm pw}=\dot{\Sigma}_{\rm pw, FUV}+\dot{\Sigma}_{\rm pw, EUV}+\dot{\Sigma}_{\rm pw, X}.
    \label{Photo}
\end{equation}
We calculated the photoevaporation rate per unit area due to EUV and X-ray radiation using the same model as in \cite{Ooyama+2025}. For the X-ray radiation, we adopted one-tenth of the rate given by the primordial disk model of \cite{Owen+2012}, following \cite{Ooyama+2025}, to correct for the overestimation caused by an underestimated cooling rate \citep[see][]{Sellek+2024}. Unlike \cite{Ooyama+2025}, we did not consider the effects of pre-main sequence stellar evolution and assumed constant irradiation. This approach makes it possible to identify the crucial physical factors that allow the infalling gas to reach the central region. The dependence on the strength of photoevaporation is discussed in Appendix~\ref{parameter study in fiducial}.

We adopted, as the fiducial radiation field, the case of a $0.2~M_{\odot}$ star at 1 Myr. This choice is motivated by our focus on disk evolution up to 2 Myr, the typical age of the Lupus star-forming region \citep{Deng_2025}, with 1 Myr representing an intermediate point in this timespan. The change in bolometric luminosity between 1 Myr and 2 Myr is within a factor of a few, and the resulting variation in UV and X-ray emission is expected to fall within the parameter range examined in Appendix~\ref{parameter study in fiducial}. The adopted bolometric luminosity \citep{Siess+2000}, FUV luminosity, EUV photon luminosity, and X-ray luminosity are expressed as
\begin{equation}
     L_{*}=1.22\times 10^{33} \unit{erg}\unit{s^{-1}},
\end{equation}
\begin{align}
    L_{\rm FUV}&=6.11\times10^{29}\unit{erg}\unit{s^{-1}},\notag\\
    \Phi_{\rm EUV}&=6.01\times10^{39} \unit{s^{-1}},\notag\\
    L_{\rm X}&=2.3\times10^{29} \unit{erg}\unit{s^{-1}}.
    \label{lum_matome}
\end{align}
We adopted the model of \citet{Gorti_2009} for the FUV and X-ray luminosities. These results were computed from the bolometric luminosity, considering only the EUV emission originating from magnetic activity, for which we used the model of \citet{Shoda_2021}. There are, however, significant uncertainties, particularly in the contributions from accretion-driven luminosity and stellar evolution. To assess the possible impact of this uncertainty on our results, we additionally explored cases with luminosities set to 0.1 and 10 times the fiducial values in Appendix~\ref{parameter study in fiducial}.

Photoevaporation rate by EUV radiation per unit area is given by
\begin{equation}
    \dot{\Sigma}_{\mathrm{pw, EUV}} (r)=
    \begin{cases}
    2m_{\mathrm{H}} c_{s, \mathrm{H II}}(T_{\mathrm{H II}}) n_{0}(r) &\text{if } r\geq r_{\mathrm{crit}} ,\\
    0 &\text{if } r<r_{\mathrm{crit}},
    \end{cases}
    \label{eq:dSigma_photo_EUV_Tanaka}
\end{equation}
where $m_{\mathrm{H}}$ is the hydrogen atom mass, $c_{s, \mathrm{H II}}$ the sound speed, and $n_{0}(r)$ the base density \citep{Hollenbach1994}. The critical radius is $r_{\mathrm{crit}}=\beta r_{\mathrm{g}}$ with $\beta=0.14$ \citep{Weder+2023, Liffman_2003}. For $n_{0}(r)$ we used \citet{Tanaka+2013}:
\begin{equation}
    n_{0}(r)=1.6 \times 10^{7} \left( \frac{\Phi_\mathrm{EUV}}{10^{49} \mathrm{s}^{-1}} \right)^{1/2}\left( \frac{r}{10^{15}\mathrm{cm}} \right)^{-3/2} \mathrm{cm^{-3}} ,
\end{equation}
where $\Phi_\mathrm{EUV}$ is the EUV photon emissivity. For further details, we refer to \citet{Ooyama+2025}. The photoevaporation rate by FUV radiation per unit area is given by
\begin{multline}
    \dot{\Sigma}_{\rm pw, FUV}(r)=\\
    \begin{cases}
        0 & \text{if } r < 4 \unit{au}  \left(\dfrac{M_{*}}{M_{\odot}}\right) ,\\
        \dot{\Sigma}_{\rm F0}\left(\dfrac{L_{\rm FUV}}{10^{31.7}\unit{erg}\unit{s}^{-1}}\right)\left(\dfrac{r}{4\unit{au}}\right)^{-2} & \text{if } r \geq 4 \unit{au} \left(\dfrac{M_{*}}{M_{\odot}}\right),
    \end{cases}
    \label{eq:FUV}
\end{multline}
where $M_*$ represents the mass of the central star \citep{Kunitomo+2021}. The base FUV surface mass loss rate was set to $\dot{\Sigma}_{\rm F0} = 10^{-12} \mathrm{g \:cm^{-2}\:s^{-1}}$ \citep{Kunitomo+2021}.

\subsection{Late infall}
\label{Method:late infall}

To separate the role of the infalling gas as a mass reservoir from
its role as a source of angular-momentum exchange, we considered two
classes of models: models without and with the effective torque
exerted by the infalling gas. We can characterize the specific angular
momentum of the infalling gas as
\begin{equation}
    j_{\rm infall}(r) = f_{\rm Kepler} j_{\rm Kepler}(r),
\end{equation}
where $j_{\rm Kepler}$ is the local Keplerian specific angular
momentum and $0 < f_{\rm Kepler} \leq 1$. The case
$f_{\rm Kepler}=1$ corresponds to gas supplied with the local
Keplerian angular momentum and, thus, it produces no effective
torque; while $f_{\rm Kepler}<1$ represents sub-Keplerian infall.
The corresponding methods are described in Sections~\ref{Method:Keplerian} and \ref{Method:Sub-Keplerian}, respectively.

\subsubsection{Models with only mass supply}
\label{Method:Keplerian}

In the mass supply-only models, we set $f_{\rm Kepler}=1$. In this limit, the infalling gas carries the local Keplerian specific angular momentum, so that no angular-momentum mismatch is produced. The evolution is therefore described by Eq.~\eqref{eq:basic_eq1}, where late infall enters only through the mass source term,
$\dot{\Sigma}_{\rm infall}$.

For simplicity, the mass supply rate from late infall, $\dot{\Sigma}_{\rm infall}$, was assumed to be constant within the infalling zone. This allows us to simplify the model and identify which mechanisms in disk evolution are critical in determining whether the infalling mass can accrete onto the central star. It is given by 
\begin{equation}
    \dot{\Sigma}_{\rm infall}(r)=
    \begin{cases}
        \Sigma_0 \equiv\dfrac{\mdoti}{\pi r_{\rm infall}^2} & \text{for } r \leq r_{\rm infall} ,\\
        0 & \text{for } r > r_{\rm infall} ,
    \end{cases}
    \label{eq:infall}
\end{equation}
where $r$, $r_{\rm infall}$, and $\mdoti$ represent the distance from the central star, the radial extent of the interstellar gas infall region, and the interstellar gas infall rate, respectively. In our late infall model without the effective torque, most of the mass is deposited around a characteristic infall radius, $r_{\rm infall}$. Based on observational detections of late infall structures \citep[e.g.,][]{Speedie+2025}, we adopted a fiducial value of $r_{\rm infall} = 250~\mathrm{au}$.

In this model, we assumed that the infalling gas settles onto the disk without disrupting it throughout the entire infall region considered. Although the ram pressure of the infalling gas at $250~{\rm au}$ exceeds the thermal pressure of the disk gas in our initial condition, the disk may still avoid substantial disruption if the supersonic late-infall flow strikes its upper and lower surfaces approximately symmetrically, forms a shock near the midplane, and dissipates its kinetic energy there. Such a configuration is plausible when the Bondi--Hoyle--Lyttleton radius is much larger than $r_{\rm infall}$,
\begin{equation}
r_{\rm BHL}
=
\frac{2GM_*}{c_s^2+v_{\rm rel}^2}
\sim 1400\,{\rm au}
\left(\frac{M_*}{0.2M_\odot}\right)
\left[
\frac{(c_s^2+v_{\rm rel}^2)^{1/2}}
{0.5\,{\rm km\,s^{-1}}}
\right]^{-2},
\end{equation}
where $v_{\rm rel}$ and $c_s$ are relative velocity between the star and the ambient gas, and sound speed, 
respectively.
For the fiducial parameters considered here, $r_{\rm BHL}$ is substantially larger than $r_{\rm infall}=250\,{\rm au}$. When $v_{\rm rel}$ is sufficiently small for this condition to hold, the ambient gas is gravitationally captured well outside the adopted infall region and can approach the disk in an approximately spherical flow. During its subsequent inward motion, the gas is accelerated by gravity and can reach supersonic velocities before encountering the disk. We additionally assumed that the gas encounters the disk approximately face-on, so that it accretes onto the upper and lower disk surfaces.

For $v_{\rm rel}\gtrsim1\,{\rm km\,s^{-1}}$, the incoming flow can still be supersonic; however, $r_{\rm BHL}$ becomes comparable to or smaller than $r_{\rm infall}$. In this regime, the assumption of approximately symmetric infall onto the upper and lower disk surfaces becomes less secure, and the interaction between the infalling gas and the outer disk may depend strongly on the direction and angular momentum of the incoming flow. Modeling these effects would require multidimensional calculations, so we have left them for a future work. The radial extent of the infall region is also uncertain because it depends on the angular momentum of the infalling gas. To investigate how such variations potentially influence the accretion rate, we ran additional models in which the infall region is set to $r < 150~\unit{au}$, $r < 50~\unit{au}$, and $225~\unit{au} < r < 250~\unit{au}$ (Appendix~\ref{ssec:apdx_rinfall}).

In the fiducial setting, we assumed a constant infall rate that continues throughout the simulation. However, a disk may leave a dense region of the interstellar medium (ISM) during its evolution, thereby shortening the infall phase. To account for this possibility, we also considered cases with limited infall durations of the initial $10^{5} \unit{yr}$ and $10^{6} \unit{yr}$. The impact of varying the infall duration is investigated separately in Appendix~\ref{ssec:apdx_tinfall}.

We treated the infall rate, $\mdoti$, as a free parameter, varying it over a range of $5\times10^{-10}$ to $5\times10^{-7}~\msunyr$. This range is motivated by estimates of Bondi–Hoyle–Lyttleton accretion rates expected in molecular cloud cores \citep{Winter+2024b}, which can be written as
\begin{equation}
\begin{aligned}
    \dot{M}_{\rm BHL}
    &=
    \frac{4\pi G^2 M_*^2 \rho_{\rm amb}}
    {\left(v_{\rm rel}^2+c_s^2\right)^{3/2}} \\
    &\simeq
    5\times10^{-8}
    \left(\frac{M_*}{0.2~\Msun}\right)^2
    \left(\frac{n_{\rm amb}}{10^4~{\rm cm}^{-3}}\right) \\
    &\quad \times
    \left(
    \frac{(v_{\rm rel}^2+c_s^2)^{1/2}}
    {0.5~{\rm km~s}^{-1}}
    \right)^{-3}
    \msunyr ,
\end{aligned}
\end{equation}
where $\rho_{\rm amb}$ and $n_{\rm amb}$ are the ambient gas density, and ambient number density, respectively.

\subsubsection{Models with mass supply and effective torque}
\label{Method:Sub-Keplerian}

Next, we considered the case with $f_{\rm Kepler}<1$, where the
infalling gas carries sub-Keplerian specific angular momentum.
Unlike the $f_{\rm Kepler}=1$ case described in Section~2.4.1,
the infalling gas then has an angular-momentum mismatch with
the Keplerian disk gas. If the infalling gas exchanges angular
momentum with the disk after deposition, this mismatch drives
radial mass transport. We included this effect as an effective
torque. In the torque-included models, we considered
$f_{\rm Kepler}=0.3$, 0.4, and 0.9. The limiting case
$f_{\rm Kepler}=1$ is reduced to the mass-supply-only model
described in Section~2.4.1.

Using the parameterization introduced above, the specific angular
momentum of the infalling gas is written as
\begin{equation}
    j_{\rm infall}(r)= f_{\rm Kepler}j_{\rm Kepler}(r),
    \label{eq:jinfall_fkep_temp}
\end{equation}
where $j_{\rm Kepler}=r^2\Omega$ is the local Keplerian specific angular momentum. To derive the effective torque associated with this angular-momentum mismatch, we first isolate the effect of infall. The viscous, MHD-wind, and photoevaporative terms are omitted in the following local derivation and are restored when we write the full surface-density evolution equation at the end of this subsection. The angular-momentum and mass conservation equations associated with infall are
\begin{equation}
    \frac{\partial}{\partial t}
    \left(\Sigma j_{\rm disk}\right)+\frac{1}{r}\frac{\partial}{\partial r}\left(r\Sigma v_r j_{\rm disk}\right)=\dot{\Sigma}_{\rm infall} j_{\rm infall},
    \label{eq:infall_angmom_cons_temp}
\end{equation}
and
\begin{equation}
    \frac{\partial \Sigma}{\partial t} + \frac{1}{r}\frac{\partial}{\partial r}\left(r\Sigma v_r\right) = \dot{\Sigma}_{\rm infall},
    \label{eq:infall_mass_cons_temp}
\end{equation}
where $v_r$, $j_{\rm disk}$, and
$j_{\rm infall}$ are the radial velocity of the disk gas, the
specific angular momentum of the disk gas, and that of the infalling gas, respectively. Combining Eqs.~\eqref{eq:infall_angmom_cons_temp} and \eqref{eq:infall_mass_cons_temp}, we obtain
\begin{equation}
    \Sigma\frac{\partial j_{\rm disk}}{\partial t}+ \Sigma v_r \frac{\partial j_{\rm disk}}{\partial r} = \dot{\Sigma}_{\rm infall}\left(j_{\rm infall} - j_{\rm disk}\right).
    \label{eq:jdisk_evolution_temp}
\end{equation}

Equation~\eqref{eq:jdisk_evolution_temp} shows that the infalling gas changes the angular momentum of the local disk gas when $j_{\rm infall}\neq j_{\rm disk}$. 
We note that, if $L \equiv \Sigma j_{\rm disk}$ denotes the angular-momentum surface density, its Lagrangian derivative satisfies
$\mathrm{D}L/\mathrm{D}t = \Sigma \mathrm{D}j_{\rm disk}/\mathrm{D}t + j_{\rm disk} \mathrm{D}\Sigma/\mathrm{D}t.$
Thus, the left-hand side of Equation~(18) is $\Sigma({\rm D}j_{\rm disk}/{\rm D}t) = {\rm D}L/{\rm D}t - j_{\rm disk}{\rm D}\Sigma/{\rm D}t$, while the right-hand side represents the torque density exerted by the infalling gas.
A simple local torque formula can then be obtained if the disk gas remains close to Keplerian rotation, $j_{\rm disk}\simeq j_{\rm Kepler}$. 
This approximation is appropriate only where the pre-existing disk gas dominates the local angular-momentum budget, so that the infalling gas acts as a perturbation. In gas-poor regions, by contrast, the angular momentum of the local gas can be strongly modified by the infalling gas itself, and the approximation $j_{\rm disk}\simeq j_{\rm Kepler}$ is no longer justified. Therefore, we treated the gas-rich and gas-poor regions separately.

We define the gas-rich, or high-$\Sigma$, region using the local infall timescale,
\begin{equation}
    t_{\rm infall}=\frac{\Sigma}{\dot{\Sigma}_0},
    \label{eq:tinfall_temp}
\end{equation}
where $\dot{\Sigma}_0$
is the uniform mass-supply rate defined in Eq.~\eqref{eq:infall}. 
If $t_{\rm infall}$ exceeds the advection timescale, the disk gas can adjust before the infalling gas significantly modifies its angular momentum, and the approximation of $j_{\rm disk} \simeq j_{\rm Kepler}$ remains valid (cf. Eq.~\eqref{eq:jdisk_evolution_temp}). Since the relevant advection timescale is not known a priori and depends on the dominant angular-momentum transport mechanism, we
introduce a dimensionless constant, $C$, and define the
high-$\Sigma$ region by
\begin{equation}
    t_{\rm infall}\geq C\Omega^{-1}.
    \label{eq:high_sigma_condition_temp}
\end{equation}
We denote the outer edge of this region by $r_{\rm tq}$. We adopted
$C=20$ as the fiducial value. Although the choice of $C$ is not
unique, the resulting disk evolution is insensitive to this parameter
because $r_{\rm tq}$ lies near the steeply declining outer edge of
the surface-density profile. We have confirmed that our results remain
essentially unchanged for $10\leq C\leq 10^6$.

In the high-$\Sigma$ region, we set $j_{\rm disk}\simeq j_{\rm Kepler}$. We did not consider the evolution of the central stellar mass, meaning that $j_{\rm Kepler}$ is time-independent at each radius, and Eq.~\eqref{eq:jdisk_evolution_temp} reduces to
\begin{equation}
    \Sigma v_r
    \frac{\partial j_{\rm Kepler}}{\partial r} = \dot{\Sigma}_{\rm infall}^{\rm loc}\left(j_{\rm infall}-j_{\rm Kepler}\right),
    \label{eq:local_mixing_temp}
\end{equation}
where $\dot{\Sigma}_{\rm infall}^{\rm loc}$ is the infall rate in the high-$\Sigma$ regime, namely the component of the infalling gas deposited locally. Using this relation, the advection term in the mass conservation Eq.~\eqref{eq:infall_mass_cons_temp} becomes
\begin{equation}
    \label{Torque_term1_temp}
    -\frac{1}{r}\frac{\partial}{\partial r}(r\Sigma v_r )=-\frac{1}{r}\frac{\partial}{\partial r}\left[r\left(\frac{1}{\partial j_{\rm Kepler}/\partial r}\right)\dot{\Sigma}_{\rm infall}^{\rm loc}(j_{\rm infall}-j_{\rm Kepler})\right].
\end{equation}
For $\dot{\Sigma}_{\rm infall}^{\rm loc}$, we replaced the step-function profile in Eq.~\eqref{eq:infall} with a smooth hyperbolic function,
\begin{equation}
    \dot{\Sigma}_{\rm infall}^{\rm loc}=\frac{\dot{\Sigma}_0}{1+\exp\left[(r-r_{\rm tq})/\Delta\right]}.
    \label{eq:infall_rate:high-sigma_temp}
\end{equation}
This was done to avoid the divergence in the derivative of Eq.~\eqref{Torque_term1_temp}. We adopted $\Delta=10\,{\rm au}$ in the calculation below. Substituting Eqs.~\eqref{eq:jinfall_fkep_temp}, \eqref{Torque_term1_temp},  and $\partial j_{\rm Kepler}/\partial r=j_{\rm Kepler}/(2r)$ into Eq.~\eqref{eq:infall_mass_cons_temp}, we obtained the surface-density evolution equation in the high-$\Sigma$ regime,
\begin{equation}
    \frac{\partial \Sigma}{\partial t}=
    \dot{\Sigma}_{\rm infall}^{\rm loc}+S_{\rm tq},
    \label{eq:infall_local_source_temp}
\end{equation}
where
\begin{equation}
    S_{\rm tq}=4\left(1-f_{\rm Kepler}\right)\dot{\Sigma}_{\rm infall}^{\rm loc} + 2r\left(1-f_{\rm Kepler}\right)\frac{\partial \dot{\Sigma}_{\rm infall}^{\rm loc}}{\partial r}.
    \label{eq:effective_torque_source_temp}
\end{equation}
The term $S_{\rm tq}$ represents the effective radial mass
transport induced by the angular-momentum mismatch between
the disk gas and the infalling gas. It vanishes in the limit
$f_{\rm Kepler}=1$.

Next, we considered the gas-poor (or low-$\Sigma$) region where $t_{\rm infall}<C\Omega^{-1}$, which typically corresponds to $r_{\rm tq}<r<r_{\rm infall}$. In this regime, applying Eq.~\eqref{eq:local_mixing_temp} would force the local gas to remain Keplerian while offsetting, through radial advection, the angular momentum sink caused by the lower-angular-momentum infalling gas. When $\Sigma$ is very small, the required mass transport can exceed the mass actually available in the annulus, leading to an unphysical behavior. \footnote{
From Eq.~\eqref{eq:local_mixing_temp}, the radial velocity required by the local-mixing prescription is $v_r = -2(1-f_{\rm Kepler})r\dot{\Sigma}_{\rm infall}^{\rm loc}/\Sigma$. Thus, for a finite local infall rate, the required inward velocity becomes arbitrarily large as $\Sigma$ decreases. The corresponding term $S_{\rm tq}$ in Eq.~\eqref{eq:effective_torque_source_temp} is a flux-divergence term: it should redistribute mass between neighbouring annuli rather than create mass. In the low-$\Sigma$ regime, however, the negative part of this redistribution can formally remove more mass than is present locally. If the numerical calculation then prevents negative surface densities by imposing a floor, the removed mass is effectively supplied back by the floor, while the positive part of the redistributed flux remains. This is equivalent to introducing an artificial mass source and can make the total disk mass increase by more than the imposed infall rate. This behavior reflects the breakdown of the Keplerian local-mixing closure, not a physical mass source.
The underlying cause is that, after assuming $j_{\rm disk} \simeq j_{\rm Kepler}$ and fixing the stellar mass, Equation~(18) is solved with $\partial j_{\rm disk}/\partial t = 0$. This amounts to seeking a local quasi-steady Keplerian solution in which the angular-momentum deficit of the infalling gas is balanced entirely by radial advection. 
This is likely not realized in the gas-poor outermost region in practice.
} Therefore, we did not apply the local-mixing torque prescription in the low-$\Sigma$ region. Instead, as a simple closure, we assumed that gas arriving in the low-$\Sigma$ region rapidly settles at its centrifugal radius,
\begin{equation}
    r_{\rm c}=f_{\rm Kepler}^2 r.
    \label{eq:centrifugal_radius_temp}
\end{equation}
Thus, material that would otherwise be deposited in $r_{\rm tq}<r<r_{\rm infall}$ is redistributed as mass supply to $f_{\rm Kepler}^2r_{\rm tq}<r<f_{\rm Kepler}^2r_{\rm infall}$. The corresponding source term is
\begin{equation}
    \dot{\Sigma}_{\rm infall}^{\rm trs}=\frac{\dot{\Sigma}_0}{f_{\rm Kepler}^4}\left[1+\exp\left(\frac{r/f_{\rm Kepler}^2-r_{\rm infall}}{\Delta}\right)\right]^{-1}, 
    \label{eq:sigmainfall_trs}
\end{equation}
which is applied for $f_{\rm Kepler}^2 r_{\rm tq}<r<f_{\rm Kepler}^2 r_{\rm infall}$, and we set $\dot{\Sigma}_{\rm infall}^{\rm trs}=0$ otherwise.
The factor $f_{\rm Kepler}^{-4}$ accounts for the change in annular area under the mapping $r\rightarrow f_{\rm Kepler}^2r$. With this prescription, no mass or angular momentum is deposited directly in the original low-$\Sigma$ region. This treatment is not unique, but it provides a mass-conserving way to represent the settling of sub-Keplerian infall within the current 1D Keplerian framework.

Thus, the total infall source term used in the torque-included models is
\begin{equation}
    \dot{\Sigma}_{\rm infall}=\dot{\Sigma}_{\rm infall}^{\rm loc}+\dot{\Sigma}_{\rm infall}^{\rm trs}.
    \label{eq:sigmainfall_total}
\end{equation}
The effective torque term, $S_{\rm tq}$, is applied only to the locally deposited component, $\dot{\Sigma}_{\rm infall}^{\rm loc}$, for which the local-mixing approximation is valid. Adding $S_{\rm tq}$ to Eq.~\eqref{eq:basic_eq1}, the surface-density evolution equation for the torque-included models becomes
\begin{equation}
\begin{split}
    \frac{\partial \Sigma}{\partial t}=&\frac{1}{r}\frac{\partial}{\partial r}\left[\frac{2}{r\Omega}\left\{\frac{\partial}{\partial r}\left(r^2\Sigma\alpha_{r\phi}c_s^2\right)+r^2\left(\rho c_s^2\right)_{\rm mid}\alpha_{\phi z}\right\}\right]\\
    &-C_{\rm w}\left(\rho c_s\right)_{\rm mid}-\dot{\Sigma}_{\rm pw}+\dot{\Sigma}_{\rm infall}+S_{\rm tq}.
\end{split}
\label{eq:basic_eq_torque}
\end{equation}
This equation is used in the models where late infall is included as both a mass source and a source of angular-momentum exchange between the infalling gas and the disk gas.

\begin{table*}
    \centering
    \caption{Model input parameters and results}
    \centering
    \begin{tabular}{ c c c c c  c c | c c|c} 
         Model parameters&&&&&&&Results&&
        \\
        $\alpha_{r\phi}$&$\alpha_{\phi z}$& Infall Period & Infall Region & UV/X-ray & $f_{\rm Kepler}$ & &  Mass & Radius &Reference \\
        \hline
        $8\times10^{-3}$& Eq.~\eqref{alpha_pz_fiducial} & $\infty$ & $\leq250\unit{au}$ & Eq.~\eqref{lum_matome} &1&&$\checkmark$&&Sect.~\ref{Result:fiducial} \\
        
        $8\times10^{-5}$& Eq.~\eqref{alpha_pz_fiducial} & $\infty$ & $\leq250\unit{au}$ & Eq.~\eqref{lum_matome}&1&& &$\checkmark$&Sect.~\ref{Result:fiducial} \\

        $8\times10^{-5}$ & $10^{-3}$& $\infty$ & $\leq250\unit{au}$ & Eq.~\eqref{lum_matome}&1&&$\checkmark$&$\checkmark$&Sect.~\ref{Results:constant_alpha} \\

        $8\times10^{-5}$& $10^{-4}$ & $\infty$ & $\leq250\unit{au}$ & Eq.~\eqref{lum_matome}&1& &$\checkmark$&&Sect.~\ref{Results:constant_alpha} \\

         $8\times10^{-3}$&  Eq.~\eqref{alpha_pz_fiducial} & $\infty$ &  $\leq250\unit{au}$ &  Eq.~\eqref{lum_matome} &  0.3--0.9&&  $\checkmark$& $\checkmark^{a}$& Sect.~\ref{Sub-Kep_StrongVis} \\

         $8\times10^{-5}$& Eq.~\eqref{alpha_pz_fiducial} &  $\infty$ &  $\leq250\unit{au}$ &  Eq.~\eqref{lum_matome}&  0.3--0.9&&  $\checkmark^{b}$&  $\checkmark$& Sect.~\ref{Sub-Kep_WeakVis} \\
        \hline

        $8\times10^{-5}$& $10^{-3}$ & $\infty$ & $\leq150\unit{au}$ & Eq.~\eqref{lum_matome}&1&&$\checkmark$&$\checkmark$&Appendix~\ref{ssec:apdx_rinfall} \\

        $8\times10^{-5}$& $10^{-3}$ & $\infty$ & $\leq50\unit{au}$ & Eq.~\eqref{lum_matome}&1&&$\checkmark$&$\checkmark$&Appendix~\ref{ssec:apdx_rinfall} \\
        
        $8\times10^{-5}$ & $10^{-3}$& $10^{6} \unit{yr}$ & $\leq250\unit{au}$ & Eq.~\eqref{lum_matome}&1&&$\checkmark$&$\checkmark$ &Appendix~\ref{ssec:apdx_tinfall}\\

        $8\times10^{-5}$ & $10^{-3}$& $10^{5} \unit{yr}$ & $\leq250\unit{au}$ & Eq.~\eqref{lum_matome}&1&&$\checkmark$&$\checkmark$ &Appendix~\ref{ssec:apdx_tinfall}\\

        $8\times10^{-5}$ & $10^{-3}$ & $\infty$ & $\leq250\unit{au}$ & Eq.~\eqref{lum_matome}$\times0.1$&1&&$\checkmark$&$\checkmark$&Appendix~\ref{ssec:apdx_pevrates}\\
        
        $8\times10^{-5}$  & $10^{-3}$& $\infty$ & $\leq250\unit{au}$ & Eq.~\eqref{lum_matome}$\times10$&1&&&$\checkmark$&Appendix~\ref{ssec:apdx_pevrates}\\

        $8\times10^{-3}$ & Eq.~\eqref{alpha_pz_fiducial}& $\infty$ & $\leq150\unit{au}$ & Eq.~\eqref{lum_matome}&1&&$\checkmark$& &Appendix~\ref{ssec:apdx_rinfall}\\

        $8\times10^{-3}$ &Eq.~\eqref{alpha_pz_fiducial} & $\infty$ & $\leq50\unit{au}$ & Eq.~\eqref{lum_matome}&1&&$\checkmark$&&Appendix~\ref{ssec:apdx_rinfall} \\

        $8\times10^{-3}$& Eq.~\eqref{alpha_pz_fiducial} & $\infty$ & $225$--$250\unit{au}$ & Eq.~\eqref{lum_matome}&1&&$\checkmark$& &Appendix~\ref{ssec:apdx_rinfall}\\
        
        $8\times10^{-3}$ & Eq.~\eqref{alpha_pz_fiducial}& $10^{6} \unit{yr}$ & $\leq250\unit{au}$ & Eq.~\eqref{lum_matome}&1&&$\checkmark$& &Appendix~\ref{ssec:apdx_tinfall}\\

        $8\times10^{-3}$& Eq.~\eqref{alpha_pz_fiducial} & $10^{5} \unit{yr}$ & $\leq250\unit{au}$ & Eq.~\eqref{lum_matome}&1&&$\checkmark$& &Appendix~\ref{ssec:apdx_tinfall}\\

        $8\times10^{-3}$ & Eq.~\eqref{alpha_pz_fiducial}& $\infty$ & $\leq250\unit{au}$ & Eq.~\eqref{lum_matome}$\times0.1$&1&&$\checkmark$& & Appendix~\ref{ssec:apdx_pevrates}\\

        $8\times10^{-3}$ &Eq.~\eqref{alpha_pz_fiducial} & $\infty$ & $\leq250\unit{au}$ & Eq.~\eqref{lum_matome}$\times10$&1&&$\checkmark$& &Appendix~\ref{ssec:apdx_pevrates}\\

    \end{tabular}
    \tablefoot{
    The interstellar gas infall rate was treated as a free parameter and varied in the range of $5\times 10^{-10}$ to $5\times 10^{-7} \msunyr$. The results indicate whether the disk mass ($10^{-6}$–$10^{-1} \Msun$) and size (50--500~au) are consistent with observations under an infall rate sufficient to sustain stellar accretion ($4\times10^{-11}$--$4\times10^{-9}\msunyr$). Details are given in Section~\ref{Result} and Appendix~\ref{sec:para_dep}.

    $^{a}$ Reproduced for $f_{\rm Kepler}\leq0.4$.
    
    $^{b}$ Reproduced for $f_{\rm Kepler}\leq0.3$.
    }
    \label{tab:model}
\end{table*}

\subsection{Initial and boundary conditions}
\label{method:ic}

The initial disk mass was assumed to be 1\% of the central star mass. We set a relatively small initial disk mass so that the effect of late infall becomes dominant, since our interest lies in the potential for late infall to enhance accretion. We assumed the same radial distribution of disk gas as in \cite{Ooyama+2025}, expressed as
\begin{equation}
    \Sigma_{\mathrm{int}}=\Sigma_{1\mathrm{au}} \left(\frac{r}{1\mathrm{au}}\right)^{-3/2}\exp{\left(-\frac{r}{r_{\mathrm{cut}}}\right)} ,
    \label{eq:sgm_ini}
\end{equation}
where the outer cut-off radius is 30~au. This cut-off radius is larger than the characteristic radii of the compact disks suggested by recent studies \citep{Tabone+2025}. However, our results are insensitive to the choice of the cut-off radius, as discussed in Section~\ref{Result:fiducial}. This initial condition corresponds to the beginning of the class II phase, when the age is on the order of $10^6\,\mathrm{yr}$. Since we are mainly discussing disks at $\sim 2\,\mathrm{Myr}$, it is reasonable to define this point as zero age. We used the same initial conditions for all the models considered. We confirmed that variations in the initial disk radius and mass lead to only minor changes in the conclusions on when late infall can enhance stellar accretion. This is because in most of the models, the disk mass is dominated by the supplied gas through late infall at $\gtrsim 1$~Myr.

We adopted the same boundary conditions at the inner and outer edges of the computational domain as in \citet{Suzuki+2016,Ooyama+2025}:
$\partial/\partial r(\Sigma r)=0$. Although this condition is not strictly equivalent to the zero-viscous-torque boundary condition considered by \citet{Lynden-Bell+1974}, it is consistent with such a boundary in the limit $r \rightarrow 0$. We adopted it following \citet{Suzuki+2016, Suzuki_2016_tum}, because the effect of the boundary condition is confined to the innermost part of the disk, so that, as shown by \citet{Suzuki_2016_tum}, the overall disk evolution is not strongly affected.
The above boundary condition imposes $\Sigma \propto r^{-1}$ but is effective only in the very vicinity of the computational boundaries. The surface density in the inner region deviates from the imposed profile through viscous evolution, disk winds, and other physical processes.

\subsection{Explored parameter space}
\label{method:parameter}

Table~\ref{tab:model} summarizes the models explored in this work. The key parameters are the viscous stress, $\alpha_{r\phi}$, the MHD disk wind torque, $\alpha_{\phi z}$, the infall duration, the infall radius, $r_{\rm infall}$, the stellar UV and X-ray luminosities, and $f_{\rm Kepler}$, which specifies the specific angular momentum of the infalling gas in units of the local Keplerian value. We explored a range of late infall rates, $\dot{M}_{\rm infall}=5\times10^{-10}$--$5\times10^{-7}~\msunyr$, and examined which combinations of parameters allow infall to enhance stellar accretion (see Appendix~\ref{parameter study in fiducial}).

\subsection{Definition of disk size}
\label{method:disksize}

To compare our models with observations, we used the CO disk radius as the gas disk size. The CO disk radius is defined as the radius at which the CO column density equals $5\times10^{15}~\unit{cm^{-2}}$ \citep{Zagaria+2023}. We measured the column density from both the outer edge and the surface of the disk, and the CO radius is defined as the location where either of them rises above the criterion. We assumed an ISM CO abundance of $x_{\rm CO} = 10^{-4}$ to calculate the CO column density. Our results are not very sensitive to the choice of this conversion factor (see Section~\ref{Rd} for details).

\section{Results}
\label{Result_all}

In this section, we present the results of our calculations. Sections~\ref{Result} and \ref{Result:Sub-Kep} focus on the models without and with the effective torque exerted by infalling gas, respectively.

\subsection{Late infall: mass supply}
\label{Result}

Here, we consider late infall only as a mass source. We first show the baseline models with the $\Sigma$-dependent MHD disk wind torque (Section~\ref{Result:fiducial}), then consider cases with stronger MHD disk wind torques (Section~\ref{Results:constant_alpha}). Finally, we discuss how the stellar accretion rate depends on the infall rate (Section~\ref{lateinfallvsacc}).

\subsubsection{Limitations of turbulent viscous torque-dominated models}
\label{Result:fiducial}

\begin{figure*}
    \sidecaption
    \includegraphics[width=12cm]{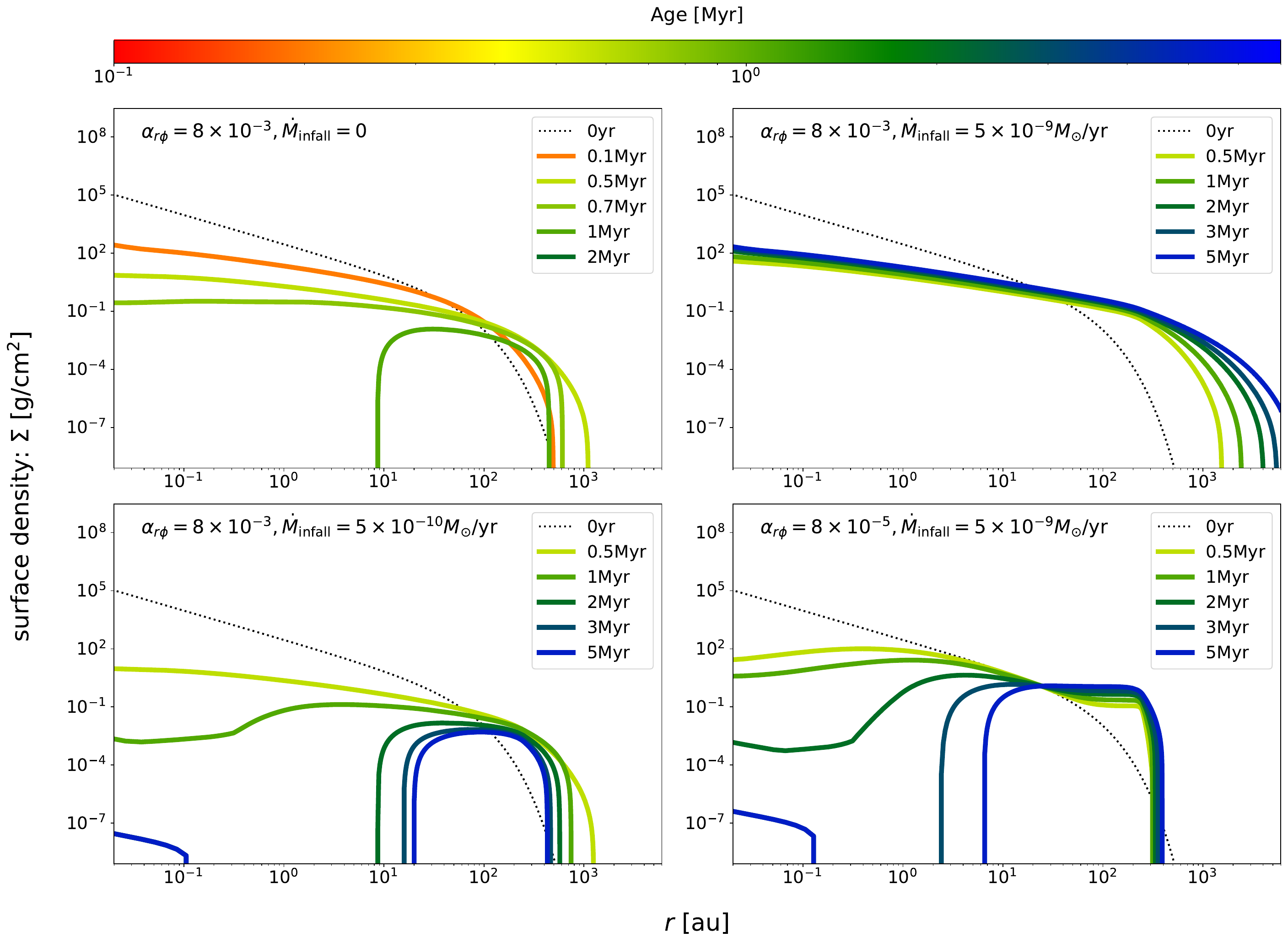}
    \caption{Time evolution of the gas surface density in the models with $\Sigma$-dependent MHD disk wind torque, highlighting the roles of the late infall and turbulent viscosity. {\it Top left}: Evolution without any late infall, assuming a disk with $\alpha_{r\phi}=8 \times 10^{-3}$. {\it Top right / Bottom left}: Models with $\alpha_{r\phi} = 8 \times 10^{-3}$ and late infall rates of $\mdoti = 5 \times 10^{-9}$ and $5 \times 10^{-10}\msunyr$, respectively. {\it Bottom right}: model with $\alpha_{r\phi}=8\times10^{-5}$ and $\mdoti = 5\times10^{-9}\:\msunyr$.}
    \label{fig:Sigma_MRIact}
\end{figure*}

We discuss how the evolution of the disk surface density differs when varying the viscosity and the late infall rate. We first examine the case without late infall to clarify the changes introduced by including it. 
The top left panel of Figure~\ref{fig:Sigma_MRIact} shows the evolution of the gas surface density in the model with $\alpha_{r\phi}=8 \times 10^{-3}$ and no late infall. In this model, the surface density gradually decreases at $r \lesssim 10^{2}~\mathrm{au}$, primarily due to accretion onto the central star and mass loss driven by MHD disk winds. After about 1~Myr, photoevaporation becomes effective in the inner region, leading to the formation of a central cavity with a radius of $\sim 10~\mathrm{au}$. The disk is almost entirely dissipated by $\sim 2$~Myr.

We compared the cases with and without late infall to examine how late infall alters the evolution of the surface density distribution. The bottom left panel of Figure~\ref{fig:Sigma_MRIact} shows the evolution of the model with $\alpha_{r\phi}=8 \times 10^{-3}$ and a relatively low late infall rate at $\mdoti = 5 \times 10^{-10}~\msunyr$. Compared to the no-infall case (top left panel), the overall evolution appears similar in the surface density profile, but the disk lifetime is substantially extended. Owing to the continuous mass supply, a considerable amount of disk material survives even at 5~Myr (see the blue line), particularly in the outer region ($r \gtrsim 10^{2}~\mathrm{au}$). 

The effect of the late infall is also seen in the inner disk at $r \lesssim 0.1~\mathrm{au}$. Even after a gap forms at around $\sim 2$~Myr, a low but finite surface density of $\sim 10^{-8}~\mathrm{g~cm^{-2}}$ remains at $r \lesssim 0.1~\unit{au}$, where the photoevaporation rate is zero. The inner disk has reached a quasi-steady state by this time, causing the lines to overlap in the figure, but the material nonetheless persists beyond 2~Myr. The persistence of the inner disk reflects our assumption of spatially uniform infall, so the inner disk itself might therefore be an artifact of the model. Moreover, its mass is too small to be significant ($4\times10^{-17}~M_\odot$) and, thus, it does not affect our conclusions.

Next, we discuss how the strength of late infall influences the disk evolution. The top right panel of Figure~\ref{fig:Sigma_MRIact} shows the evolution of the model with $\alpha_{r\phi}=8 \times 10^{-3}$ and a higher infall rate of $\mdoti = 5 \times 10^{-9}~\msunyr$.  \footnote{The radial surface density profile is only slightly shallower than $r^{-1}$, and thus remains broadly consistent with commonly adopted disk models.}
In contrast to the previous two models (top and bottom left panels),  no gap forms even after 2~Myr, despite the ongoing photoevaporation driven by the central star. This is because a higher infall rate increases the surface density and builds up a larger mass reservoir, thereby producing a higher radial mass flux that exceeds the local photoevaporation rate. The surface density increases at all radii after $\sim$1~Myr, reflecting the accumulation of the infalling gas. The disk continues to grow in size, as the outward viscous transport exceeds the dispersal caused by photoevaporation, particularly at the outer edge.

As demonstrated so far, in the model with $\alpha_{r\phi}=8 \times 10^{-3}$, the disk expands significantly due to the strong turbulent viscosity. In contrast, when $\alpha_{r\phi}$ is reduced to $8 \times 10^{-5}$, the disk remains compact even with the same infall rate of $\mdoti = 5 \times 10^{-9}~\msunyr$ (bottom right panel), due to the suppression of the radial spreading of the disk material. The long viscous timescale also prevents the infalling gas from reaching the inner region. As a result, photoevaporation opens a gap at $r \sim 10~\mathrm{au}$ after $\sim$2~Myr, which halts accretion onto the central star. Most of the gas that manages to reach the inner disk is ultimately removed via photoevaporation at $r \sim 1$--$10~\mathrm{au}$. Moreover, a substantial fraction of the infalling gas remains near its initial arrival radius without being transferred inward. \footnote{The limited inward mass transport also contributes to the relatively flat surface density profile.}

\begin{figure}
    \centering
    \includegraphics[width=8cm]{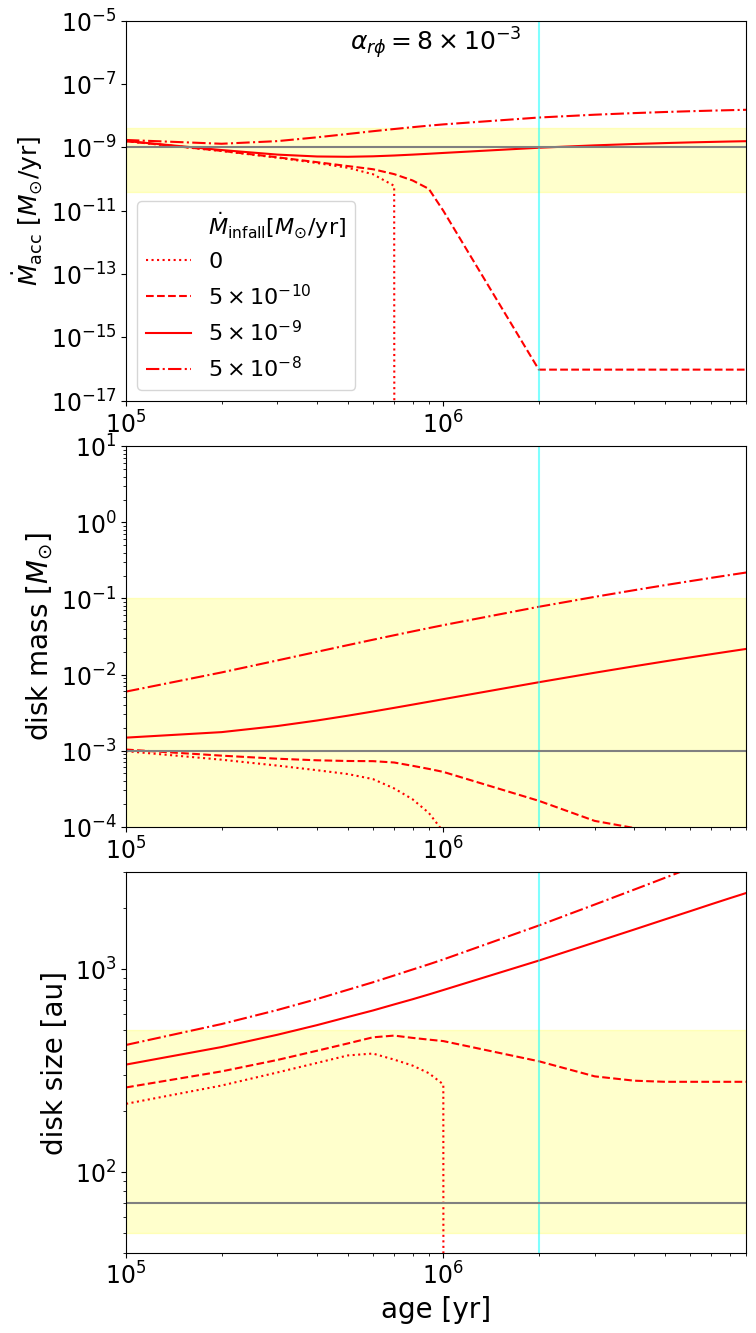}
    \caption{Time evolution of the stellar accretion rate, $\mdota$ ({\it upper panel}), disk mass ({\it middle panel}), and disk size ({\it lower panel}) in the models with $\alpha_{r\phi}=8 \times 10^{-3}$ and different $\mdoti$. In each panel, the yellow-shaded area represents the ranges of $\mdota$, disk mass, or disk size observationally inferred for the Lupus region \citep{Winter+2024a, Winter+2024b, Ansdell+2018}. The disk mass is estimated as 100 times the observed dust mass. In each panel, the dotted, dashed, solid, and dash-dot lines represent the different infall rates of $\mdoti = 0$, $5 \times 10^{-10}$, $5 \times 10^{-9}$, and $5 \times 10^{-8} \msunyr$. The horizontal gray lines show the median values reported by recent AGE-PRO observations \citep{Tabone+2025}.  The vertical aqua line represents the epoch of 2 Myr, which corresponds to the age of Lupus.    
    }
    \label{fig:Acc_M_MRIact}
\end{figure}

We turn to the time evolution of the stellar accretion rate, disk mass, and disk size, which provide direct benchmarks for comparison with observations. The upper panel of Figure~\ref{fig:Acc_M_MRIact} shows the evolution of the stellar accretion rate, $\mdota$, for a series of models with $\alpha_{r\phi} = 8 \times 10^{-3}$ and different values of the late infall rate, $\mdoti$. Following Figure 2(a) of \citet{Winter+2024a}, the observed values of $\mdota$ in Lupus 3 range from $4 \times 10^{-11}$ to $4 \times 10^{-9}~\msunyr$, which we indicate as a yellow shaded region in the figure. This range serves as a benchmark to evaluate whether the simulated accretion is significantly influenced by late infall.
For high infall rates ($\mdoti \gtrsim 5 \times 10^{-9}~\msunyr$; solid line), the modeled $\mdota$ falls within the observational range, indicating that late infall can effectively sustain stellar accretion. In these cases, $\mdota$ continues to increase after $\sim$1~Myr, as the infalling gas continuously replenishes the inner disk. 
However, it should be noted that if the infall rate becomes too large ($\mdoti \gtrsim 5 \times 10^{-8}~\msunyr$; dash-dot line), the stellar accretion rate also becomes excessively high. In contrast, at lower infall rates (dashed and dotted lines), $\mdota$ declines rapidly by $\sim$2~Myr due to the formation of a gap that impedes mass transport from the outer disk. For example, in the case with $\mdoti = 5 \times 10^{-10}~\msunyr$ (dashed line), the accretion rate drops to $\sim 10^{-15}~\msunyr$.

The middle panel of Figure~\ref{fig:Acc_M_MRIact} shows the time evolution of the disk mass for different values of the infall rate $\mdoti$. Since disks with masses exceeding $0.1~M_\odot$ are observationally rare \citep{Winter+2024b,Manara_2023}, we adopt $0.1~M_\odot$ as an upper threshold to assess whether the simulated disks remain consistent with observed systems in Lupus.
For example, with $\mdoti = 5 \times 10^{-9}~\msunyr$ (solid line), the disk mass stays below this threshold for more than $\sim$10~Myr, suggesting that prolonged accretion can occur without violating observational constraints. In contrast, for higher infall rates such as $\mdoti \gtrsim 5 \times 10^{-8}~\msunyr$ (dash-dot line), the disk mass rapidly increases and exceeds $0.1~M_\odot$ within the first $\sim$1~Myr, indicating that such high infall rates can lead to overly massive disks inconsistent with observations.

The lower panel of Figure~\ref{fig:Acc_M_MRIact} presents the time evolution of the disk size (i.e., CO disk radius) for different $\mdoti$. Observationally, disks larger than $500~\unit{au}$ are rare in the Lupus region \citep{Ansdell+2018,Sanchis_2021}, so we adopted $500~\unit{au}$ as a representative upper limit (shown as the yellow shaded region).
At low infall rates ($\mdoti \leq 5 \times 10^{-10}~\msunyr$; dashed line), the disk size stays below this limit for more than $\sim$10~Myr, consistent with observations. Conversely, for higher infall rates ($\mdoti \gtrsim 5 \times 10^{-9}~\msunyr$; solid line), the disk rapidly expands beyond 500~au within the first $\sim$1~Myr.
These results suggest that while models with $\alpha_{r\phi}=8\times10^{-3}$ and $\mdoti \simeq 5 \times 10^{-9}~\msunyr$ can successfully reproduce the observed stellar accretion rates and disk masses, they tend to overpredict disk sizes. However, this result follows from the assumption that infall does not exert an effective torque (Section~\ref{Sub-Kep_StrongVis}).

\begin{figure}
    \centering
    \includegraphics[width=8cm]{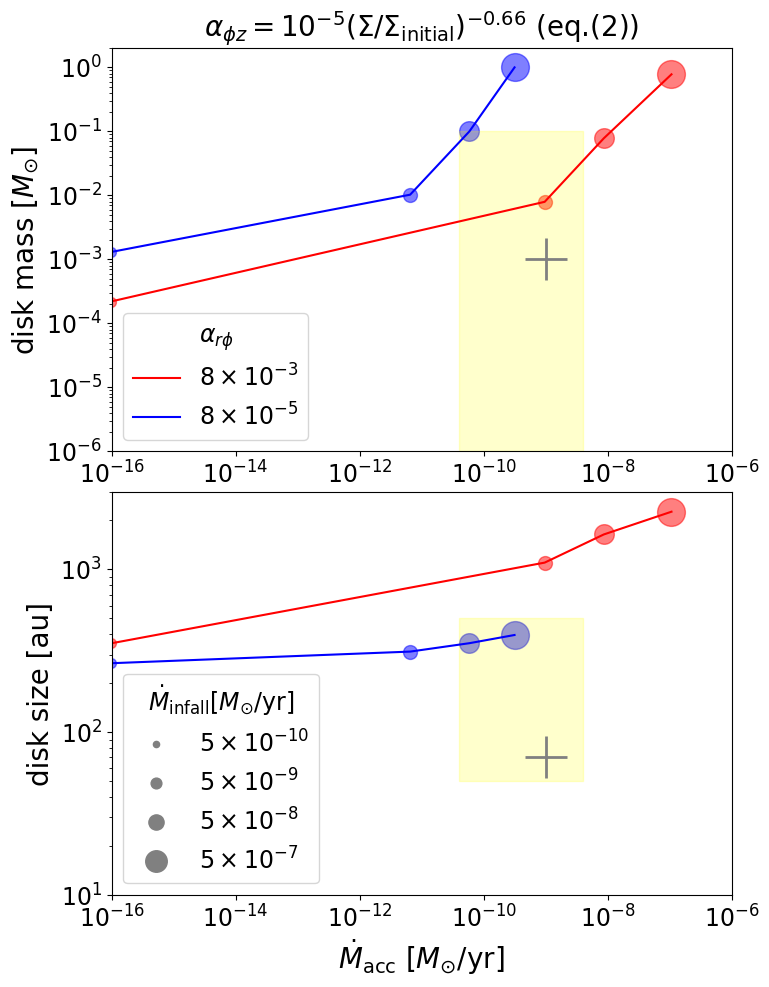}
\caption{Comparison of the stellar accretion rate, $\mdota$, disk mass, and disk radius between observations and the models with $\Sigma$-dependent MHD disk wind torque at the epoch of 2~Myr. The top panel shows the relation between $\mdota$ and disk mass, while the bottom shows that between $\mdota$ and disk size.
In each panel, the yellow shaded region marks the observed distributions in the Lupus star-forming region \citep{Winter+2024a, Winter+2024b, Ansdell+2018}, and the gray crosses indicate the median values reported by recent AGE-PRO observations \citep{Tabone+2025}. Model results are shown by red ($\alpha_{r\phi}=8\times10^{-3}$) and blue ($\alpha_{r\phi}=8\times10^{-5}$) lines. Different marker sizes correspond to different infall rates of $\mdoti = 5\times10^{-10}$, $5\times10^{-9}$, $5\times10^{-8}$, and $5\times10^{-7}\msunyr$, respectively, with smaller markers representing lower rates.
}
    \label{fig:Acc_Diskmass_size_MRI}
\end{figure}

To discuss whether the scenario in which late infall enhances accretion onto the central star can be reproduced, we compared the stellar accretion rate, disk mass, and disk size with the results of \citet{Winter+2024a} as well as with observations. The upper panel of Figure~\ref{fig:Acc_Diskmass_size_MRI} shows the correlation between the stellar accretion rate, $\mdota$, and disk mass at 2~Myr for both models with $\alpha_{r\phi}=8\times10^{-3}$ and $8\times10^{-5}$. The yellow rectangle represents the observationally favored range of disk masses (as in Figure~\ref{fig:Acc_M_MRIact}). 

Models with $\alpha_{r\phi}=8\times10^{-3}$ (red line) with $\mdoti \simeq 5 \times 10^{-9}~\msunyr$ (the second smallest marker) fall within the observational box, indicating that this combination yields simultaneous agreement with both observed $\mdota$ and disk mass.
In contrast, models with $\alpha_{r\phi}=8\times10^{-5}$ (blue line) require high infall rates ($\mdoti \gtrsim 5 \times 10^{-8}~\msunyr$; the two largest blue markers) to maintain stellar accretion rates within the observational range. However, under such conditions, the disk mass becomes significantly larger than the observed values. This suggests that models with $\alpha_{r\phi}=8\times10^{-5}$ struggle to simultaneously reproduce both $\mdota$ and disk mass within the observed constraints.
If the disk mass exceeds $0.1M_{\odot}$, gravitational instability may enhance the effective viscous $\alpha$ to increase the accretion rate, which can mitigate this inconsistency. 
However, even in such cases, excessively large disk mass remains an issue, and thus 
gravitational instability alone is unlikely to account for the discrepancy with observations.

The lower panel of Figure~\ref{fig:Acc_Diskmass_size_MRI} shows the relationship between the stellar accretion rate, $\mdota$, and the disk radius at 2~Myr. The observed disk radius in the Lupus region typically falls within the range 50–500~au \citep{Ansdell+2018}. 
The median disk size reported by the AGE-PRO observations is 70 au \citep{Deng_2025,Tabone+2025}, located in the lower region of the yellow-shaded area (represented by a gray cross). 
The models with $\alpha_{r\phi}=8\times10^{-3}$ tend to produce significantly larger disks, often exceeding 500~au, due to strong turbulent viscosity and efficient angular momentum transport. The results do not change even when the infall region, the infall duration, or the strength of photoevaporation is varied (Appendix~\ref{sec:para_dep}). In contrast, the models with $\alpha_{r\phi}=8\times10^{-5}$ and $\mdoti = 5 \times 10^{-8}$–$10^{-7}~\msunyr$ (the two largest blue markers) yield disk sizes that lie within the observed range, indicating better agreement with the Lupus data, although they are larger than the median value reported by the AGE-PRO observations.

One possible way to reduce the excessively large disk sizes predicted by the models, particularly those with $\alpha_{r\phi} = 8 \times 10^{-3}$, is to adopt a smaller cut-off radius for the initial disk. Indeed, \citet{Tabone+2025} suggested that disks in Lupus may typically originate as compact disks with initial radii of approximately $10\,{\rm au}$, whereas disks with larger initial radii may represent only a long-lived subset of the disk population. Motivated by this possibility, we repeated our calculations adopting an initial cut-off radius of $10\,{\rm au}$. However, the results remained almost unchanged because the gas supplied by late infall dominates over the gas initially present in the disk, making the subsequent evolution largely insensitive to the initial gas distribution. Thus, adopting a more compact initial disk does not resolve the discrepancy in disk size. This also confirms that our conclusions are not an artifact of adopting an initially large disk representative of only an atypical subset of the disk population.

To summarize, neither the strong ($\alpha_{r\phi} = 8 \times 10^{-3}$) nor the weak ($\alpha_{r\phi} = 8 \times 10^{-5}$) turbulent viscosity models can simultaneously reproduce the observed $\mdota$, disk masses, and disk sizes in the Lupus region, without the effective torque. This highlights the difficulty of explaining all observational constraints within the framework of the models where turbulent viscosity is responsible for angular momentum transfer, and infall is incorporated purely as a mass reservoir. 
This suggests either that efficient angular momentum extraction through MHD disk wind torque is preferred, or that the assumption of no effective torque from the infalling gas is not valid. We explore these possibilities in the following sections.

\subsubsection{Enhancing torque exerted by MHD disk wind}
\label{Results:constant_alpha}
\begin{figure}
    \centering
    \includegraphics[width=8cm]{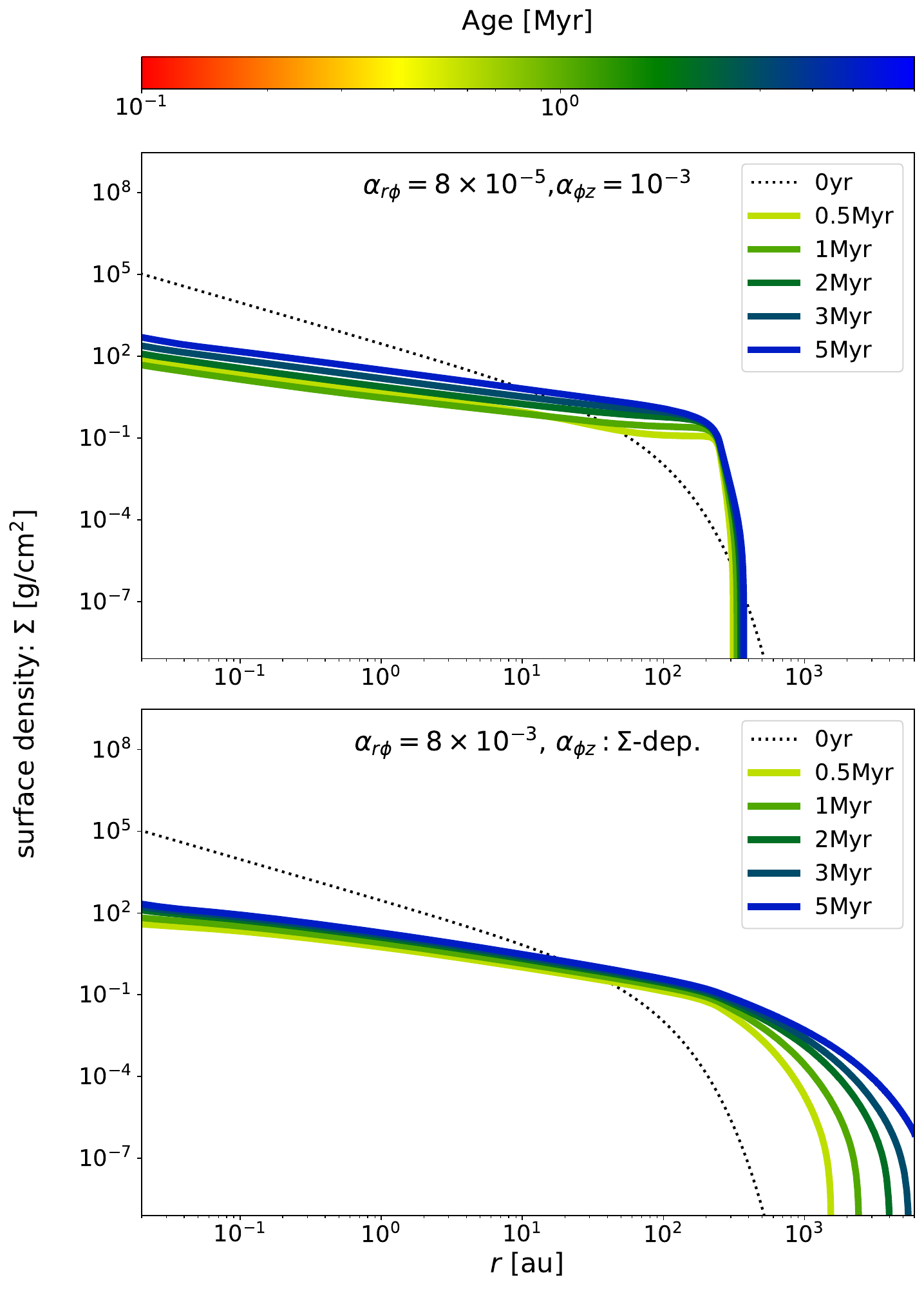}
    \caption{
Comparison of gas surface density evolution between a wind-torque-dominated model and a turbulent-viscosity-dominated model under the same infall rate of $\mdoti = 5 \times 10^{-9}~\msunyr$. 
{\it Upper panel}: wind-torque-dominated model with a strong constant MHD disk wind torque ($\alpha_{\phi z} = 10^{-3}$) and weak turbulent viscosity ($\alpha_{r\phi} = 8 \times 10^{-5}$). 
{\it Lower panel}: turbulent-viscosity-dominated model with strong turbulent viscosity ($\alpha_{r\phi} = 8 \times 10^{-3}$) and the $\Sigma$-dependent MHD disk wind torque prescribed by Eq.~\eqref{alpha_pz_fiducial}, corresponding to the upper right panel of Figure~\ref{fig:Sigma_MRIact}.
    }
    \label{fig:Sigma_alphapz}
\end{figure}

To examine a regime in which angular momentum is removed more efficiently by MHD disk winds, we considered a wind-torque-dominated model, where we set $\alpha_{r\phi}=8\times10^{-5}$ and replace the $\Sigma$-dependent MHD disk wind torque prescription in Eq.~\eqref{alpha_pz_fiducial} with a constant value of $\alpha_{\phi z}=10^{-3}$.
Figure~\ref{fig:Sigma_alphapz} compares the surface density evolution of this wind-torque-dominated model with that of the turbulent-viscosity-dominated model discussed in Section~\ref{Result:fiducial}, characterized by $\alpha_{r\phi}=8\times10^{-3}$ and the $\Sigma$-dependent wind torque (see also the upper right panel of Figure~\ref{fig:Sigma_MRIact}). The same infall rate of $\mdoti = 5 \times 10^{-9}~\msunyr$ was adopted for these models.

The top panel of Figure~\ref{fig:Sigma_alphapz} shows the wind-torque-dominated model.
\footnote{In the wind-torque-dominated model, the surface density profile within the disk remains slightly shallower than $r^{-1}$.} 
No gap is seen to form, which stands in contrast to the case with the same $\alpha_{r\phi}$ but weaker $\Sigma$-dependent MHD disk wind torque $\alpha_{\phi z}$ (see also the bottom right panel of Figure~\ref{fig:Sigma_MRIact}).
This is because the wind removes angular momentum efficiently enough to drive an inward radial mass flux that overcomes photoevaporative mass loss.
The lower panel of Figure~\ref{fig:Sigma_alphapz} shows the corresponding run with $\alpha_{r\phi}=8\times10^{-3}$, the $\Sigma$-dependent wind torque, and the same infall rate. Its surface density evolution is broadly similar to that in the upper panel, especially in the inner disk. However, the outer disk continues to expand, whereas the disk radius remains nearly constant when $\alpha_{r\phi}=8\times10^{-5}$ and $\alpha_{\phi z}=10^{-3}$.
This suggests that strong angular momentum extraction by MHD disk winds may help meet the observational constraints on disk sizes with a single parameter set.

\begin{figure}
    \centering
    \includegraphics[width=8cm]{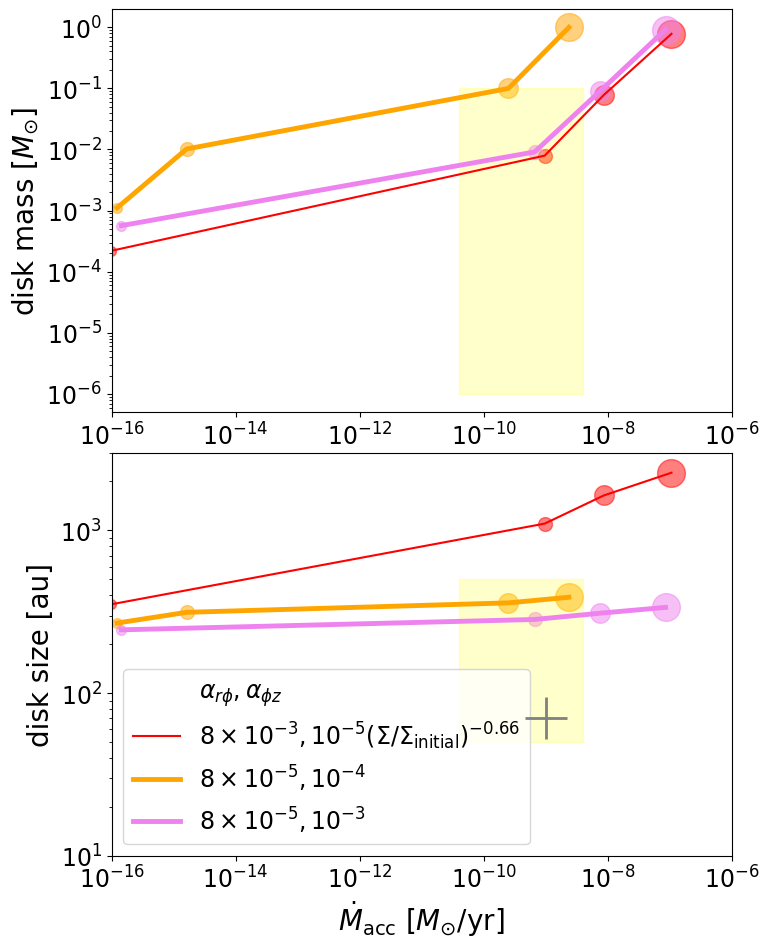}
\caption{Same as Figure~\ref{fig:Acc_Diskmass_size_MRI}, but for different prescriptions of MHD disk wind torque $\alpha_{\phi z}$, under the models with $\alpha_{r\phi}=8\times10^{-5}$. The red, orange, and violet lines represent models with $\alpha_{r\phi} = 8\times10^{-3}$ and $\Sigma$-dependent $\alpha_{\phi z}$, with $\alpha_{r\phi} = 8\times10^{-5}$ and $\alpha_{\phi z} = 10^{-4}$, and with $\alpha_{r\phi} = 8\times10^{-5}$ and $\alpha_{\phi z} = 10^{-3}$, respectively.
Different marker sizes correspond to different infall rates of $\mdoti = 5\times10^{-10}$, $5\times10^{-9}$, $5\times10^{-8}$, and $5\times10^{-7}\msunyr$, respectively, with smaller markers representing lower rates.
}
    \label{fig:Acc_Infall_alphapz}
\end{figure}

Indeed, adopting strong angular momentum extraction yields a consistent disk size. 
Figure~\ref{fig:Acc_Infall_alphapz} illustrates how the enhanced MHD disk wind torque affects the relationships between $\mdota$ and disk mass (upper panel), and between $\mdota$ and disk size (lower panel). The lower panel shows that the problem of excessive disk expansion in the models with $\alpha_{r\phi} = 8\times10^{-3}$ is mitigated when the MHD disk wind torque exceeds $\alpha_{\phi z} \gtrsim  10^{-4}$ (orange and violet lines). At $\alpha_{\phi z} \simeq  10^{-3}$ (violet line), both $\mdota$ and disk mass simultaneously fall within the observational range, as shown in the upper panel. Among the models tested, the most favored infall rate is $\mdoti = 5 \times 10^{-9}~\msunyr$ (the second smallest violet marker). 
These results suggest that even within the framework of the models without the effective torque, efficient angular momentum extraction by MHD disk winds allows late infall to significantly enhance stellar accretion without causing disk expansion. 
This provides a possible theoretical basis for the scenario proposed by \citet{Winter+2024a}, whereby continued infall is invoked to explain the observed correlation between ISM density and $\mdota$ in Lupus disks.

\subsubsection{Relationship between late infall rate and accretion rate}
\label{lateinfallvsacc}
\begin{figure}
    \centering
    \includegraphics[width=8cm]{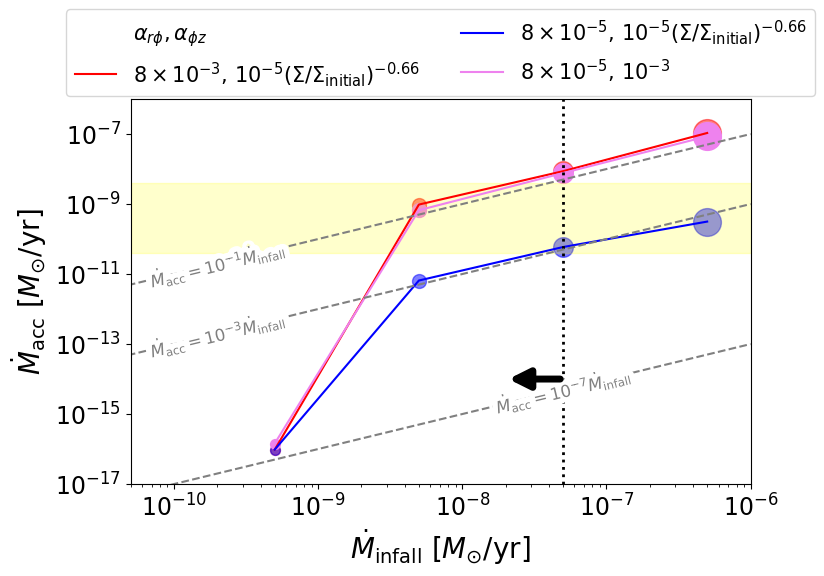}
    \caption{
Correlation between infall rates onto the disk $\mdoti$ and accretion rate onto the central star $\mdota$ at the epoch of 2~Myr. The horizontal yellow region indicates the values of $\mdota$ observationally inferred for the Lupus region \citep{Winter+2024a, Winter+2024b}. 
The red, violet, and blue lines represent the turbulent-viscosity-dominated models ($\alpha_{r\phi}=8\times10^{-3}$ and the $\Sigma$-dependent $\alpha_{\phi z}$), the wind-torque-dominated models ($\alpha_{r\phi}=8\times10^{-5}$ and $\alpha_{\phi z}=10^{-3}$), and the models where both the turbulent viscosity and wind torque are relatively weak ($\alpha_{r\phi} = 8\times10^{-5}$ and the $\Sigma$-dependent $\alpha_{\phi z}$), respectively. Different marker sizes correspond to different infall rates of $\mdoti = 5\times10^{-10}$, $5\times10^{-9}$, $5\times10^{-8}$, and $5\times10^{-7}\msunyr$, respectively, with smaller markers representing lower rates.
The dotted vertical line represents the upper limit of $\mdoti$, which is estimated by Bondi–Hoyle–Lyttleton accretion.
}
    \label{fig:Acc_Infall}
\end{figure}

The question of whether late infall can enhance accretion depends critically on the fraction that reaches the central star. In this section, we interpret our results by analytically considering what fraction of the infalling mass actually reaches the central star.  Figure~\ref{fig:Acc_Infall} shows that in the absence of a gap, slightly more than 10\% of the infalling mass reaches the star in both the turbulent-viscosity-dominated models ($\alpha_{r\phi}=8\times10^{-3}$ and $\Sigma$-dependent $\alpha_{\phi z}$) and the wind-torque-dominated models ($\alpha_{r\phi}=8\times10^{-5}$ and $\alpha_{\phi z}=10^{-3}$).
In contrast, only about 0.1\% reaches the star when both the turbulent viscosity and wind torque are inefficient  ($\alpha_{r\phi}=8\times10^{-5}$ with $\Sigma$-dependent $\alpha_{\phi z}$). Once a gap forms, this fraction drops to only $10^{-5}$\% of the infalling mass regardless of the value of $\alpha_{r\phi}$. 

Figure~\ref{fig:Acc_Infall} clarifies why high infall rates are required in the models with $\alpha_{r\phi}=8\times10^{-5}$ and $\Sigma$-dependent $\alpha_{\phi z}$ to sustain stellar accretion at the observed level: only material deposited where accretion timescales are short ($\lesssim 2$~Myr) can reach the star, so most of the infalling gas remains in the outer disk. The accretion timescale is given by the smaller of the viscous timescale, $t_{\nu}$, and the timescale of advection driven by the MHD disk wind torque, $t_{\rm mw}$. The former is given by
\begin{equation}
t_{\nu}(r)=\frac{r^2}{\alpha_{r\phi} c_s^2/\Omega}\sim1\unit{Myr}\left(\frac{\alpha_{r\phi}}{8\times10^{-5}}\right)^{-1}\left(\frac{r}{1\unit{au}}\right),
\label{eq:t_nu0}
\end{equation}
and the latter is given by
\begin{equation}
\label{eq:adv_timescale0}
t_{\rm mw}(r)=\frac{r}{\sqrt{\frac{2}{\pi}}c_{\rm s}\alpha_{\phi z}}\sim0.7\unit{Myr}\left(\frac{\alpha_{\phi z}}{10^{-5}}\right)^{-1}\left(\frac{r}{1\unit{au}}\right)^{\frac{3}{4}}.
\end{equation}

In the turbulent-viscosity-dominated models ($\alpha_{r\phi}=8\times10^{-3}$ and $\Sigma$-dependent $\alpha_{\phi z}$), $t_{\nu}$ is smaller than $t_{\rm mw}$ and approximates the accretion timescale. For $t_\nu<2\unit{Myr}$, the corresponding region is $r<200\unit{au}$. Assuming that half of the gas infalling into this region loses angular momentum to accrete onto the central star (while the other half gains angular momentum and moves outward), roughly 30\% of the infalling material is expected to accrete onto the star. This estimate differs from the simulation results by a factor of about three, which is caused by mass loss via disk winds and the photoevaporation.

In the wind-torque-dominated models ($\alpha_{r\phi}=8\times10^{-5}$ and $\alpha_{\phi z}=10^{-3}$),  $t_{\rm mw}$ ($ < t_{\nu}$) approximates the accretion timescale. For $t_{\rm mw}<2\unit{Myr}$, the corresponding region covers the entire infall region. Therefore, the infall rate would be expected to equal the accretion rate, but due to the effects of MHD disk winds and photoevaporation, only a little over 10\% of the late-infalling gas reaches the central star. This is because the mass loss rate of the MHD disk wind scales with $\Sigma$: stronger late infall leads to higher $\Sigma$, and thus proportionally stronger mass loss. Consequently, even for high late infall rates, the fraction of material that ultimately accretes onto the star remains essentially unchanged.

In models with weak turbulent viscosity ($\alpha_{r\phi}=8\times10^{-5}$) and $\Sigma$-dependent $\alpha_{\phi z}$, $t_{\rm mw}$ ($ < t_{\rm \nu}$) approximates the accretion timescale. For $t_{\rm mw}<2\unit{Myr}$, the corresponding region is $r<4\unit{au}$. If we assume that all of the mass infalling into this region eventually accretes, the fraction amounts to only about 0.03\% of the total. Although this estimate differs from the numerical results by a factor of a few, the discrepancy can be attributed to uncertainties such as the $\Sigma$-dependence of $\alpha_{\phi z}$.

If a gap is formed, only the material infalling within $0.1\unit{au}$ is expected to accrete. Consequently, only about $10^{-7}$ of the total infalling mass is expected to accrete onto the star, which is in good agreement with the numerical results.
Consequently, the $\alpha_{r\phi}=8\times10^{-5}$ models need unrealistically high infall rates ($\gtrsim 10^{-7}\msunyr$; the largest blue marker) to match the observed $\mdota$ range; lower rates instead lead to gap formation and suppressed accretion (Figure~\ref{fig:Sigma_MRIact}). In contrast, the $\alpha_{r\phi}=8\times10^{-3}$ models reproduce the observed $\mdota$ with more moderate infall rates ($\sim 5 \times 10^{-9}~\msunyr$; the second smallest red marker). 

From the above discussion, we find that the accretion rate can be estimated (to an order of magnitude) from the infall rate and the accretion timescale. The difference between the analytical estimate and the numerical results is mainly attributable to mass loss through MHD disk winds and photoevaporation.

\subsection{Late infall: Mass supply and effective torque}
\label{Result:Sub-Kep}
This section presents the results for the models with the effective torque exerted by the infalling gas. The effective torque arising from the angular momentum difference between the disk gas and the infalling gas can affect both the stellar accretion rate and the disk size. Therefore, introducing the effective torque may revise the conclusion discussed in Section~\ref{Result:fiducial}---that, in models without an effective torque, accretion cannot be enhanced while reproducing the observed disk masses and radii unless a strong MHD disk wind torque is invoked. In the following, we discuss the conditions under which the effective torque can revise this conclusion. The cases with a strong ($\alpha_{r\phi}=8\times10^{-3}$) and weak ($8\times10^{-5}$) viscosity are presented in Section~\ref{Sub-Kep_StrongVis} and Section~\ref{Sub-Kep_WeakVis}, respectively.

\subsubsection{Cases with $\alpha_{r\phi}=8\times10^{-3}$}
\label{Sub-Kep_StrongVis}
\begin{figure}
    \centering
    \includegraphics[width=8cm]{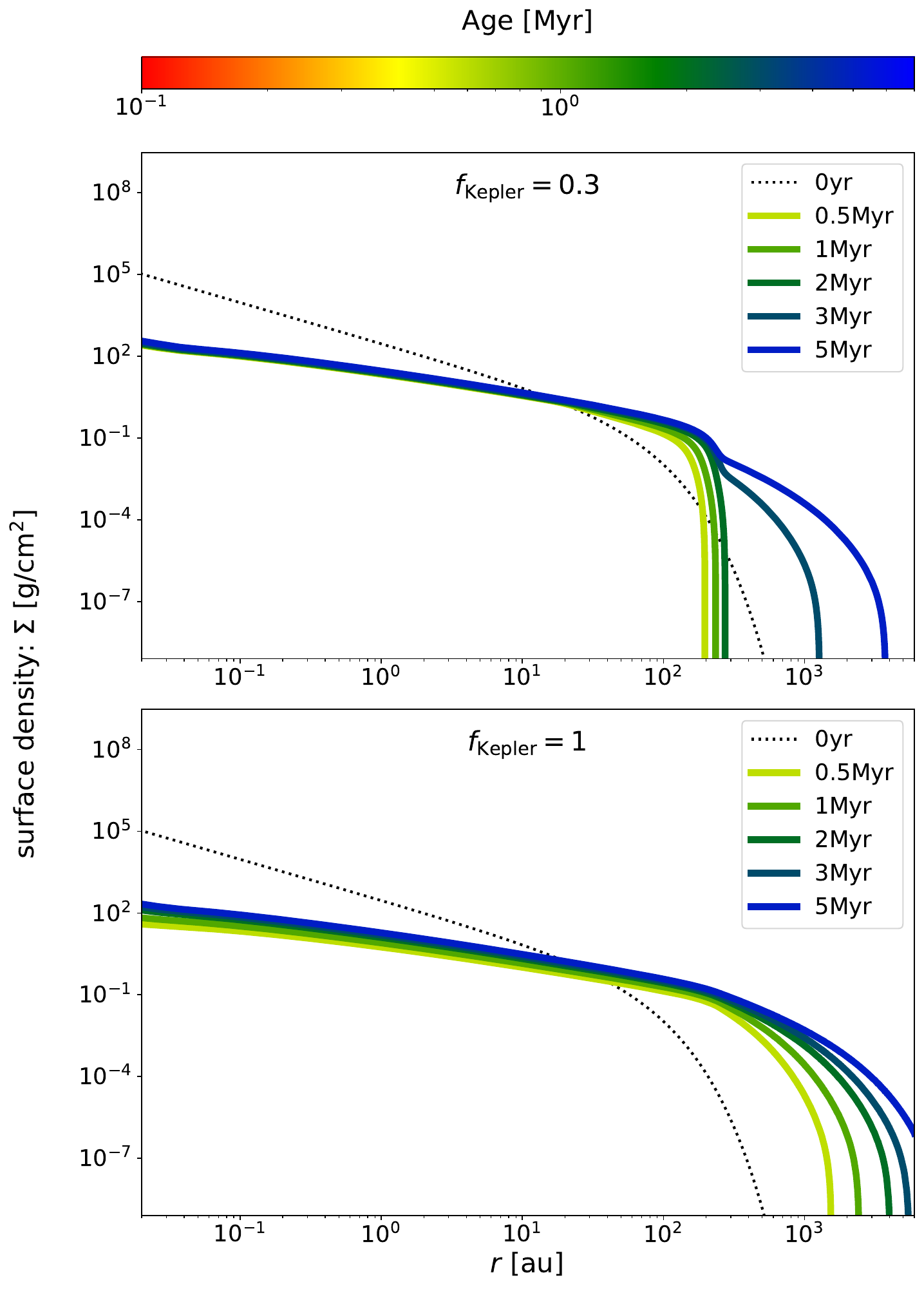}
    \caption{
Comparison of gas surface density evolution between models with the effective torque ({\it upper panel}, $f_{\rm Kepler}=0.3$) and without the effective torque ({\it lower panel}). Both models adopt $\alpha_{r\phi}=8\times10^{-3}$, the $\Sigma$-dependent MHD disk wind torque $\alpha_{\phi z}$, and the same infall rate,  $\mdoti = 5 \times 10^{-9}~\msunyr$. 
    }
    \label{fig:Sigma_Torque_MRIact}
\end{figure}
As discussed in Section~\ref{Result}, the models without the effective torque and with $\alpha_{r\phi}=8\times10^{-3}$ produce disk sizes that are too large to be consistent with the observations (Figures~\ref{fig:Acc_Diskmass_size_MRI} and \ref{fig:Acc_Infall_alphapz}). 
We thus first examine whether the effective torque of the infalling gas reconciles this issue. 
Figure~\ref{fig:Sigma_Torque_MRIact} compares the evolution of the gas surface density between models with and without the effective torque. The upper panel shows the case with the effective torque, where the infalling gas carries $0.3$ times the Keplerian angular momentum, while the lower panel shows the case without the effective torque, as discussed in Section~\ref{Result}.

While the inner disk slope remains nearly unchanged, the effective torque 
keeps the disk size confined within $\sim 250$~au, corresponding to the outer edge of the infall region.
This behavior arises because the low surface density there means that the effective torque is more efficient (cf. Eq.~\eqref{eq:local_mixing_temp}), and thus the gas is transported inward before viscously spreading outward.
Similar trends are observed for different $f_{\rm Kepler}$.

\begin{figure}
    \centering
    \includegraphics[width=8cm]{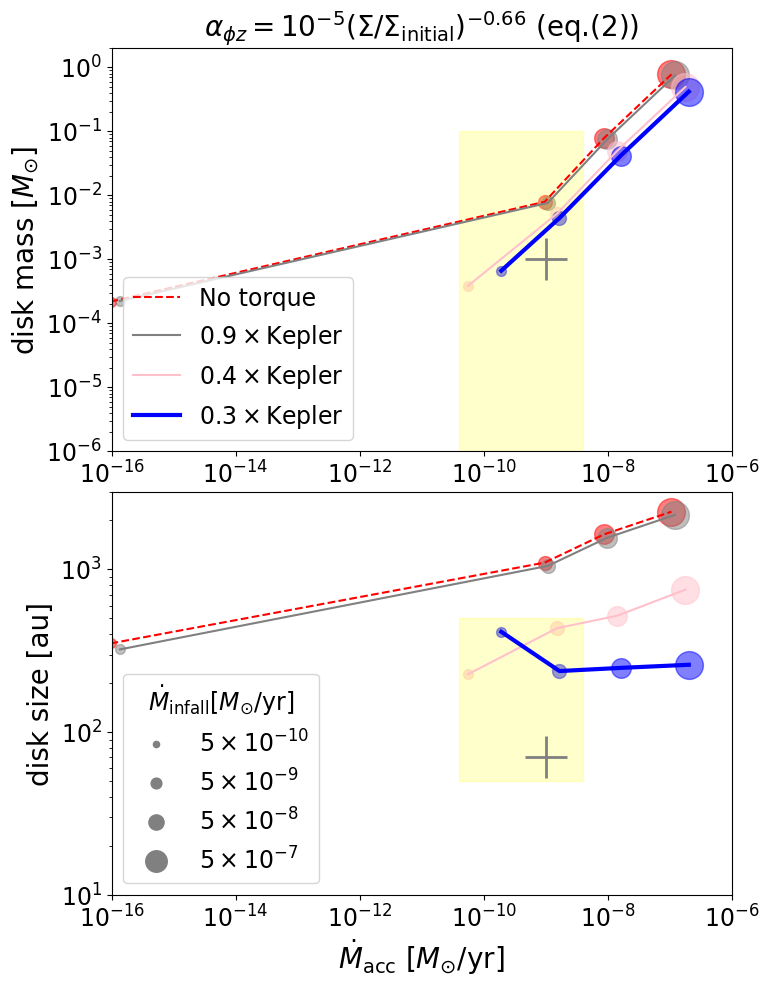}
\caption{Same as Figure~\ref{fig:Acc_Diskmass_size_MRI}, but for the models including the effective torque exerted by the infalling gas. All models in this figure adopt $\alpha_{r\phi}=8\times10^{-3}$ and the $\Sigma$-dependent $\alpha_{\phi z}$. The red dotted lines show the models without the effective torque. The gray, pink, and blue lines show the cases in which the infalling gas has angular momentum equal to $0.9$, $0.4$, and $0.3$ times the Keplerian value, respectively.
Different marker sizes correspond to different infall rates of $\mdoti = 5\times10^{-10}$, $5\times10^{-9}$, $5\times10^{-8}$, and $5\times10^{-7}\msunyr$, respectively, with smaller markers representing lower rates.
}
    \label{fig:Acc_Infall_Torque_MRI}
\end{figure}

Figure~\ref{fig:Acc_Infall_Torque_MRI} compares the disk masses, sizes, and stellar accretion rates predicted by our effective-torque models for various $f_{\rm Kepler}$ with the observed ranges.
The effective torque becomes stronger for lower $f_{\rm Kepler}$, and thus the disk size decreases. The stellar accretion rate slightly increases, whereas the disk mass slightly decreases, but these effects are secondary.
The stellar accretion rate is set by the inner disk, while the disk mass reflects the global gas distribution. These quantities are therefore less sensitive to the influence from the effective torque. 
The model with $f_{\rm Kepler}=0.3$ shows 
a smaller reduction in disk size, likely because the disk size is primarily set by $r_{\rm infall}$. Therefore, in contrast to the case with $f_{\rm Kepler}=1$ (Section~\ref{Result}), the disk size is sensitive to the choice of $r_{\rm infall}$.

In contrast to the models without the effective torque (Section~\ref{Result}), Figure~\ref{fig:Acc_Infall_Torque_MRI} shows that the models with the effective torque produce smaller disk sizes even with a high $\alpha_{r\phi}$. In particular, the cases with $f_{\rm Kepler}\leq0.4$ yield disk sizes consistent with the observations. Hence, models with a strong effective torque, or equivalently smaller angular momentum injection, can allow late infall to enhance stellar accretion while keeping the disk mass and size within the observed ranges.

\subsubsection{Cases with $\alpha_{r\phi}=8\times10^{-5}$}
\label{Sub-Kep_WeakVis}
\begin{figure}
    \centering
    \includegraphics[width=8cm]{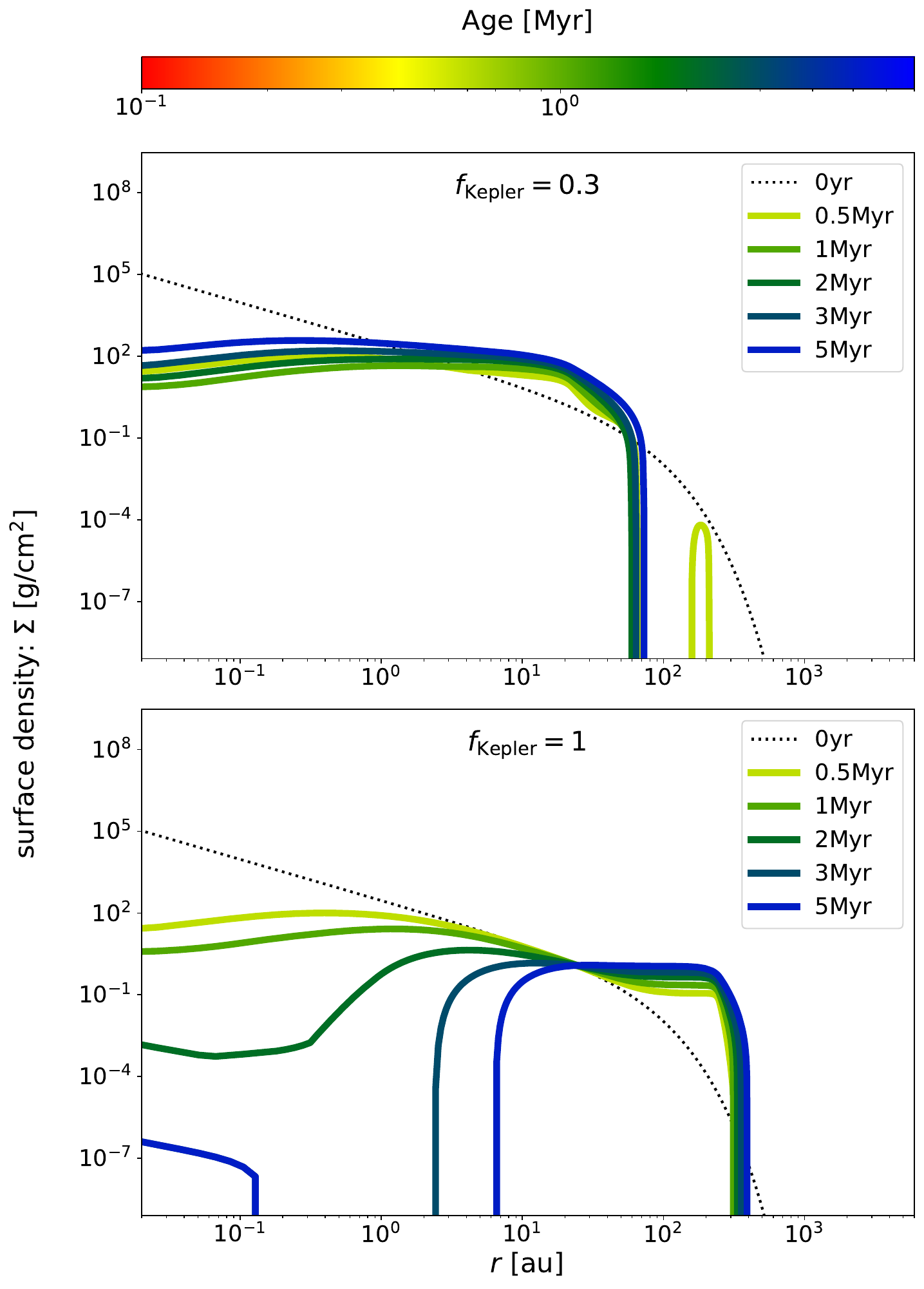}
    \caption{
Same as Figure~\ref{fig:Sigma_Torque_MRIact}, but for models with the relatively weaker turbulent viscosity, $\alpha_{r\phi}=8\times10^{-5}$.   
    }
    \label{fig:Sigma_Torque_MRIinact}
\end{figure}

As discussed in Section~\ref{Result}, models without the effective torque and with $\alpha_{r\phi}=8\times10^{-5}$ result in unrealistically massive disks for a given accretion rate due to a long viscous timescale. 
We examine whether the effective torque of the infalling gas mitigates this inconsistency in this section. 

Figure~\ref{fig:Sigma_Torque_MRIinact} compares the gas surface density evolution for the models with and without the effective torque. 
Similar to the previous cases with $\alpha_{r\phi}=8\times10^{-3}$ (Sec.~\ref{Sub-Kep_StrongVis}), the effective torque reduces the disk size. However, in the case of $\alpha_{r\phi}=8\times10^{-5}$, gas near the outer edge of the late-infall region is depleted because the effective torque drives inward transport while viscous spreading from smaller radii cannot efficiently resupply this region.

On the other hand, the surface density profile in the inner disk remains relatively flat, as in the model without the effective torque. This is because the effective torque mainly operates near the outer edge of the gas disk, around $\sim r_{\rm tq}$, while its contribution in the inner disk is relatively limited.
Therefore, although the effective torque reduces the disk size, it does not significantly change the slope of the inner surface density profile before the gap forms (\(t\lesssim1\) Myr).

\begin{figure}
    \centering
    \includegraphics[width=8cm]{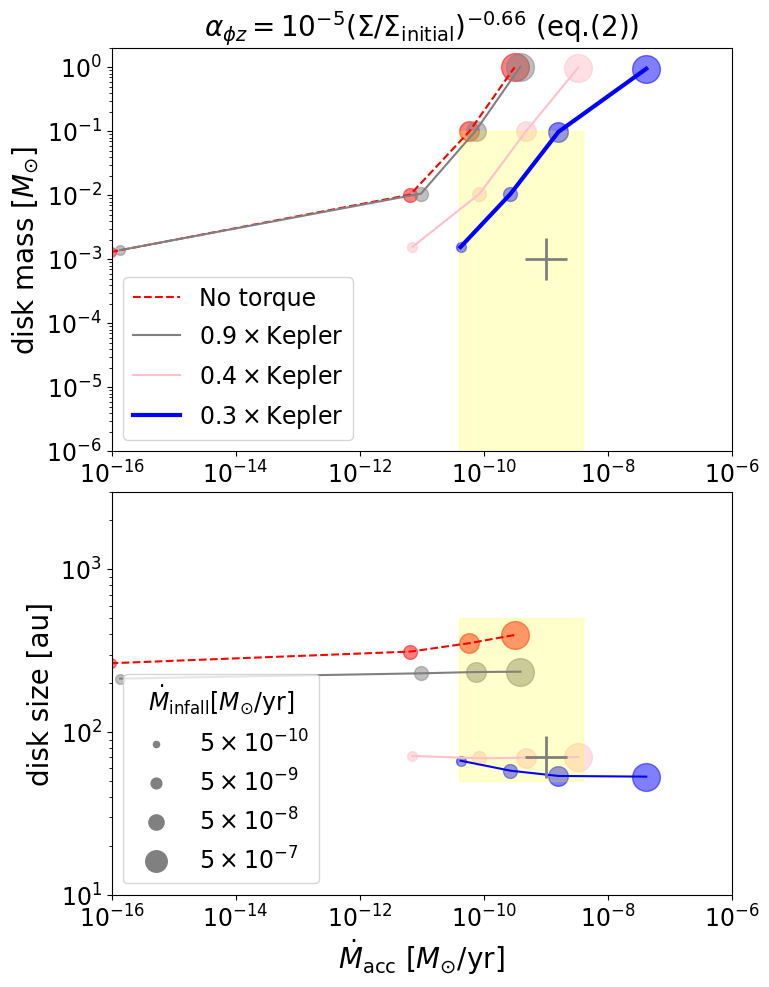}
\caption{
Same as Figure~\ref{fig:Acc_Infall_Torque_MRI}, but for the models with the relatively weak turbulent viscosity, $\alpha_{r\phi}=8\times10^{-5}$.  
}
    \label{fig:Acc_Infall_Torque_MRIinact}
\end{figure}

Figure~\ref{fig:Acc_Infall_Torque_MRIinact} compares the observed disk masses, sizes, and stellar accretion rates with our model results.
Again, the disk size decreases for lower $f_{\rm Kepler}$ values due to stronger effective torques. 
Furthermore, compared with the high-viscosity case ($\alpha_{r\phi}=8\times10^{-3}$), this model can reproduce smaller disks, which may be relevant for explaining the compact disks required to account for the CO-faint disk population in Lupus \citep{Miotello_2021}.
However, in contrast to the high-viscosity case, 
the disk mass decreases substantially with decreasing $f_{\rm Kepler}$, when compared at the same accretion rate.
This is because, in this regime, viscous spreading from smaller radii cannot efficiently resupply gas to the outer edge of the infall region, where the effective torque efficiently reduces the angular momentum around $r_{\rm infall}$ (see Figure~\ref{fig:Sigma_Torque_MRIinact}). 
This results in inward transport of the gas, since gas infalling near $r_{\rm infall}$ is assumed not to mix with the local disk gas, but instead to move instantaneously to the centrifugal radius (see Section~\ref{Method:Sub-Keplerian}).  
As a result, the stellar accretion rate is enhanced for a given infall rate. Equivalently, a given stellar accretion rate can be reproduced with a lower infall rate. Since the lower infall rate supplies a smaller mass reservoir, the disk mass required to reproduce the observed stellar accretion rate is also reduced.
This result directly derives from our model assumption of the instantaneous transport of infall gas arriving in the low-$\Sigma$ region to the corresponding centrifugal radius, and therefore uncertainty remains. Nevertheless, this might be realized in certain circumstances: For example, this would occur when infalling gas arriving near $r_{\rm infall}$ is directed inward at the shock front and subsequently reaches its centrifugal radius on a free-fall timescale. However, whether this actually occurs depends on the shape of the shock front, the velocity and trajectory of infalling gas, and thermochemical processes. This requires multidimensional simulations, but such detailed modeling is outside the scope of this study and therefore left for future work.

In contrast to the cases without the effective torque, the inclusion of the effective torque allows the models to remain within the observed range of disk masses. As $f_{\rm Kepler}$ decreases, the disk mass can remain within the observed range while still maintaining a sufficiently high stellar accretion rate. In particular, for $f_{\rm Kepler} \leq 0.3$, the models avoid producing unrealistically massive disks while still reproducing the observed stellar accretion rates.

This criterion for $f_{\rm Kepler}$ can be analytically derived by considering the condition under which the infalling gas can efficiently accrete onto the star. Specifically, we compare the accretion timescale, $t_{\rm acc}$, with the system age, $2~{\rm Myr}$. Let $r_{\rm max}$ denote the radius at which these timescales coincide, namely,
\begin{equation}
    \label{t_nu=2Myr}
    t_{\rm acc}(r_{\rm max})=2\rm Myr.
\end{equation}
For $\alpha_{r\phi}=8\times10^{-5}$, this gives
\begin{equation}
    \label{r_max_value}
    r_{\rm max}\sim4\rm au.
\end{equation}
When the viscosity is weak, the surface density at the outer edge of the infall region becomes zero. In this case, our model assumes infalling gas to instantly move inward to radii $\leq f_{\rm Kepler}^2 r_{\rm infall}$. Therefore, if
\begin{equation}
    r_{\rm max}\geq f_{\rm Kepler}^2 r_{\rm infall},
\end{equation}
most of the infalling mass can reach the central star. This yields
\begin{equation}
    f_{\rm Kepler} \lesssim 0.13.
\end{equation}
If this condition is satisfied, the infalling gas can be efficiently accreted onto the star, so that an unrealistically massive disk is no longer required to sustain a high accretion rate. This is roughly consistent with the numerical result that, for $f_{\rm Kepler} \lesssim 0.3$, the models avoid unrealistically massive disks while still reproducing the observed stellar accretion rates. A discrepancy of a factor on the order of unity is not surprising, because this analytic argument only compares $r_{\rm max}$ with the radius reached by gas that falls onto the disk at $r=r_{\rm infall}$. Our estimate is thus useful at the order-of-magnitude level.

As discussed above, for $f_{\rm Kepler} \leq 0.3$, the disk masses can remain consistent with observations while still sustaining sufficiently high stellar accretion rates.
Moreover, although the disk radius is reduced relative to the model without the effective torque, it remains within a realistic range. Therefore, even for the weak viscosity model with $\alpha_{r\phi}=8\times10^{-5}$, the observational trend reported by \citet{Winter+2024a} can be explained if the angular momentum injection is small ($f_{\rm Kepler} \leq 0.3$).

\subsubsection{Relationship between late infall and accretion rates}
\label{lateinfallvsacc_subKep}
\begin{figure}
    \centering
    \includegraphics[width=8cm]{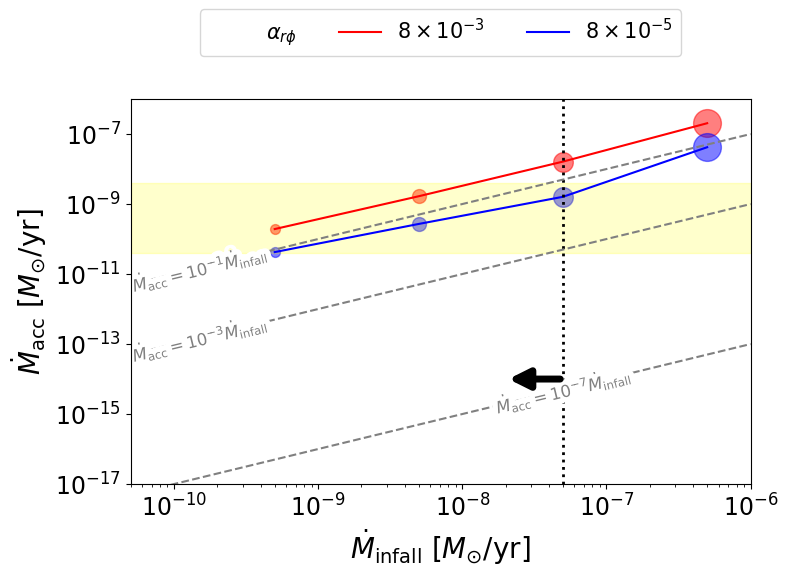}
    \caption{
Same as Figure~\ref{fig:Acc_Infall}, but for the models including the effective torque exerted by the infalling gas with $f_{\rm Kepler}=0.3$.
The red and blue lines represent the models with $\alpha_{r\phi}=8\times10^{-3}$ and $8\times10^{-5}$, respectively.
Both models adopt the $\Sigma$-dependent $\alpha_{\phi z}$.
Different marker sizes correspond to different infall rates, $\mdoti = 5\times10^{-10}$, $5\times10^{-9}$, $5\times10^{-8}$, and $5\times10^{-7}\msunyr$, with smaller markers representing lower infall rates.
    }
    \label{fig:Acc_Infall_subkep}
\end{figure}
Here, we interpret the results including the effective torque by analytically estimating the fraction of infalling gas that reaches the central star, following the approach described in Section~\ref{lateinfallvsacc}. Figure~\ref{fig:Acc_Infall_subkep} shows the relation between the infall rate onto the disk, $\mdoti$, and $\mdota$, for the models with $f_{\rm Kepler}=0.3$ at 2~Myr. This shows that in the absence of a gap, slightly over 10\% of the infalling mass reaches the star when $\alpha_{r\phi}=8\times10^{-3}$ with $f_{\rm Kepler}=0.3$, similar to the case with $f_{\rm Kepler}=1$ (see the red line in Figure~\ref{fig:Acc_Infall}). This is because the effective torque acts mainly around $r_{\rm infall}$ and does not significantly affect the inner disk, as discussed in Section~\ref{Sub-Kep_StrongVis}. 
The value of $\sim 10\%$ is determined in the same manner as described in Section~\ref{lateinfallvsacc}.

For the model with $\alpha_{r\phi}=8\times10^{-5}$ and $f_{\rm Kepler}=0.3$, slightly less than 10\% of the infalling mass reaches the central star. This is approximately 100 times larger than in the case without the effective torque (see the blue line in Figure~\ref{fig:Acc_Infall}). 
This follows directly from our model assumption that 
gas falling onto a gas-depleted region at $r=250~{\rm au}$ is instantly transported to
the centrifugal radius, namely
$f_{\rm Kepler}^2 r = 22.5~{\rm au}$. 
In other words, the infalling gas effectively arrives within $\approx 22.5{\rm \,au}$ in these models.
Of the gas deposited within this radius, only the fraction located in the region where the accretion timescale is shorter than $2~{\rm Myr}$ can ultimately reach the star; this region extends to $\lesssim 4~{\rm au}$.
The fraction of the infalling mass that
reaches the star is therefore estimated as $\left(4/22.5\right)^2 \sim 0.03$.
This is consistent with the numerical result within a factor on the order of unity.

The above discussion shows that, in the high-viscosity model with $\alpha_{r\phi}=8\times10^{-3}$, the inclusion of the effective torque only slightly increases the ratio of the stellar accretion rate to the infall rate, whereas in the low-viscosity model with $\alpha_{r\phi}=8\times10^{-5}$, this ratio is substantially enhanced. However, the latter result should be interpreted with caution, as it may be affected by uncertainties in the model assumptions.

\section{Discussion}
\label{Discussion}

This section presents the implications of our results. We first assess the key question of this paper: whether the late infall can enhance stellar accretion as suggested for the Lupus region (Section~\ref{discussion:late infall enhance}). Then, we examine the conditions for gap formation (Section~\ref{discussion:gap}), which provides a general criterion for the required infall rate to correlate with the accretion rate.
Additionally, we show that the results are insensitive to the definition of disk size (Section~\ref{Rd}). We also discuss the impact of the energy and angular momentum supplied by late infall, as well as its effects on planet formation (Section~\ref{E_and_AM} and \ref{planet_form}).

\subsection{Considering whether late infall can enhance stellar accretion}
\label{discussion:late infall enhance}

Here, we assess the plausibility of the scenario that late infall leads to high stellar accretion rates, $\dot{M}_{\ast}$. We note that this could offer an explanation of the observed trend in the Lupus region; namely, that stellar accretion is increased in the dense ISM gas of the central part of Lupus 3 \citep{Winter+2024a}.

\subsubsection{Late infall: mass supply}

In Section~\ref{Result},
we explore whether the disk mass and size remain within observationally plausible ranges, while late infall enhances accretion onto the central star, using the models where late infall is incorporated purely as a mass reservoir. In the models with a high viscosity of $\alpha_{r\phi}=8\times10^{-3}$, we find that $\mdota$ is generally high enough and disk masses remain within realistic values, but the disk sizes tend to become excessively large (Figures~\ref{fig:Acc_M_MRIact} and \ref{fig:Acc_Diskmass_size_MRI}). This occurs because angular momentum is efficiently transported outward via viscous diffusion, leading to continuous disk expansion. In contrast, the models with a low viscosity, $\alpha_{r\phi}=8\times10^{-5}$ and the $\Sigma$-dependent $\alpha_{\phi z}$ (Eq.~\eqref{alpha_pz_fiducial}) can avoid excessively large disk sizes with similar $\mdoti$, but they typically result in overly massive disks ($>0.1M_{\odot}$). In these cases, the inefficient angular momentum transport suppresses mass accretion, and material accumulates in the disk rather than being accreted onto the star.
However, if a strong wind torque is introduced ($\alpha_{\phi z} = 10^{-3}$), the low-viscosity models have succeeded in reproducing all three properties: high $\mdota$, realistic disk masses, and disk sizes that are reasonable compared to observations (Figure~\ref{fig:Acc_Infall_alphapz}). In these models, angular momentum is extracted by disk winds rather than being redistributed within the disk, allowing infalling material to accrete efficiently without causing disk expansion.

As discussed in Section~\ref{Results:constant_alpha}, a large $\alpha_{\phi z}$ naturally leads to late infall enhancing stellar accretion. Here, we discuss the plausibility of this effect.
\citet{Tabone+2025} performed a population synthesis modeling study based on observational data from several star-forming regions including Lupus. They suggested that $\alpha_{\rm DW} = 4.6 \times 10^{-4}$ best reproduces the observed distributions of $\mdota$, disk masses, and disk sizes. They used the parameter $\alpha_{\rm DW}$, which is similar to $\alpha_{\phi z}$, to characterize the strength of angular momentum removal by MHD disk winds \citep{Tabone+2022}. The relationship between $\alpha_{\phi z}$ and $\alpha_{\rm DW}$ is given by
\begin{equation}
    \label{eq:alphapz_DW}
    \alpha_{\phi z} = \frac{3\sqrt{2\pi}}{4} \frac{c_{\rm s}}{r\Omega} \alpha_{\rm DW}.
\end{equation}
Assuming that the disk temperature is entirely determined by stellar irradiation (i.e., neglecting viscous heating) and the central star has a mass of $0.2M_{\odot}$, Eq.~\eqref{eq:alphapz_DW} simplifies to
\begin{equation}
    \label{eq:alphapz_DW2}
    \alpha_{\phi z} = 0.12 \left( \frac{r}{1\unit{au}} \right)^{1/4} \alpha_{\rm DW}.
\end{equation}
By substituting $\alpha_{\rm DW} = 4.6 \times 10^{-4}$ into Eq.~\eqref{eq:alphapz_DW2}, we obtain
\begin{equation}
    \label{eq:alphapz_DW3}
    \alpha_{\phi z} = 5.5 \times 10^{-5} \left( \frac{r}{1\unit{au}} \right)^{1/4}.
\end{equation}
This yields $\alpha_{\phi z} \simeq 1.7 \times 10^{-4}$ at $r = 10^{2}~\mathrm{au}$, which is about a factor of five smaller than our favored value $\alpha_{\phi z} = 10^{-3}$. 
This apparent discrepancy likely arises from the absence of late infall 
in the MHD disk wind models of \citet{Tabone+2025}; including late infall would require a systematically higher $\alpha_{\rm DW}$ to match the observations. 
\footnote{When we performed calculations with $\alpha_{\phi z}$ set to the best-fit value of \citet{Tabone+2025} but without including late infall, both the disk mass and stellar accretion rate fell below the values observed in Lupus. This outcome reflects that our model evolves more rapidly, which would naturally shift the favored value of $\alpha_{\phi z}$ toward smaller values. Therefore, the larger $\alpha_{\phi z}$ favored in this study cannot be attributed to differences in the models themselves, but rather arises from the inclusion of late infall as an additional physical process.}

There are also simulation-based studies investigating the strength of MHD disk wind torques. \citet{Bai+2013} explored this using local shearing-box simulations, showing that when 
the plasma beta of the net vertical field at the midplane,
defined as $\beta_0 \equiv 2\rho_0 c_{\rm s}^2 / B_{z0}^2$, is sufficiently small (i.e., $\beta_0 \leq 10^4$), $\alpha_{\phi z}$ can exceed $10^{-3}$. 
It is known that for the Solar System, $\beta\sim10^{4}$ from meteorite data \citep{Fu_2014,Pascucci_2023}. Additionally, observations using the Zeeman effect have also reported that the plasma beta in TW Hya is $\beta \gtrsim 4 \times 10^{3}$ \citep{Vlemmings_2019}. Therefore, our favored value $\alpha_{\phi z} = 10^{-3}$ is within a reasonable range, though it is somewhat larger than that typically assumed in previous 1D models \citep[e.g.,][]{Suzuki+2016, Kunitomo+2020, Weder+2023}. Importantly, independent theoretical studies have also adopted values of a similar order \citep{Mori_2025}, further verifying the reasonability of the preferred $\alpha_{\phi z}$.

\citet{Winter+2024a} also suggest that there is no correlation between ISM density and dust disk mass. However, in our models, late infall increases the total disk gas mass (e.g., Figure~\ref{fig:Acc_Infall_alphapz}), implying the increase of dust disk mass. We cannot provide a definitive explanation for this discrepancy, but one possible explanation is variation in the gas-to-dust ratio among disks. \citet{Winter+2024a} derived the disk masses from dust observations, whereas we estimated them from gas distributions. In the Lupus region, the gas-to-dust ratio varies by about an order of magnitude \citep{Miotello_2017,Ruaud_2022,Trapman_2025,Deng_2025}, while late infall changes the disk mass by roughly two orders of magnitude. However, this alone may be insufficient, and other mechanisms must be taken into account. 

In summary, the value of $\alpha_{\phi z}$ required for infall to enhance stellar accretion without the effective torque is not in serious conflict with previous studies. Therefore, the scenario in which strong MHD disk winds enable infalling material to accrete efficiently onto the star appears to be physically plausible. However, a gap may still remain in explaining the absence of a clear correlation between the ISM density and the dust disk mass.

\subsubsection{Late infall: Mass supply and torque}

As shown in Section~\ref{Result:Sub-Kep}, the effective torque exerted by infalling gas provides another way to enhance stellar accretion, while keeping the disk mass and size within observationally plausible ranges. Even with $\alpha_{\phi z}$ given by Eq.~\eqref{alpha_pz_fiducial}, this can be achieved if the angular momentum injection of the infalling gas is sufficiently small. In the high-viscosity case with $\alpha_{r\phi}=8\times10^{-3}$, both the disk mass and size remain consistent with observations while stellar accretion is enhanced if the angular momentum is $\lesssim 0.4$ times the Keplerian value. In the low-viscosity case with $\alpha_{r\phi}=8\times10^{-5}$, a similar outcome is obtained for $\lesssim 0.3$ times the Keplerian value.

The underlying mechanisms differ between the two regimes. In the high-viscosity case, the torque arising from the angular momentum difference between the infalling gas and the disk suppresses radial expansion, allowing the accretion rate to increase without producing an excessively large disk. 
In contrast, in the low-viscosity case, the disk is already depleted in the relevant region. Therefore, in our models, the infalling gas is assumed to settle at the radius where its centrifugal support balances gravity, thereby maintaining a high accretion rate without significantly increasing the disk mass.

It remains uncertain whether such instantaneous settling onto the centrifugal radius occurs.
Indeed, \cite{Visser_2009} suggested that, if the finite thickness of the disk is considered, the infalling gas may hit the disk before reaching the centrifugal radius and settle at the impact location.
Previous studies have shown, both analytically \citep[e.g.,][]{Padoan_2025} and numerically \citep[e.g.,][]{Moeckel_2009}, that disk sizes depend on the angular momentum of the infalling gas. Moreover, depending on the geometry of the inflow, angular momenta as low as $0.3$--$0.4$ times the Keplerian value may be realized. However, the detailed distribution of landing locations and angular momentum of the infalling material is still poorly constrained and remains an important topic for future work.

In summary, in both the high- and low-viscosity models, infalling gas with sufficiently low angular momentum can reproduce a situation in which late infall enhances stellar accretion while keeping the disk properties observationally plausible. However, especially in the low-viscosity case, this result depends more strongly on the model assumptions, such as the treatment of gas settling onto the centrifugal radius. Therefore, verifying this scenario requires multidimensional simulations and is left for future work.

\subsection{Required infall rates}
\label{discussion:gap}

In this section, we discuss the conditions required for the late infall to enhance the stellar accretion rate. As shown in Section~\ref{Result:fiducial}, 
the absence of a gap is required for the late infall to enhance the accretion rate. Therefore, we give a rough estimate of the criterion for stable gap formation. Whether or not a stable gap opens depends on the competition between the mass flux driven by angular momentum transport and the local photoevaporation rate: If the angular momentum transport is too slow (and the infall too weak), the gas tends to be photoevaporated before it can be transported to fill the gap. 

To evaluate the competition, we performed two sets of comparisons to examine the effects of differences in the infall rate and in the mechanisms driving accretion. In the former case, we compared the models with $\alpha_{r\phi} = 8\times10^{-3}$ and $\dot{M}_{\rm infall} = 5 \times 10^{-9}$ and $5\times10^{-10} \msunyr$ with respect to the physical conditions between the models without (former) and with (latter) gap opening, as well as the effects of differences in the infall rate. In the latter case, we compared the models with $\alpha_{r\phi} = 8\times10^{-3}$, the $\Sigma$-dependent $\alpha_{\phi z}$ (Eq.~\eqref{alpha_pz_fiducial}), and with $\alpha_{r\phi} = 8\times10^{-5}$ and $\alpha_{\phi z} = 10^{-3}$ to examine how the difference in the mechanisms driving accretion affect the criterion for gap opening.

In the viscosity-dominated models, the mass loss rate by photoevaporation tends to dominate over the mass supply by late infall in the inner regions of the disk. As a result, there exists a characteristic radius, which we define as $r_{\mathrm{eq}}$, where the mass-loss rate equals the mass supply rate: $\dot{\Sigma}_{\mathrm{pw}}(r_{\mathrm{eq}})=\dot{\Sigma}_{\mathrm{infall}}(r_{\mathrm{eq}})$. Inside $r_{\mathrm{eq}}$, photoevaporation dominates and removes material faster than it is supplied, whereas outside $r_{\mathrm{eq}}$, infall dominates, and the disk can gain mass. In our fiducial photoevaporation model, $r_{\mathrm{eq}}$ is approximately 20~au for $\mdoti = 5 \times 10^{-9}~\msunyr$, and about 100~au for $\mdoti = 5 \times 10^{-10}~\msunyr$. These values illustrate that weaker infall leads to a broader region where photoevaporation dominates, making gap formation more likely.

Since mass tends to accumulate outside $r_{\mathrm{eq}}$, we evaluate whether all of it is dispersed by photoevaporation during its inward transport from $r_{\mathrm{eq}}$ to the central star. When accretion is driven by viscosity, the timescale to reach the central star is estimated by the viscous timescale, $t_{\nu}$, which is defined as Eq.~\eqref{eq:t_nu0}.
The actual values of the viscous timescale are $t_{\nu}(r_{\mathrm{eq}}=20 \unit{au}) \simeq 0.2 \unit{Myr}$ for $\mdoti = 5 \times 10^{-9}~\msunyr$, and $t_{\nu}(r_{\mathrm{eq}}=10^{2} \unit{au}) \simeq 1\unit{Myr}$ for $\mdoti = 5 \times 10^{-10}~\msunyr$.
The condition for gap formation, under viscosity-dominated angular momentum transport (i.e., when all the mass is removed before the accretion), is
\begin{equation}
    \Sigma(r_{\mathrm{eq}})
    <\int_{0}^{r_{\mathrm{eq}}}\dot{\Sigma}_{\rm pw}(r') \frac{dt_{\nu}(r')}{dr'}dr' 
     \sim\dot{\Sigma}_{\rm pw}(r_{\mathrm{eq}})t_{\nu}(r_{\mathrm{eq}}).
    \label{eq:condition_visc}
\end{equation}
The surface density is evaluated at the time of interest. $\frac{dt_{\nu}(r')}{dr'}dr'$ in the integral represents the duration for which the gas resides at radius $r'$. We ignore the infall rate because infall is not dominant compared to photoevaporation for $r< r_{\rm eq}$. For $\mdoti = 5 \times 10^{-9}~\msunyr$,  we have $\dot{\Sigma}_{\rm pw}(r_{\mathrm{eq}})t_{\nu}(r_{\mathrm{eq}})\sim4\times10^{-2}\unit{g/cm^{2}}$; thus, the inequality \eqref{eq:condition_visc} is not satisfied, 
\begin{equation}
\Sigma(r_{\mathrm{eq}})=8\unit{g/cm^2}>\dot{\Sigma}_{\rm pw}(r_{\mathrm{eq}})t_{\nu}(r_{\mathrm{eq}}).
\end{equation}
Indeed, this case does not show gap formation throughout the evolution (top right panel of Figure~\ref{fig:Sigma_MRIact}). 
In contrast, $\mdoti = 5 \times 10^{-10}~\msunyr$ yields $\dot{\Sigma}_{\rm pw}(r_{\mathrm{eq}})t_{\nu}(r_{\mathrm{eq}})\sim2\times10^{-2}\unit{g/cm^{2}}$, and the gap-opening condition is met via
\begin{equation}
\Sigma(r_{\mathrm{eq}})=1.1\times10^{-2}\unit{g/cm^2}<\dot{\Sigma}_{\rm pw}(r_{\mathrm{eq}})t_{\nu}(r_{\mathrm{eq}}), 
\end{equation}
which is consistent with the presence of a gap in this model (bottom left panel of Figure~\ref{fig:Sigma_MRIact}).

Next, we compared the MHD-wind-dominated model ($\alpha_{r\phi} = 8\times10^{-5}$ and $\alpha_{\phi z}=10^{-3}$)
and the turbulent-viscosity-dominated model ($\alpha_{r\phi} = 8\times10^{-3}$ and $\Sigma$-dependent $\alpha_{\phi z}$) under the same infall rate of $\mdoti = 5 \times 10^{-9}~\msunyr$. When the advection driven by the MHD disk wind torque is dominant over the turbulent viscosity, Eq.~\eqref{eq:condition_visc} is modified to
\begin{equation}
    \Sigma(r_{\mathrm{eq}})<\int_{0}^{r_{\mathrm{eq}}}\dot{\Sigma}_{\rm pw}(r') \frac{dt_{\rm mw}(r')}{dr'}dr'  \sim\dot{\Sigma}_{\rm pw}(r_{\mathrm{eq}})t_{\rm mw}(r_{\mathrm{eq}}),
    \label{eq:condition_mw}
\end{equation}
where $t_{\rm mw}$ is the timescale of the advection driven by the MHD disk wind torque.
We assume the same $\mdoti$ of $5\times10^{-9}\msunyr$ for these models. 
The timescale, $t_{\rm mw}$, is given by Eq.~\eqref{eq:adv_timescale0}
and its value at the characteristic radius, $r_{\mathrm{eq}}$, which is 20~au with $\mdoti=5\times10^{-9}\msunyr$, is given by 
\begin{equation}
    \label{eq:adv_timescale1}
    t_{\rm mw}(r_{\mathrm{eq}}=20\unit{au}) \sim 6.6\unit{Myr}\times\left(\frac{\alpha_{\phi z}}{10^{-5}}\right)^{-1}.
\end{equation}

In the model with $\alpha_{r\phi} = 8\times10^{-5}$ and $\alpha_{\phi z} = 10^{-3}$, the advection timescale is $t_{\rm mw} \simeq 6.6\times10^{4} \unit{yr}$ (while $t_{\nu} \simeq 20 \unit{Myr}$). In contrast to the case with $\alpha_{r\phi} = 8\times10^{-3}$ and $\Sigma$-dependent $\alpha_{\phi z}$, the accretion timescale is well approximated by $t_{\rm mw}$. Consequently, the condition for gap opening is determined by Eq.~\eqref{eq:condition_mw}.
The right-hand side of the condition (Eq.~\eqref{eq:condition_mw}) is estimated as $\dot{\Sigma}_{\rm pw}(r_{\mathrm{eq}})t_{\rm mw} (r_{\mathrm{eq}}) \sim 1.4 \times10^{-2} \unit{g/cm^2}$ and the gap-opening condition is not met ($\Sigma(r_{\mathrm{eq}}) \sim 1 \unit{g/cm^2} > \dot{\Sigma}_{\rm pw}(r_{\mathrm{eq}}) t_{\rm mw}(r_{\mathrm{eq}})$).
This agrees with the absence of a gap in the upper panel of Figure~\ref{fig:Sigma_alphapz}.

Next, we discuss the criterion for the models with the effective torque. In some cases, the torque induced by the angular momentum difference between the infalling gas and the disk gas drives mass transport, thereby filling the gap. Following the previous discussion, the criterion is estimated as follows.
From Eq.~\eqref{eq:effective_torque_source_temp}, the advection velocity driven by the torque exerted by the infalling gas, $v_{\rm tq}$, is estimated as
\begin{equation}
    v_{\rm tq}\sim2r(1-f_{\rm Kepler})\frac{\dot{\Sigma}_{\rm infall}}{\Sigma}.
\end{equation}
The corresponding advection timescale is then
\begin{equation}
    t_{\rm tq}=\frac{r}{v_{\rm tq}}\sim\frac{\Sigma}{2(1-f_{\rm Kepler})\dot{\Sigma}_{\rm infall}}.
\end{equation}
Using the same criterion as in the case without the effective torque, we obtain
\begin{multline}
    \Sigma(r_{\rm eq})\geq \int_{0}^{r_{\rm eq}}\dot{\Sigma}_{\rm pw}(r') \frac{dt_{\rm tq}(r')}{dr'}dr' \\
    \sim t_{\rm tq}\dot{\Sigma}_{\rm pw}(r_{\rm eq})
    \sim\frac{\Sigma(r_{\rm eq})}{2(1-f_{\rm Kepler})\dot{\Sigma}_{\rm infall}(r_{\rm eq})}\dot{\Sigma}_{\rm pw}(r_{\rm eq}).
\end{multline}
At $r=r_{\rm eq}$, we have $\dot{\Sigma}_{\rm infall}(r_{\rm eq})=\dot{\Sigma}_{\rm pw}(r_{\rm eq})$. This yields the approximate criterion, expressed as
\begin{equation}
    \label{criterion:Torque}
    f_{\rm Kepler}\leq0.5.
\end{equation}
Indeed, under this condition, the gap that existed in the case without the effective torque disappears (see Figures~\ref{fig:Acc_Infall_Torque_MRI} and~\ref{fig:Acc_Infall_Torque_MRIinact}).

The criteria (Eqs.~\eqref{eq:condition_visc}, \eqref{eq:condition_mw} and \eqref{criterion:Torque}) can also be applied to estimate gap formation in cases not explicitly examined here. Their predictive accuracy diminishes when the left- and right-hand sides are comparable in magnitude (e.g., in the model with $\mdoti=5\times10^{-9}\msunyr$ and $10\times$ the fiducial UV and X-ray radiation), owing to the approximation used in the integral. Nonetheless, they provide a useful diagnostic tool for assessing gap formation, illustrating the underlying physics at work.

\subsection{Impact of disk size definition}
\label{Rd}
\begin{figure}
    \centering
    \includegraphics[width=8cm]{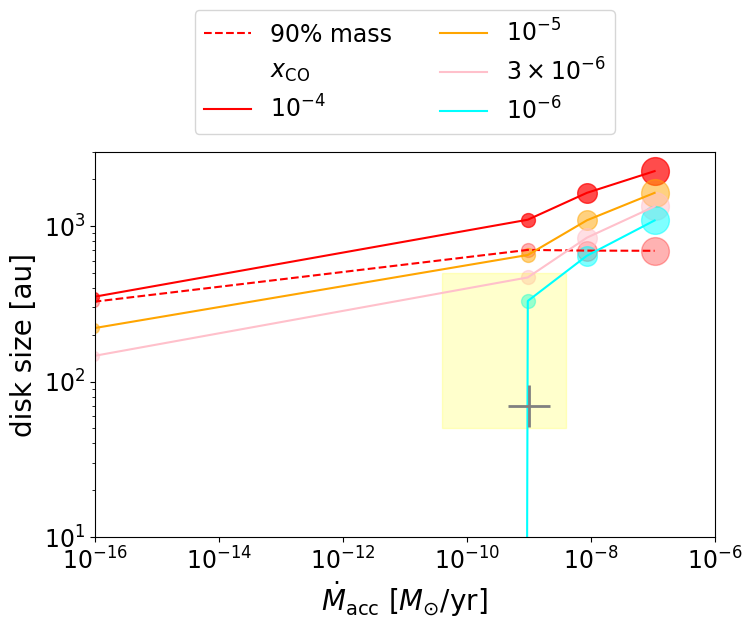}
    \caption{Same as the lower panel of Figure~\ref{fig:Acc_Diskmass_size_MRI}, but showing the effects of increasing CO gas depletion. The models with $\alpha_{r\phi} = 8\times10^{-3}$ and the $\Sigma$-dependent MHD disk wind torque are assumed. The dashed red line shows the radius enclosing 90\% of the total disk mass. The solid red, orange, pink, and aqua lines show the reduction in disk size when the gas-phase CO abundance is decreased to $x_{\rm CO} = 10^{-4}, 10^{-5}, 3 \times 10^{-6}$, and $10^{-6}$, respectively. Different marker sizes correspond to different infall rates of $\mdoti = 5\times10^{-10}$, $5\times10^{-9}$, $5\times10^{-8}$, and $5\times10^{-7}\msunyr$, respectively, with smaller markers representing lower rates.
}
    \label{fig:Acc_CO_x_CO}
\end{figure}

The observed disk size corresponds to the CO-emitting region and is therefore sensitive to the CO abundance within the disk. In the following, we examine how the inferred disk radius depends on both the CO abundance and the definition of disk size, and assess their impacts on our results. 
We focus on the $\alpha_{r\phi} = 8\times10^{-3}$ model, which reproduces disk masses and accretion rates well but predicts too large disk sizes (Section~\ref{Result:fiducial}; for example, see red lines in Figure~\ref{fig:Acc_Diskmass_size_MRI}), and evaluate whether adjusting these parameters allows this model to yield disk sizes consistent with observations.

Figure~\ref{fig:Acc_CO_x_CO} presents the correlation between $\mdota$ and the CO disk radius for various CO abundances, along with the radius enclosing 90\% of the total mass. The latter (dashed red line) is generally smaller than the CO disk radius with $x_{\rm CO}=10^{-4}$ (red line).

The red, orange, pink, and cyan lines represent results computed using CO abundances of $10^{-4}$, $10^{-5}$, $3\times10^{-6}$, and $10^{-6}$, respectively, to define the disk radius. When $x_{\rm CO} > 3\times10^{-6}$, the resulting disk radius in the models with $\alpha_{r\phi} = 8\times10^{-3}$ becomes too large to be consistent with observations. Agreement with observed disk sizes is achieved only when $x_{\rm CO} \sim 10^{-6}$; however, this value represents the lower end of observational estimates \citep{Deng_2025} and likely corresponds to an extreme scenario. Moreover, our models incorporate late infall from the interstellar medium, where the CO abundance is expected to be close to the canonical ISM value ($x_{\rm CO} \sim 10^{-4}$) \citep{Frerking_1982}. Therefore, such a low CO abundance is unrealistic in the context of the models with $\alpha_{r\phi} = 8\times10^{-3}$, making it unlikely that their predicted disk sizes are consistent with observations. In \cite{Zagaria+2023}, a fudge factor of $\sim 1/40$ (for Lupus) was introduced, which is effectively equivalent to reducing $x_{\rm CO}$ by the same factor. This corresponds to $x_{\rm CO} \simeq 3 \times 10^{-6}$. Even in this case, our results indicate that it remains difficult to reproduce the observations for high viscosity.

Another possible definition is to identify the disk region as the range where the azimuthal velocity, $v_\phi$, exceeds twice the radial velocity, $v_r$, and to define the disk radius accordingly \citep{Joos_2012}. However, this method is not applicable to the present study, because we have assumed Keplerian rotation throughout the entire computational domain and treated the radial velocity solely as that arising from angular momentum transport from the outset.

From these results, even when varying the method of calculating the disk radius, it remains difficult to reproduce disk sizes consistent with the Lupus observations for $\alpha_{r\phi}=8\times10^{-3}$. Therefore,
strong viscosity models are unfavored to yield the correlation between late infall and stellar accretion. 

\subsection{Impact of the energy and angular momentum supplied by late infall on disk evolution}
\label{E_and_AM}
Late infall not only adds gas mass but also injects energy and angular momentum, both of which can affect disk evolution and mass transport. An important consequence of the injected energy is the amplification of turbulence. The energy added to the disk can be transferred to the gas, enhancing the turbulence and potentially increasing $\alpha_{r\phi}$. This enhancement is expected to be confined to the region where late infall occurs. Numerical results by \citet{Winter+2024b} showed that the energy from late infall can increase $\alpha_{r\phi}$ to values that are on the order of $10^{-3}$–$10^{-2}$, comparable to the strong turbulent viscosity adopted in this study. Therefore, turbulence enhanced by late infall may sustain a sufficiently high accretion rate. Moreover, since the turbulence is strengthened only locally, infall limited to the inner disk may avoid excessive disk expansion. It should be noted, however, that if $\alpha_{r\phi}$ becomes sufficiently large in the outer regions (>500~au), the material there tends to spread outward, leading to disk expansion. In addition, for efficient accretion onto the central star, turbulence must also be strengthened in the innermost regions of the disk. In particular, the latter implies that a substantial amount of energy must be injected even into the innermost regions near the central star. Reproducing such a situation in real astrophysical environments would require rather contrived conditions.

Beyond these energy-driven effects and the effective torque considered above, the angular momentum carried by late infall may also have important dynamical consequences. One possible outcome is the formation of warps or misalignments between the inner and outer disk, arising from the difference in orientation between the angular momentum of the late infall and that of the pre-existing disk gas. This phenomenon has already been pointed out both observationally and theoretically in previous studies \citep[e.g.,][]{Dullemond_2019,Ginski_2021}. Warps or misalignments can change how stellar radiation impinges on the disk, which in turn modifies the photoevaporation rate and thus the likelihood of gap formation. Although this aspect lies beyond the scope of our analysis, it may be interesting to explore these additional effects of late infall on the long-term disk evolution. 

\subsection{Effects on planet formation}
\label{planet_form}

We discuss the implications of our results for planet formation.
One important consequence of the late infall is the additional supply of dust, which provides the raw material for building planets. 
Assuming a dust-to-gas mass ratio of 0.01 with our most favored infall rate, $\mdoti=5\times10^{-9}\msunyr$, the dust mass supplied over 2~Myr is about $30M_\oplus$, which is roughly consistent with the masses of exoplanetary systems around $0.2\Msun$ stars \citep{Manara_2018}. Thus, late infall has the potential to provide a solution to the issue of insufficient disk dust mass for planet formation.

Another important consequence is a potential explanation for the observed correlation between stellar kinematics and planet properties \citep{Hamer_2019,Winter_2020}. \citet{Winter+2024b} proposed that variations in relative velocity with the ISM lead to differences in the amount of late infall, thereby giving rise to the correlation. This trend can be understood in the context of BHL accretion, where a higher relative velocity between the ISM and the disk leads to a lower infall rate. Our simulations show that such a reduction in the late infall rate promotes the formation of a gap, whose presence strongly constrains the regions available for planet formation. Consequently, our study strengthens the case that late infall provides a physical mechanism linking stellar kinematics to planetary properties.

Additionally, if a gap is present, infalling gas fails to go into the inner regions. This can cause the molecular compositions of inner and outer planets to diverge. This provides a potential link between infall dynamics and planetary chemistry.

\section{Conclusion}
\label{Conclusion}

In this study, we investigated the effects of late infall on long-term ($\sim$ 1--10~Myr) protoplanetary disk evolution. We employed a set of 1D disk evolution models incorporating the effects of photoevaporation, MHD disk winds, and late infall. As a first step, we adopted a fiducial infall model in which mass is uniformly deposited within a certain radius ($r_{\rm infall} = 250\,\unit{au}$) and supplied at a constant rate throughout the simulation. By systematically varying the infall parameters and comparing the outcomes with observational constraints from the Lupus region \citep{Winter+2024a}, we clarified the conditions under which late infall can substantially enhance stellar accretion rates, while still remaining consistent with the observed range of disk properties, such as masses and sizes. Additionally, by performing calculations with and without the effective torque exerted by the infalling gas, we disentangled the effect of the infalling gas as a source of torque from its mass-reservoir effect.

Our results identify two viable pathways by which late infall can sustain stellar accretion, while maintaining disk masses and radii consistent with observational constraints: (i) efficient angular-momentum extraction by MHD disk winds; or (ii) low-angular-momentum infall, whose angular-momentum mismatch with the disk promotes inward gas transport without driving excessive viscous spreading. Our main findings are summarized as follows:
\begin{enumerate}
    \item 
In the case with no effective torque and the commonly assumed MHD disk wind torque (Eq.~\eqref{alpha_pz_fiducial}), the turbulent viscous torque parametrized by $\alpha_{r \phi}$ is the primary driver of the mass and angular momentum transport for the gas supplied by the late infall. However, in such models, the infall rate, $\mdoti$, required to sustain sufficient stellar accretion leads to inconsistent outcomes with observations: 
high-viscosity disks ($\alpha_{r \phi} = 8 \times 10^{-3}$) become overly extended ($>1000~\unit{au}$), whereas low-viscosity disks ($\alpha_{r \phi} = 8 \times 10^{-5}$) become overly massive ($>0.1~M_\odot$; Figure~\ref{fig:Acc_Diskmass_size_MRI}). 
This reflects the opposite limitations of the two regimes: strong viscosity enhances outward spreading, while weak viscosity makes angular momentum transport inefficient and leaves a large disk mass for a given accretion rate. 
Thus, viscosity alone cannot make late infall an efficient mechanism for enhancing stellar accretion while keeping disk masses and sizes consistent with observations.

\item 
Introducing a strong MHD disk wind torque ($\alpha_{\phi z} = 10^{-3}$) can resolve the issue described above, without introducing the effective torque exerted by infalling gas. 
In this case, only late infall with $\mdoti = 5 \times 10^{-9}~\msunyr$ can sustain stellar accretion, while keeping the disk mass and size within the observed ranges (Figure~\ref{fig:Acc_Infall_alphapz}). In such cases, a higher $\mdoti$ results in overly massive disks. The favored $\alpha_{\phi z}$ value is plausible from both observational and theoretical perspectives (Section~\ref{discussion:late infall enhance}). Importantly, if such an infall rate is maintained for 2 Myr, the supplied dust mass becomes sufficient for exoplanet formation, suggesting that late infall may also provide a potential solution to the issue of insufficient disk dust mass for planet formation.

\item 
Considering the effective torque exerted by the infalling gas provides a simple way to reconcile the models with the observed disk sizes, disk masses, and stellar accretion rates, even without invoking a strong MHD disk wind torque. For $\alpha_{r\phi}=8\times10^{-3}$, the effective torque suppresses disk expansion, yielding disk sizes that are consistent with observations when the supplied angular momentum is sub-Keplerian ($\lesssim 0.4$ times the Keplerian value). For $\alpha_{r\phi}=8\times10^{-5}$, the inward transport of the infalling gas enhances stellar accretion and allows the observed accretion rates to be reproduced with realistic disk masses when the supplied angular momentum is somewhat smaller ($\lesssim 0.3$ times the Keplerian value). However, the latter result should be interpreted with caution, because it relies on the assumption that gas in the low-$\Sigma$ region immediately moves to its centrifugal radius. Verifying this process requires multidimensional simulations and is left for future work.

\end{enumerate}

The above results are largely insensitive to initial conditions, as the disk mass in our models is primarily governed by the infalling material rather than the initial setup. Varying the late infall properties, such as radial distribution and duration, also has little impact on the conclusions.
Furthermore, the results are also insensitive to the definition of the disk size (Section~\ref{Rd}). When the disk size is defined by the radius enclosing 90\% of the disk mass instead of the CO radius, the conclusions remain unchanged. Moreover, in the cases with $\alpha_{r\phi} = 8\times10^{-3}$, the disk size becomes excessively large regardless of variations in CO abundance within observationally plausible ranges. This means that cases with high turbulent viscosity and without the effective torque consistently yield an excessively large disk size, irrespective of CO abundance. This strengthens our conclusion that viscous disks alone are unlikely to yield a correlation between infall and accretion rates.

If sufficient mass transport or late infall is absent, photoevaporation opens a large central gap in the disk, quenching accretion onto the central star at a level that spans several orders of magnitude. The presence or absence of the gap can be analytically estimated by comparing the accretion timescale, the photoevaporation rate, and the surface density. This offers an apt explanation of our numerical results (Section~\ref{discussion:gap}).

Our results indicate that either strong wind torque or effective infall torque is necessary for late infall to enhance stellar accretion, while keeping the disk masses and sizes within the observed ranges of Lupus disks. A key challenge that remains is related to our model prediction of a positive correlation between gas disk masses and accretion rates, whereas observations show no correlation between dust disk masses and accretion rates. This suggests that further investigations into the diversity of infall histories and angular momentum transport are needed. Notwithstanding, our study provides a concrete step toward improving our understanding of the role of late infall in long-term disk evolution.

\begin{acknowledgements}
We are grateful to Shoji Mori, Takeru K. Suzuki, Gregory Herczeg, Emilie Habart, Giovanni Rosotti, Lorenzo Malanga, Luigi Zallio, Cornelis Dullemond, Hiroshi Kobayashi, Shuichiro Inutsuka, Kiyoaki Doi, Jorge P\'erez Gonz\'alez, and Le\'on-Alexander H\"uhn for many stimulating discussions. We also thank Takahiro Tanaka for offering helpful suggestions on broader aspects of this study. We are grateful to the anonymous referee for their insightful comments, which have significantly improved the quality of the manuscript.
This research was made possible thanks to support from the Japan Society for the Promotion of Science through Grants-in-Aid for Scientific Research (TH: 19KK0353, 22H00149, HM: 25K17432). This work was also supported by JST SPRING (WO: JPMJSP2110). W.O. is supported by the Graduate School of Science, Kyoto University, under the Ginpu Fund. R.N. acknowledges support from the European Union (ERC Starting Grant DiscEvol, project number 101039651) and from Fondazione Cariplo, grant No. 2022-1217. T.H. appreciates the financial support from the Kyoto University Foundation. H.M. acknowledges the support of the JSPS Overseas Research Fellowship.
\end{acknowledgements}

\bibliography{reference}

@ARTICLE{Winter+2018,
       author = {{Winter}, A.~J. and {Clarke}, C.~J. and {Rosotti}, G. and {Ih}, J. and {Facchini}, S. and {Haworth}, T.~J.},
        title = "{Protoplanetary disc truncation mechanisms in stellar clusters: comparing external photoevaporation and tidal encounters}",
      journal = {\mnras},
         year = 2018,
        month = aug,
       volume = {478},
       number = {2},
        pages = {2700-2722},
          doi = {10.1093/mnras/sty984},
archivePrefix = {arXiv},
       eprint = {1804.00013},
 primaryClass = {astro-ph.SR},
       adsurl = {https://ui.adsabs.harvard.edu/abs/2018MNRAS.478.2700W}
}

@ARTICLE{Nakamoto+1994,
       author = {{Nakamoto}, Taishi and {Nakagawa}, Yoshitsugo},
        title = "{Formation, Early Evolution, and Gravitational Stability of Protoplanetary Disks}",
      journal = {\apj},
         year = 1994,
        month = feb,
       volume = {421},
        pages = {640},
          doi = {10.1086/173678},
       adsurl = {https://ui.adsabs.harvard.edu/abs/1994ApJ...421..640N}
}

@ARTICLE{Sellek+2024,
       author = {{Sellek}, A.~D. and {Grassi}, T. and {Picogna}, G. and {Rab}, Ch. and {Clarke}, C.~J. and {Ercolano}, B.},
        title = "{Photoevaporation of protoplanetary discs with PLUTO+PRIZMO: I. Lower X-ray{\textendash}driven mass-loss rates due to enhanced cooling}",
      journal = {\aap},
         year = 2024,
        month = oct,
       volume = {690},
          eid = {A296},
        pages = {A296},
          doi = {10.1051/0004-6361/202450171},
archivePrefix = {arXiv},
       eprint = {2408.00848},
 primaryClass = {astro-ph.EP},
       adsurl = {https://ui.adsabs.harvard.edu/abs/2024A&A...690A.296S}
}

@ARTICLE{Tabone+2025,
       author = {{Tabone}, Beno{\^\i}t and {Rosotti}, Giovanni P. and {Trapman}, Leon and {Pinilla}, Paola and {Pascucci}, Ilaria and {Somigliana}, Alice and {Alexander}, Richard and {Vioque}, Miguel and {Anania}, Rossella and {Kuznetsova}, Aleksandra and {Zhang}, Ke and {P{\'e}rez}, Laura M. and {Cieza}, Lucas A. and {Carpenter}, John and {Deng}, Dingshan and {Agurto-Gangas}, Carolina and {Ruiz-Rodriguez}, Dary A. and {Sierra}, Anibal and {Kurtovic}, Nicol{\'a}s T. and {Miley}, James and {Gonz{\'a}lez-Ruilova}, Camilo and {TorresVillanueva}, Estephani and {Hogerheijde}, Michiel R. and {Schwarz}, Kamber and {Toci}, Claudia and {Testi}, Leonardo and {Lodato}, Giuseppe},
        title = "{The ALMA Survey of Gas Evolution of PROtoplanetary Disks (AGE-PRO). VII. Testing Accretion Mechanisms from Disk Population Synthesis}",
      journal = {\apj},
         year = 2025,
        month = aug,
       volume = {989},
       number = {1},
          eid = {7},
        pages = {7},
          doi = {10.3847/1538-4357/adc7b1},
archivePrefix = {arXiv},
       eprint = {2506.10742},
 primaryClass = {astro-ph.EP},
       adsurl = {https://ui.adsabs.harvard.edu/abs/2025ApJ...989....7T}
}

@ARTICLE{Clarke+2001,
       author = {{Clarke}, C.~J. and {Gendrin}, A. and {Sotomayor}, M.},
        title = "{The dispersal of circumstellar discs: the role of the ultraviolet switch}",
      journal = {\mnras},
         year = 2001,
        month = dec,
       volume = {328},
       number = {2},
        pages = {485-491},
          doi = {10.1046/j.1365-8711.2001.04891.x},
       adsurl = {https://ui.adsabs.harvard.edu/abs/2001MNRAS.328..485C}
}

@ARTICLE{Lynden-Bell+1974,
       author = {{Lynden-Bell}, D. and {Pringle}, J.~E.},
        title = "{The evolution of viscous discs and the origin of the nebular variables.}",
      journal = {\mnras},
         year = 1974,
        month = sep,
       volume = {168},
        pages = {603-637},
          doi = {10.1093/mnras/168.3.603},
       adsurl = {https://ui.adsabs.harvard.edu/abs/1974MNRAS.168..603L}
}

@ARTICLE{Winter+2024b,
       author = {{Winter}, Andrew J. and {Benisty}, Myriam and {Andrews}, Sean M.},
        title = "{Planet Formation Regulated by Galactic-scale Interstellar Turbulence}",
      journal = {\apjl},
         year = 2024,
        month = sep,
       volume = {972},
       number = {1},
          eid = {L9},
        pages = {L9},
          doi = {10.3847/2041-8213/ad6d5d},
archivePrefix = {arXiv},
       eprint = {2405.08451},
 primaryClass = {astro-ph.EP},
       adsurl = {https://ui.adsabs.harvard.edu/abs/2024ApJ...972L...9W}
}

@ARTICLE{Winter+2024a,
       author = {{Winter}, Andrew J. and {Benisty}, Myriam and {Manara}, Carlo F. and {Gupta}, Aashish},
        title = "{Spatially correlated stellar accretion in the Lupus star-forming region: Evidence for ongoing infall from the interstellar medium}",
      journal = {\aap},
         year = 2024,
        month = nov,
       volume = {691},
          eid = {A169},
        pages = {A169},
          doi = {10.1051/0004-6361/202452120},
archivePrefix = {arXiv},
       eprint = {2409.17220},
 primaryClass = {astro-ph.EP},
       adsurl = {https://ui.adsabs.harvard.edu/abs/2024A&A...691A.169W}
}

@ARTICLE{Kunitomo+2020,
       author = {{Kunitomo}, Masanobu and {Suzuki}, Takeru K. and {Inutsuka}, Shu-ichiro},
        title = "{Dispersal of protoplanetary discs by the combination of magnetically driven and photoevaporative winds}",
      journal = {\mnras},
         year = 2020,
        month = mar,
       volume = {492},
       number = {3},
        pages = {3849-3858},
          doi = {10.1093/mnras/staa087},
archivePrefix = {arXiv},
       eprint = {2001.03949},
 primaryClass = {astro-ph.EP},
       adsurl = {https://ui.adsabs.harvard.edu/abs/2020MNRAS.492.3849K}
}

@ARTICLE{Komaki+2023,
       author = {{Komaki}, A. and {Fukuhara}, S. and {Suzuki}, T.~K. and {Yoshida}, N.},
        title = "{Simulations of Protoplanetary Disk Dispersal: Stellar Mass Dependence of the Disk Lifetime}",
      journal = {arXiv e-prints},
         year = 2023,
        month = apr,
          eid = {arXiv:2304.13316},
        pages = {arXiv:2304.13316},
          doi = {10.48550/arXiv.2304.13316},
archivePrefix = {arXiv},
       eprint = {2304.13316},
 primaryClass = {astro-ph.SR},
       adsurl = {https://ui.adsabs.harvard.edu/abs/2023arXiv230413316K}
}

@ARTICLE{Weder+2023,
       author = {{Weder}, Jesse and {Mordasini}, Christoph and {Emsenhuber}, Alexandre},
        title = "{Population study on MHD wind-driven disc evolution. Confronting theory and observation}",
      journal = {\aap},
         year = 2023,
        month = jun,
       volume = {674},
          eid = {A165},
        pages = {A165},
          doi = {10.1051/0004-6361/202243453},
archivePrefix = {arXiv},
       eprint = {2304.12380},
 primaryClass = {astro-ph.EP},
       adsurl = {https://ui.adsabs.harvard.edu/abs/2023A&A...674A.165W}
}

@ARTICLE{Ooyama+2025,
       author = {{Ooyama}, Wataru and {Nakatani}, Riouhei and {Hosokawa}, Takashi and {Mitani}, Hiroto and {Turner}, Neal J.},
        title = "{Secret of Longevity: Protoplanetary Disks as a Source of Gas in Debris Disks}",
      journal = {\apj},
         year = 2025,
        month = apr,
       volume = {983},
       number = {2},
          eid = {153},
        pages = {153},
          doi = {10.3847/1538-4357/adbbb8},
       adsurl = {https://ui.adsabs.harvard.edu/abs/2025ApJ...983..153O}
}

@ARTICLE{Suzuki+2016,
       author = {{Suzuki}, Takeru K. and {Ogihara}, Masahiro and {Morbidelli}, Alessandro and {Crida}, Aur{\'e}lien and {Guillot}, Tristan},
        title = "{Evolution of protoplanetary discs with magnetically driven disc winds}",
      journal = {\aap},
         year = 2016,
        month = dec,
       volume = {596},
          eid = {A74},
        pages = {A74},
          doi = {10.1051/0004-6361/201628955},
archivePrefix = {arXiv},
       eprint = {1609.00437},
 primaryClass = {astro-ph.EP},
       adsurl = {https://ui.adsabs.harvard.edu/abs/2016A&A...596A..74S}
}

@ARTICLE{Bai+2013,
       author = {{Bai}, Xue-Ning},
        title = "{Wind-driven Accretion in Protoplanetary Disks. II. Radial Dependence and Global Picture}",
      journal = {\apj},
         year = 2013,
        month = aug,
       volume = {772},
       number = {2},
          eid = {96},
        pages = {96},
          doi = {10.1088/0004-637X/772/2/96},
archivePrefix = {arXiv},
       eprint = {1305.7232},
 primaryClass = {astro-ph.EP},
       adsurl = {https://ui.adsabs.harvard.edu/abs/2013ApJ...772...96B}
}

@ARTICLE{Kunitomo+2021,
       author = {{Kunitomo}, Masanobu and {Ida}, Shigeru and {Takeuchi}, Taku and {Pani{\'c}}, Olja and {Miley}, James M. and {Suzuki}, Takeru K.},
        title = "{Photoevaporative Dispersal of Protoplanetary Disks around Evolving Intermediate-mass Stars}",
      journal = {\apj},
         year = 2021,
        month = mar,
       volume = {909},
       number = {2},
          eid = {109},
        pages = {109},
          doi = {10.3847/1538-4357/abdb2a},
archivePrefix = {arXiv},
       eprint = {2103.07673},
 primaryClass = {astro-ph.EP},
       adsurl = {https://ui.adsabs.harvard.edu/abs/2021ApJ...909..109K}
}

@ARTICLE{Tanaka+2013,
       author = {{Tanaka}, Kei E.~I. and {Nakamoto}, Taishi and {Omukai}, Kazuyuki},
        title = "{Photoevaporation of Circumstellar Disks Revisited: The Dust-free Case}",
      journal = {\apj},
         year = 2013,
        month = aug,
       volume = {773},
       number = {2},
          eid = {155},
        pages = {155},
          doi = {10.1088/0004-637X/773/2/155},
archivePrefix = {arXiv},
       eprint = {1306.6623},
 primaryClass = {astro-ph.SR},
       adsurl = {https://ui.adsabs.harvard.edu/abs/2013ApJ...773..155T}
}

@ARTICLE{Owen+2012,
       author = {{Owen}, James E. and {Clarke}, Cathie J. and {Ercolano}, Barbara},
        title = "{On the theory of disc photoevaporation}",
      journal = {\mnras},
         year = 2012,
        month = may,
       volume = {422},
       number = {3},
        pages = {1880-1901},
          doi = {10.1111/j.1365-2966.2011.20337.x},
archivePrefix = {arXiv},
       eprint = {1112.1087},
 primaryClass = {astro-ph.SR},
       adsurl = {https://ui.adsabs.harvard.edu/abs/2012MNRAS.422.1880O}
}

@ARTICLE{Siess+2000,
       author = {{Siess}, L. and {Dufour}, E. and {Forestini}, M.},
        title = "{An internet server for pre-main sequence tracks of low- and intermediate-mass stars}",
      journal = {\aap},
         year = 2000,
        month = jun,
       volume = {358},
        pages = {593-599},
          doi = {10.48550/arXiv.astro-ph/0003477},
archivePrefix = {arXiv},
       eprint = {astro-ph/0003477},
 primaryClass = {astro-ph},
       adsurl = {https://ui.adsabs.harvard.edu/abs/2000A&A...358..593S}
}

@ARTICLE{Ansdell+2018,
       author = {{Ansdell}, M. and {Williams}, J.~P. and {Trapman}, L. and {van Terwisga}, S.~E. and {Facchini}, S. and {Manara}, C.~F. and {van der Marel}, N. and {Miotello}, A. and {Tazzari}, M. and {Hogerheijde}, M. and {Guidi}, G. and {Testi}, L. and {van Dishoeck}, E.~F.},
        title = "{ALMA Survey of Lupus Protoplanetary Disks. II. Gas Disk Radii}",
      journal = {\apj},
         year = 2018,
        month = may,
       volume = {859},
       number = {1},
          eid = {21},
        pages = {21},
          doi = {10.3847/1538-4357/aab890},
archivePrefix = {arXiv},
       eprint = {1803.05923},
 primaryClass = {astro-ph.EP},
       adsurl = {https://ui.adsabs.harvard.edu/abs/2018ApJ...859...21A}
}

@ARTICLE{Speedie+2025,
       author = {{Speedie}, Jessica and {Dong}, Ruobing and {Teague}, Richard and {Segura-Cox}, Dominique and {Pineda}, Jaime E. and {Calcino}, Josh and {Longarini}, Cristiano and {Hall}, Cassandra and {Tang}, Ya-Wen and {Hashimoto}, Jun and {Paneque-Carre{\~n}o}, Teresa and {Lodato}, Giuseppe and {Veronesi}, Bennedetta},
        title = "{Mapping the Merging Zone of Late Infall in the AB Aur Planet-forming System}",
      journal = {\apjl},
         year = 2025,
        month = mar,
       volume = {981},
       number = {2},
          eid = {L30},
        pages = {L30},
          doi = {10.3847/2041-8213/adb7d5},
archivePrefix = {arXiv},
       eprint = {2503.01957},
 primaryClass = {astro-ph.EP},
       adsurl = {https://ui.adsabs.harvard.edu/abs/2025ApJ...981L..30S}
}

@ARTICLE{Tabone+2022,
       author = {{Tabone}, Beno{\^\i}t and {Rosotti}, Giovanni P. and {Cridland}, Alexander J. and {Armitage}, Philip J. and {Lodato}, Giuseppe},
        title = "{Secular evolution of MHD wind-driven discs: analytical solutions in the expanded {\ensuremath{\alpha}}-framework}",
      journal = {\mnras},
         year = 2022,
        month = may,
       volume = {512},
       number = {2},
        pages = {2290-2309},
          doi = {10.1093/mnras/stab3442},
archivePrefix = {arXiv},
       eprint = {2111.10145},
 primaryClass = {astro-ph.SR},
       adsurl = {https://ui.adsabs.harvard.edu/abs/2022MNRAS.512.2290T}
}

@ARTICLE{Zagaria+2023,
       author = {{Zagaria}, Francesco and {Facchini}, Stefano and {Miotello}, Anna and {Manara}, Carlo F. and {Toci}, Claudia and {Clarke}, Cathie J.},
        title = "{Testing protoplanetary disc evolution with CO fluxes. A proof of concept in Lupus and Upper Sco}",
      journal = {\aap},
         year = 2023,
        month = apr,
       volume = {672},
          eid = {L15},
        pages = {L15},
          doi = {10.1051/0004-6361/202346164},
archivePrefix = {arXiv},
       eprint = {2304.01760},
 primaryClass = {astro-ph.EP},
       adsurl = {https://ui.adsabs.harvard.edu/abs/2023A&A...672L..15Z}
}

@ARTICLE{Dullemond_2019,
       author = {{Dullemond}, C.~P. and {K{\"u}ffmeier}, M. and {Goicovic}, F. and {Fukagawa}, M. and {Oehl}, V. and {Kramer}, M.},
        title = "{Cloudlet capture by transitional disk and FU Orionis stars}",
      journal = {\aap},
         year = 2019,
        month = aug,
       volume = {628},
          eid = {A20},
        pages = {A20},
          doi = {10.1051/0004-6361/201832632},
archivePrefix = {arXiv},
       eprint = {1911.05158},
 primaryClass = {astro-ph.EP},
       adsurl = {https://ui.adsabs.harvard.edu/abs/2019A&A...628A..20D}
}

@ARTICLE{Kuffmeier_2020,
       author = {{Kuffmeier}, M. and {Goicovic}, F.~G. and {Dullemond}, C.~P.},
        title = "{Late encounter events as source of disks and spiral structures. Forming second generation disks}",
      journal = {\aap},
         year = 2020,
        month = jan,
       volume = {633},
          eid = {A3},
        pages = {A3},
          doi = {10.1051/0004-6361/201936820},
archivePrefix = {arXiv},
       eprint = {1911.04833},
 primaryClass = {astro-ph.SR},
       adsurl = {https://ui.adsabs.harvard.edu/abs/2020A&A...633A...3K}
}

@ARTICLE{Kuffmeier_2021,
       author = {{Kuffmeier}, M. and {Dullemond}, C.~P. and {Reissl}, S. and {Goicovic}, F.~G.},
        title = "{Misaligned disks induced by infall}",
      journal = {\aap},
         year = 2021,
        month = dec,
       volume = {656},
          eid = {A161},
        pages = {A161},
          doi = {10.1051/0004-6361/202039614},
archivePrefix = {arXiv},
       eprint = {2110.04309},
 primaryClass = {astro-ph.SR},
       adsurl = {https://ui.adsabs.harvard.edu/abs/2021A&A...656A.161K}
}

@ARTICLE{Hanawa_2024,
       author = {{Hanawa}, Tomoyuki and {Garufi}, Antonio and {Podio}, Linda and {Codella}, Claudio and {Segura-Cox}, Dominique},
        title = "{Cloudlet capture model for the accretion streamer onto the disc of DG Tau}",
      journal = {\mnras},
         year = 2024,
        month = mar,
       volume = {528},
       number = {4},
        pages = {6581-6592},
          doi = {10.1093/mnras/stae338},
archivePrefix = {arXiv},
       eprint = {2402.02706},
 primaryClass = {astro-ph.GA},
       adsurl = {https://ui.adsabs.harvard.edu/abs/2024MNRAS.528.6581H}
}

@ARTICLE{Adamas_1987,
       author = {{Adams}, Fred C. and {Lada}, Charles J. and {Shu}, Frank H.},
        title = "{Spectral Evolution of Young Stellar Objects}",
      journal = {\apj},
         year = 1987,
        month = jan,
       volume = {312},
        pages = {788},
          doi = {10.1086/164924},
       adsurl = {https://ui.adsabs.harvard.edu/abs/1987ApJ...312..788A}
}

@INPROCEEDINGS{Pineda_2023,
       author = {{Pineda}, J.~E. and {Arzoumanian}, D. and {Andre}, P. and {Friesen}, R.~K. and {Zavagno}, A. and {Clarke}, S.~D. and {Inoue}, T. and {Chen}, C. and {Lee}, Y. and {Soler}, J.~D. and {Kuffmeier}, M.},
        title = "{From Bubbles and Filaments to Cores and Disks: Gas Gathering and Growth of Structure Leading to the Formation of Stellar Systems}",
    booktitle = {Protostars and Planets VII},
         year = 2023,
       editor = {{Inutsuka}, S. and {Aikawa}, Y. and {Muto}, T. and {Tomida}, K. and {Tamura}, M.},
       series = {Astronomical Society of the Pacific Conference Series},
       volume = {534},
        month = jul,
        pages = {233},
          doi = {10.48550/arXiv.2205.03935},
archivePrefix = {arXiv},
       eprint = {2205.03935},
 primaryClass = {astro-ph.GA},
       adsurl = {https://ui.adsabs.harvard.edu/abs/2023ASPC..534..233P}
}

@ARTICLE{Manara_2018,
       author = {{Manara}, C.~F. and {Morbidelli}, A. and {Guillot}, T.},
        title = "{Why do protoplanetary disks appear not massive enough to form the known exoplanet population?}",
      journal = {\aap},
         year = 2018,
        month = oct,
       volume = {618},
          eid = {L3},
        pages = {L3},
          doi = {10.1051/0004-6361/201834076},
archivePrefix = {arXiv},
       eprint = {1809.07374},
 primaryClass = {astro-ph.EP},
       adsurl = {https://ui.adsabs.harvard.edu/abs/2018A&A...618L...3M}
}

@ARTICLE{Nakajima_1995,
       author = {{Nakajima}, Tadashi and {Golimowski}, David A.},
        title = "{Coronagraphic Imaging of Pre-Main-Sequence Stars: Remnant Envelopes of Star Formation Seen in Reflection}",
      journal = {\aj},
         year = 1995,
        month = mar,
       volume = {109},
        pages = {1181},
          doi = {10.1086/117351},
       adsurl = {https://ui.adsabs.harvard.edu/abs/1995AJ....109.1181N}
}

@ARTICLE{Grady_1999,
       author = {{Grady}, C.~A. and {Woodgate}, B. and {Bruhweiler}, F.~C. and {Boggess}, A. and {Plait}, Philip and {Lindler}, Don J. and {Clampin}, M. and {Kalas}, P.},
        title = "{Hubble Space Telescope Space Telescope Imaging Spectrograph Coronagraphic Imaging of the Herbig AE Star AB Aurigae}",
      journal = {\apjl},
         year = 1999,
        month = oct,
       volume = {523},
       number = {2},
        pages = {L151-L154},
          doi = {10.1086/312270},
       adsurl = {https://ui.adsabs.harvard.edu/abs/1999ApJ...523L.151G}
}

@ARTICLE{Tang_2012,
       author = {{Tang}, Y. -Wen and {Guilloteau}, S. and {Pi{\'e}tu}, V. and {Dutrey}, A. and {Ohashi}, N. and {Ho}, P.~T.~P.},
        title = "{The circumstellar disk of AB Aurigae: evidence for envelope accretion at late stages of star formation?}",
      journal = {\aap},
         year = 2012,
        month = nov,
       volume = {547},
          eid = {A84},
        pages = {A84},
          doi = {10.1051/0004-6361/201219414},
archivePrefix = {arXiv},
       eprint = {1209.1299},
 primaryClass = {astro-ph.GA},
       adsurl = {https://ui.adsabs.harvard.edu/abs/2012A&A...547A..84T}
}

@ARTICLE{Grady_2001,
       author = {{Grady}, C.~A. and {Polomski}, E.~F. and {Henning}, Th. and {Stecklum}, B. and {Woodgate}, B.~E. and {Telesco}, C.~M. and {Pi{\~n}a}, R.~K. and {Gull}, T.~R. and {Boggess}, A. and {Bowers}, C.~W. and {Bruhweiler}, F.~C. and {Clampin}, M. and {Danks}, A.~C. and {Green}, R.~F. and {Heap}, S.~R. and {Hutchings}, J.~B. and {Jenkins}, E.~B. and {Joseph}, C. and {Kaiser}, M.~E. and {Kimble}, R.~A. and {Kraemer}, S. and {Lindler}, D. and {Linsky}, J.~L. and {Maran}, S.~P. and {Moos}, H.~W. and {Plait}, P. and {Roesler}, F. and {Timothy}, J.~G. and {Weistrop}, D.},
        title = "{The Disk and Environment of the Herbig Be Star HD 100546}",
      journal = {\aj},
         year = 2001,
        month = dec,
       volume = {122},
       number = {6},
        pages = {3396-3406},
          doi = {10.1086/324447},
       adsurl = {https://ui.adsabs.harvard.edu/abs/2001AJ....122.3396G}
}

@ARTICLE{Ardila_2007,
       author = {{Ardila}, D.~R. and {Golimowski}, D.~A. and {Krist}, J.~E. and {Clampin}, M. and {Ford}, H.~C. and {Illingworth}, G.~D.},
        title = "{Hubble Space Telescope Advanced Camera for Surveys Coronagraphic Observations of the Dust Surrounding HD 100546}",
      journal = {\apj},
         year = 2007,
        month = aug,
       volume = {665},
       number = {1},
        pages = {512-534},
          doi = {10.1086/519296},
       adsurl = {https://ui.adsabs.harvard.edu/abs/2007ApJ...665..512A}
}

@INPROCEEDINGS{Pascucci_2023,
       author = {{Pascucci}, I. and {Cabrit}, S. and {Edwards}, S. and {Gorti}, U. and {Gressel}, O. and {Suzuki}, T.~K.},
        title = "{The Role of Disk Winds in the Evolution and Dispersal of Protoplanetary Disks}",
    booktitle = {Protostars and Planets VII},
         year = 2023,
       editor = {{Inutsuka}, S. and {Aikawa}, Y. and {Muto}, T. and {Tomida}, K. and {Tamura}, M.},
       series = {Astronomical Society of the Pacific Conference Series},
       volume = {534},
        month = jul,
        pages = {567},
          doi = {10.48550/arXiv.2203.10068},
archivePrefix = {arXiv},
       eprint = {2203.10068},
 primaryClass = {astro-ph.EP},
       adsurl = {https://ui.adsabs.harvard.edu/abs/2023ASPC..534..567P}
}

@ARTICLE{Deng_2025,
       author = {{Deng}, Dingshan and {Vioque}, Miguel and {Pascucci}, Ilaria and {P{\'e}rez}, Laura M. and {Zhang}, Ke and {Kurtovic}, Nicol{\'a}s and {Trapman}, Leon and {TorresVillanueva}, Estephani E. and {Agurto-Gangas}, Carolina and {Carpenter}, John and {Pinilla}, Paola and {Gorti}, Uma and {Tabone}, Beno{\^\i}t and {Sierra}, Anibal and {Rosotti}, Giovanni P. and {Cieza}, Lucas A. and {Anania}, Rossella and {Gonz{\'a}lez-Ruilova}, Camilo and {Hogerheijde}, Michiel R. and {Miley}, James and {Ruiz-Rodriguez}, Dary A. and {Ruaud}, Maxime and {Schwarz}, Kamber},
        title = "{The ALMA Survey of Gas Evolution of PROtoplanetary Disks (AGE-PRO). III. Dust and Gas Disk Properties in the Lupus Star-forming Region}",
      journal = {\apj},
         year = 2025,
        month = aug,
       volume = {989},
       number = {1},
          eid = {3},
        pages = {3},
          doi = {10.3847/1538-4357/add43a},
archivePrefix = {arXiv},
       eprint = {2506.10734},
 primaryClass = {astro-ph.EP},
       adsurl = {https://ui.adsabs.harvard.edu/abs/2025ApJ...989....3D}
}

@ARTICLE{Hanawa_2022,
       author = {{Hanawa}, Tomoyuki and {Sakai}, Nami and {Yamamoto}, Satoshi},
        title = "{Cloudlet Capture Model for Asymmetric Molecular Emission Lines Observed in TMC-1A with ALMA}",
      journal = {\apj},
         year = 2022,
        month = jun,
       volume = {932},
       number = {2},
          eid = {122},
        pages = {122},
          doi = {10.3847/1538-4357/ac6e6a},
archivePrefix = {arXiv},
       eprint = {2205.04742},
 primaryClass = {astro-ph.GA},
       adsurl = {https://ui.adsabs.harvard.edu/abs/2022ApJ...932..122H}
}

@ARTICLE{Shoda_2021,
       author = {{Shoda}, Munehito and {Takasao}, Shinsuke},
        title = "{Corona and XUV emission modelling of the Sun and Sun-like stars}",
      journal = {\aap},
         year = 2021,
        month = dec,
       volume = {656},
          eid = {A111},
        pages = {A111},
          doi = {10.1051/0004-6361/202141563},
archivePrefix = {arXiv},
       eprint = {2106.08915},
 primaryClass = {astro-ph.SR},
       adsurl = {https://ui.adsabs.harvard.edu/abs/2021A&A...656A.111S}
}

@ARTICLE{Fu_2014,
       author = {{Fu}, Roger R. and {Weiss}, Benjamin P. and {Lima}, Eduardo A. and {Harrison}, Richard J. and {Bai}, Xue-Ning and {Desch}, Steven J. and {Ebel}, Denton S. and {Suavet}, Cl{\'e}ment and {Wang}, Huapei and {Glenn}, David and {Le Sage}, David and {Kasama}, Takeshi and {Walsworth}, Ronald L. and {Kuan}, Aaron T.},
        title = "{Solar nebula magnetic fields recorded in the Semarkona meteorite}",
      journal = {Science},
         year = 2014,
        month = nov,
       volume = {346},
       number = {6213},
        pages = {1089-1092},
          doi = {10.1126/science.1258022},
       adsurl = {https://ui.adsabs.harvard.edu/abs/2014Sci...346.1089F}
}

@ARTICLE{Vlemmings_2019,
       author = {{Vlemmings}, W.~H.~T. and {Lankhaar}, B. and {Cazzoletti}, P. and {Ceccobello}, C. and {Dall'Olio}, D. and {van Dishoeck}, E.~F. and {Facchini}, S. and {Humphreys}, E.~M.~L. and {Persson}, M.~V. and {Testi}, L. and {Williams}, J.~P.},
        title = "{Stringent limits on the magnetic field strength in the disc of TW Hya. ALMA observations of CN polarisation}",
      journal = {\aap},
         year = 2019,
        month = apr,
       volume = {624},
          eid = {L7},
        pages = {L7},
          doi = {10.1051/0004-6361/201935459},
archivePrefix = {arXiv},
       eprint = {1904.01632},
 primaryClass = {astro-ph.SR},
       adsurl = {https://ui.adsabs.harvard.edu/abs/2019A&A...624L...7V}
}

@ARTICLE{Padoan_2005,
       author = {{Padoan}, Paolo and {Kritsuk}, Alexei and {Norman}, Michael L. and {Nordlund}, {\r{A}}ke},
        title = "{A Solution to the Pre-Main-Sequence Accretion Problem}",
      journal = {\apjl},
         year = 2005,
        month = mar,
       volume = {622},
       number = {1},
        pages = {L61-L64},
          doi = {10.1086/429562},
archivePrefix = {arXiv},
       eprint = {astro-ph/0411129},
 primaryClass = {astro-ph},
       adsurl = {https://ui.adsabs.harvard.edu/abs/2005ApJ...622L..61P}
}

@ARTICLE{Klessen_2010,
       author = {{Klessen}, R.~S. and {Hennebelle}, P.},
        title = "{Accretion-driven turbulence as universal process: galaxies, molecular clouds, and protostellar disks}",
      journal = {\aap},
         year = 2010,
        month = sep,
       volume = {520},
          eid = {A17},
        pages = {A17},
          doi = {10.1051/0004-6361/200913780},
archivePrefix = {arXiv},
       eprint = {0912.0288},
 primaryClass = {astro-ph.CO},
       adsurl = {https://ui.adsabs.harvard.edu/abs/2010A&A...520A..17K}
}

@ARTICLE{Troop_2008,
       author = {{Throop}, Henry B. and {Bally}, John},
        title = "{Tail-End Bondi-Hoyle Accretion in Young Star Clusters: Implications for Disks, Planets, and Stars}",
      journal = {\aj},
         year = 2008,
        month = jun,
       volume = {135},
       number = {6},
        pages = {2380-2397},
          doi = {10.1088/0004-6256/135/6/2380},
archivePrefix = {arXiv},
       eprint = {0804.0438},
 primaryClass = {astro-ph},
       adsurl = {https://ui.adsabs.harvard.edu/abs/2008AJ....135.2380T}
}

@ARTICLE{Padoan_2025,
       author = {{Padoan}, Paolo and {Pan}, Liubin and {Pelkonen}, Veli-Matti and {Haugb{\o}lle}, Troels and {Nordlund}, {\^a}. {\guillemotleft}ke},
        title = "{The formation of protoplanetary disks through pre-main-sequence Bondi-Hoyle accretion}",
      journal = {Nature Astronomy},
         year = 2025,
        month = jun,
       volume = {9},
        pages = {862-871},
          doi = {10.1038/s41550-025-02529-3},
archivePrefix = {arXiv},
       eprint = {2405.07334},
 primaryClass = {astro-ph.GA},
       adsurl = {https://ui.adsabs.harvard.edu/abs/2025NatAs...9..862P}
}

@ARTICLE{Scicluna_2014,
       author = {{Scicluna}, P. and {Rosotti}, G. and {Dale}, J.~E. and {Testi}, L.},
        title = "{Old pre-main-sequence stars. Disc reformation by Bondi-Hoyle accretion}",
      journal = {\aap},
         year = 2014,
        month = jun,
       volume = {566},
          eid = {L3},
        pages = {L3},
          doi = {10.1051/0004-6361/201423654},
archivePrefix = {arXiv},
       eprint = {1405.6051},
 primaryClass = {astro-ph.SR},
       adsurl = {https://ui.adsabs.harvard.edu/abs/2014A&A...566L...3S}
}

@ARTICLE{Gorti_2009,
       author = {{Gorti}, U. and {Hollenbach}, D.},
        title = "{Photoevaporation of Circumstellar Disks By Far-Ultraviolet, Extreme-Ultraviolet and X-Ray Radiation from the Central Star}",
      journal = {\apj},
         year = 2009,
        month = jan,
       volume = {690},
       number = {2},
        pages = {1539-1552},
          doi = {10.1088/0004-637X/690/2/1539},
archivePrefix = {arXiv},
       eprint = {0809.1494},
 primaryClass = {astro-ph},
       adsurl = {https://ui.adsabs.harvard.edu/abs/2009ApJ...690.1539G}
}

@ARTICLE{Hollenbach1994,
       author = {{Hollenbach}, David and {Johnstone}, Doug and {Lizano}, Susana and {Shu}, Frank},
        title = "{Photoevaporation of Disks around Massive Stars and Application to Ultracompact H II Regions}",
      journal = {\apj},
         year = 1994,
        month = jun,
       volume = {428},
        pages = {654},
          doi = {10.1086/174276},
       adsurl = {https://ui.adsabs.harvard.edu/abs/1994ApJ...428..654H}
}

@ARTICLE{Fukagawa_2004,
       author = {{Fukagawa}, Misato and {Hayashi}, Masahiko and {Tamura}, Motohide and {Itoh}, Yoichi and {Hayashi}, Saeko S. and {Oasa}, Yumiko and {Takeuchi}, Taku and {Morino}, Jun-ichi and {Murakawa}, Koji and {Oya}, Shin and {Yamashita}, Takuya and {Suto}, Hiroshi and {Mayama}, Satoshi and {Naoi}, Takahiro and {Ishii}, Miki and {Pyo}, Tae-Soo and {Nishikawa}, Takayuki and {Takato}, Naruhisa and {Usuda}, Tomonori and {Ando}, Hiroyasu and {Iye}, Masanori and {Miyama}, Shoken M. and {Kaifu}, Norio},
        title = "{Spiral Structure in the Circumstellar Disk around AB Aurigae}",
      journal = {\apjl},
         year = 2004,
        month = apr,
       volume = {605},
       number = {1},
        pages = {L53-L56},
          doi = {10.1086/420699},
       adsurl = {https://ui.adsabs.harvard.edu/abs/2004ApJ...605L..53F}
}

@ARTICLE{Mesa_2022,
       author = {{Mesa}, D. and {Ginski}, C. and {Gratton}, R. and {Ertel}, S. and {Wagner}, K. and {Bonavita}, M. and {Fedele}, D. and {Meyer}, M. and {Henning}, T. and {Langlois}, M. and {Garufi}, A. and {Antoniucci}, S. and {Claudi}, R. and {Defr{\`e}re}, D. and {Desidera}, S. and {Janson}, M. and {Pawellek}, N. and {Rigliaco}, E. and {Squicciarini}, V. and {Zurlo}, A. and {Boccaletti}, A. and {Bonnefoy}, M. and {Cantalloube}, F. and {Chauvin}, G. and {Feldt}, M. and {Hagelberg}, J. and {Hugot}, E. and {Lagrange}, A. -M. and {Lazzoni}, C. and {Maurel}, D. and {Perrot}, C. and {Petit}, C. and {Rouan}, D. and {Vigan}, A.},
        title = "{Signs of late infall and possible planet formation around DR Tau using VLT/SPHERE and LBTI/LMIRCam}",
      journal = {\aap},
         year = 2022,
        month = feb,
       volume = {658},
          eid = {A63},
        pages = {A63},
          doi = {10.1051/0004-6361/202142219},
archivePrefix = {arXiv},
       eprint = {2111.01702},
 primaryClass = {astro-ph.EP},
       adsurl = {https://ui.adsabs.harvard.edu/abs/2022A&A...658A..63M}
}

@ARTICLE{Ginski_2021,
       author = {{Ginski}, Christian and {Facchini}, Stefano and {Huang}, Jane and {Benisty}, Myriam and {Vaendel}, Dennis and {Stapper}, Lucas and {Dominik}, Carsten and {Bae}, Jaehan and {M{\'e}nard}, Fran{\c{c}}ois and {Muro-Arena}, Gabriela and {Hogerheijde}, Michiel R. and {McClure}, Melissa and {van Holstein}, Rob G. and {Birnstiel}, Tilman and {Boehler}, Yann and {Bohn}, Alexander and {Flock}, Mario and {Mamajek}, Eric E. and {Manara}, Carlo F. and {Pinilla}, Paola and {Pinte}, Christophe and {Ribas}, {\'A}lvaro},
        title = "{Disk Evolution Study Through Imaging of Nearby Young Stars (DESTINYS): Late Infall Causing Disk Misalignment and Dynamic Structures in SU Aur}",
      journal = {\apjl},
         year = 2021,
        month = feb,
       volume = {908},
       number = {2},
          eid = {L25},
        pages = {L25},
          doi = {10.3847/2041-8213/abdf57},
archivePrefix = {arXiv},
       eprint = {2102.08781},
 primaryClass = {astro-ph.EP},
       adsurl = {https://ui.adsabs.harvard.edu/abs/2021ApJ...908L..25G}
}

@ARTICLE{Garufi_2024,
       author = {{Garufi}, A. and {Ginski}, C. and {van Holstein}, R.~G. and {Benisty}, M. and {Manara}, C.~F. and {P{\'e}rez}, S. and {Pinilla}, P. and {Ribas}, {\'A}. and {Weber}, P. and {Williams}, J. and {Cieza}, L. and {Dominik}, C. and {Facchini}, S. and {Huang}, J. and {Zurlo}, A. and {Bae}, J. and {Hagelberg}, J. and {Henning}, Th. and {Hogerheijde}, M.~R. and {Janson}, M. and {M{\'e}nard}, F. and {Messina}, S. and {Meyer}, M.~R. and {Pinte}, C. and {Quanz}, S.~P. and {Rigliaco}, E. and {Roccatagliata}, V. and {Schmid}, H.~M. and {Szul{\'a}gyi}, J. and {van Boekel}, R. and {Wahhaj}, Z. and {Antichi}, J. and {Baruffolo}, A. and {Moulin}, T.},
        title = "{The SPHERE view of the Taurus star-forming region. The full census of planet-forming disks with GTO and DESTINYS programs}",
      journal = {\aap},
         year = 2024,
        month = may,
       volume = {685},
          eid = {A53},
        pages = {A53},
          doi = {10.1051/0004-6361/202347586},
archivePrefix = {arXiv},
       eprint = {2403.02158},
 primaryClass = {astro-ph.GA},
       adsurl = {https://ui.adsabs.harvard.edu/abs/2024A&A...685A..53G}
}

@ARTICLE{Huang_2023,
       author = {{Huang}, Jane and {Bergin}, Edwin A. and {Bae}, Jaehan and {Benisty}, Myriam and {Andrews}, Sean M.},
        title = "{Molecular Mapping of DR Tau's Protoplanetary Disk, Envelope, Outflow, and Large-scale Spiral Arm}",
      journal = {\apj},
         year = 2023,
        month = feb,
       volume = {943},
       number = {2},
          eid = {107},
        pages = {107},
          doi = {10.3847/1538-4357/aca89c},
archivePrefix = {arXiv},
       eprint = {2301.02674},
 primaryClass = {astro-ph.SR},
       adsurl = {https://ui.adsabs.harvard.edu/abs/2023ApJ...943..107H}
}

@ARTICLE{Garufi_2020,
       author = {{Garufi}, A. and {Avenhaus}, H. and {P{\'e}rez}, S. and {Quanz}, S.~P. and {van Holstein}, R.~G. and {Bertrang}, G.~H. -M. and {Casassus}, S. and {Cieza}, L. and {Principe}, D.~A. and {van der Plas}, G. and {Zurlo}, A.},
        title = "{Disks Around T Tauri Stars with SPHERE (DARTTS-S). II. Twenty-one new polarimetric images of young stellar disks}",
      journal = {\aap},
         year = 2020,
        month = jan,
       volume = {633},
          eid = {A82},
        pages = {A82},
          doi = {10.1051/0004-6361/201936946},
archivePrefix = {arXiv},
       eprint = {1911.10853},
 primaryClass = {astro-ph.EP},
       adsurl = {https://ui.adsabs.harvard.edu/abs/2020A&A...633A..82G}
}

@ARTICLE{Huang_2021,
       author = {{Huang}, Jane and {Bergin}, Edwin A. and {{\"O}berg}, Karin I. and {Andrews}, Sean M. and {Teague}, Richard and {Law}, Charles J. and {Kalas}, Paul and {Aikawa}, Yuri and {Bae}, Jaehan and {Bergner}, Jennifer B. and {Booth}, Alice S. and {Bosman}, Arthur D. and {Calahan}, Jenny K. and {Cataldi}, Gianni and {Cleeves}, L. Ilsedore and {Czekala}, Ian and {Ilee}, John D. and {Le Gal}, Romane and {Guzm{\'a}n}, Viviana V. and {Long}, Feng and {Loomis}, Ryan A. and {M{\'e}nard}, Fran{\c{c}}ois and {Nomura}, Hideko and {Qi}, Chunhua and {Schwarz}, Kamber R. and {Tsukagoshi}, Takashi and {van't Hoff}, Merel L.~R. and {Walsh}, Catherine and {Wilner}, David J. and {Yamato}, Yoshihide and {Zhang}, Ke},
        title = "{Molecules with ALMA at Planet-forming Scales (MAPS). XIX. Spiral Arms, a Tail, and Diffuse Structures Traced by CO around the GM Aur Disk}",
      journal = {\apjs},
         year = 2021,
        month = nov,
       volume = {257},
       number = {1},
          eid = {19},
        pages = {19},
          doi = {10.3847/1538-4365/ac143e},
archivePrefix = {arXiv},
       eprint = {2109.06224},
 primaryClass = {astro-ph.EP},
       adsurl = {https://ui.adsabs.harvard.edu/abs/2021ApJS..257...19H}
}

@INPROCEEDINGS{Benisty_2023,
       author = {{Benisty}, M. and {Dominik}, C. and {Follette}, K. and {Garufi}, A. and {Ginski}, C. and {Hashimoto}, J. and {Keppler}, M. and {Kley}, W. and {Monnier}, J.},
        title = "{Optical and Near-infrared View of Planet-forming Disks and Protoplanets}",
    booktitle = {Protostars and Planets VII},
         year = 2023,
       editor = {{Inutsuka}, S. and {Aikawa}, Y. and {Muto}, T. and {Tomida}, K. and {Tamura}, M.},
       series = {Astronomical Society of the Pacific Conference Series},
       volume = {534},
        month = jul,
        pages = {605},
          doi = {10.48550/arXiv.2203.09991},
archivePrefix = {arXiv},
       eprint = {2203.09991},
 primaryClass = {astro-ph.EP},
       adsurl = {https://ui.adsabs.harvard.edu/abs/2023ASPC..534..605B}
}

@ARTICLE{Kuffmeier_2023,
       author = {{Kuffmeier}, Michael and {Jensen}, Sigurd S. and {Haugb{\o}lle}, Troels},
        title = "{Rejuvenating infall: a crucial yet overlooked source of mass and angular momentum}",
      journal = {European Physical Journal Plus},
         year = 2023,
        month = mar,
       volume = {138},
       number = {3},
          eid = {272},
        pages = {272},
          doi = {10.1140/epjp/s13360-023-03880-y},
archivePrefix = {arXiv},
       eprint = {2303.05261},
 primaryClass = {astro-ph.SR},
       adsurl = {https://ui.adsabs.harvard.edu/abs/2023EPJP..138..272K}
}

@ARTICLE{Tychoniec_2018,
       author = {{Tychoniec}, {\L}ukasz and {Tobin}, John J. and {Karska}, Agata and {Chandler}, Claire and {Dunham}, Michael M. and {Harris}, Robert J. and {Kratter}, Kaitlin M. and {Li}, Zhi-Yun and {Looney}, Leslie W. and {Melis}, Carl and {P{\'e}rez}, Laura M. and {Sadavoy}, Sarah I. and {Segura-Cox}, Dominique and {van Dishoeck}, Ewine F.},
        title = "{The VLA Nascent Disk and Multiplicity Survey of Perseus Protostars (VANDAM). IV. Free-Free Emission from Protostars: Links to Infrared Properties, Outflow Tracers, and Protostellar Disk Masses}",
      journal = {\apjs},
         year = 2018,
        month = oct,
       volume = {238},
       number = {2},
          eid = {19},
        pages = {19},
          doi = {10.3847/1538-4365/aaceae},
archivePrefix = {arXiv},
       eprint = {1806.02434},
 primaryClass = {astro-ph.SR},
       adsurl = {https://ui.adsabs.harvard.edu/abs/2018ApJS..238...19T}
}

@ARTICLE{Ward-Duong_2018,
       author = {{Ward-Duong}, K. and {Patience}, J. and {Bulger}, J. and {van der Plas}, G. and {M{\'e}nard}, F. and {Pinte}, C. and {Jackson}, A.~P. and {Bryden}, G. and {Turner}, N.~J. and {Harvey}, P. and {Hales}, A. and {De Rosa}, R.~J.},
        title = "{The Taurus Boundary of Stellar/Substellar (TBOSS) Survey. II. Disk Masses from ALMA Continuum Observations}",
      journal = {\aj},
         year = 2018,
        month = feb,
       volume = {155},
       number = {2},
          eid = {54},
        pages = {54},
          doi = {10.3847/1538-3881/aaa128},
archivePrefix = {arXiv},
       eprint = {1712.07669},
 primaryClass = {astro-ph.EP},
       adsurl = {https://ui.adsabs.harvard.edu/abs/2018AJ....155...54W}
}

@ARTICLE{Blandford_1982,
       author = {{Blandford}, R.~D. and {Payne}, D.~G.},
        title = "{Hydromagnetic flows from accretion disks and the production of radio jets.}",
      journal = {\mnras},
         year = 1982,
        month = jun,
       volume = {199},
        pages = {883-903},
          doi = {10.1093/mnras/199.4.883},
       adsurl = {https://ui.adsabs.harvard.edu/abs/1982MNRAS.199..883B}
}

@ARTICLE{Trapman_2025,
       author = {{Trapman}, Leon and {Zhang}, Ke and {Rosotti}, Giovanni P. and {Pinilla}, Paola and {Tabone}, Beno{\^\i}t and {Pascucci}, Ilaria and {Agurto-Gangas}, Carolina and {Anania}, Rossella and {Carpenter}, John and {Cieza}, Lucas A. and {Deng}, Dingshan and {Gonz{\'a}lez-Ruilova}, Camilo and {Hogerheijde}, Michiel R. and {Kurtovic}, Nicol{\'a}s T. and {Kuznetsova}, Aleksandra and {Miley}, James and {P{\'e}rez}, Laura M. and {Ruiz-Rodriguez}, Dary A. and {Schwarz}, Kamber and {Sierra}, Anibal and {TorresVillanueva}, Estephani and {Vioque}, Miguel},
        title = "{The ALMA Survey of Gas Evolution of PROtoplanetary Disks (AGE-PRO). V. Protoplanetary Gas Disk Masses}",
      journal = {\apj},
         year = 2025,
        month = aug,
       volume = {989},
       number = {1},
          eid = {5},
        pages = {5},
          doi = {10.3847/1538-4357/adcd6e},
archivePrefix = {arXiv},
       eprint = {2506.10738},
 primaryClass = {astro-ph.EP},
       adsurl = {https://ui.adsabs.harvard.edu/abs/2025ApJ...989....5T}
}

@ARTICLE{Miotello_2017,
       author = {{Miotello}, A. and {van Dishoeck}, E.~F. and {Williams}, J.~P. and {Ansdell}, M. and {Guidi}, G. and {Hogerheijde}, M. and {Manara}, C.~F. and {Tazzari}, M. and {Testi}, L. and {van der Marel}, N. and {van Terwisga}, S.},
        title = "{Lupus disks with faint CO isotopologues: low gas/dust or high carbon depletion?}",
      journal = {\aap},
         year = 2017,
        month = mar,
       volume = {599},
          eid = {A113},
        pages = {A113},
          doi = {10.1051/0004-6361/201629556},
archivePrefix = {arXiv},
       eprint = {1612.01538},
 primaryClass = {astro-ph.SR},
       adsurl = {https://ui.adsabs.harvard.edu/abs/2017A&A...599A.113M}
}

@ARTICLE{Ruaud_2022,
       author = {{Ruaud}, Maxime and {Gorti}, Uma and {Hollenbach}, David J.},
        title = "{C$^{18}$O Emission as an Effective Measure of Gas Masses of Protoplanetary Disks}",
      journal = {\apj},
         year = 2022,
        month = jan,
       volume = {925},
       number = {1},
          eid = {49},
        pages = {49},
          doi = {10.3847/1538-4357/ac3826},
archivePrefix = {arXiv},
       eprint = {2111.05833},
 primaryClass = {astro-ph.GA},
       adsurl = {https://ui.adsabs.harvard.edu/abs/2022ApJ...925...49R}
}

@ARTICLE{Frerking_1982,
       author = {{Frerking}, M.~A. and {Langer}, W.~D. and {Wilson}, R.~W.},
        title = "{The relationship between carbon monoxide abundance and visual extinction in interstellar clouds.}",
      journal = {\apj},
         year = 1982,
        month = nov,
       volume = {262},
        pages = {590-605},
          doi = {10.1086/160451},
       adsurl = {https://ui.adsabs.harvard.edu/abs/1982ApJ...262..590F}
}

@ARTICLE{Shu_1993,
       author = {{Shu}, Frank H. and {Johnstone}, Doug and {Hollenbach}, David},
        title = "{Photoevaporation of the Solar Nebula and the Formation of the Giant Planets}",
      journal = {\icarus},
         year = 1993,
        month = nov,
       volume = {106},
       number = {1},
        pages = {92-101},
          doi = {10.1006/icar.1993.1160},
       adsurl = {https://ui.adsabs.harvard.edu/abs/1993Icar..106...92S}
}

@ARTICLE{Suzuki_2009,
       author = {{Suzuki}, Takeru K. and {Inutsuka}, Shu-ichiro},
        title = "{Disk Winds Driven by Magnetorotational Instability and Dispersal of Protoplanetary Disks}",
      journal = {\apjl},
         year = 2009,
        month = jan,
       volume = {691},
       number = {1},
        pages = {L49-L54},
          doi = {10.1088/0004-637X/691/1/L49},
archivePrefix = {arXiv},
       eprint = {0812.0844},
 primaryClass = {astro-ph},
       adsurl = {https://ui.adsabs.harvard.edu/abs/2009ApJ...691L..49S}
}

@ARTICLE{Hayashi_1993,
       author = {{Hayashi}, Masahiko and {Ohashi}, Nagayoshi and {Miyama}, Shoken M.},
        title = "{A Dynamically Accreting Gas Disk around HL Tauri}",
      journal = {\apjl},
         year = 1993,
        month = dec,
       volume = {418},
        pages = {L71},
          doi = {10.1086/187119},
       adsurl = {https://ui.adsabs.harvard.edu/abs/1993ApJ...418L..71H}
}

@ARTICLE{Welch_2000,
       author = {{Welch}, Wm. J. and {Hartmann}, Lee and {Helfer}, Tamara and {Brice{\~n}o}, Cesar},
        title = "{High-Resolution, Wide-Field Imaging of the HL Tauri Environment in $^{13}$CO (1-0)}",
      journal = {\apj},
         year = 2000,
        month = sep,
       volume = {540},
       number = {1},
        pages = {362-371},
          doi = {10.1086/309290},
       adsurl = {https://ui.adsabs.harvard.edu/abs/2000ApJ...540..362W}
}

@ARTICLE{Yen_2019,
       author = {{Yen}, Hsi-Wei and {Gu}, Pin-Gao and {Hirano}, Naomi and {Koch}, Patrick M. and {Lee}, Chin-Fei and {Liu}, Hauyu Baobab and {Takakuwa}, Shigehisa},
        title = "{HL Tau Disk in HCO$^{+}$ (3-2) and (1-0) with ALMA: Gas Density, Temperature, Gap, and One-arm Spiral}",
      journal = {\apj},
         year = 2019,
        month = aug,
       volume = {880},
       number = {2},
          eid = {69},
        pages = {69},
          doi = {10.3847/1538-4357/ab29f8},
archivePrefix = {arXiv},
       eprint = {1906.05535},
 primaryClass = {astro-ph.SR},
       adsurl = {https://ui.adsabs.harvard.edu/abs/2019ApJ...880...69Y}
}

@ARTICLE{Garufi_2022,
       author = {{Garufi}, A. and {Podio}, L. and {Codella}, C. and {Segura-Cox}, D. and {Vander Donckt}, M. and {Mercimek}, S. and {Bacciotti}, F. and {Fedele}, D. and {Kasper}, M. and {Pineda}, J.~E. and {Humphreys}, E. and {Testi}, L.},
        title = "{ALMA chemical survey of disk-outflow sources in Taurus (ALMA-DOT). VI. Accretion shocks in the disk of DG Tau and HL Tau}",
      journal = {\aap},
         year = 2022,
        month = feb,
       volume = {658},
          eid = {A104},
        pages = {A104},
          doi = {10.1051/0004-6361/202141264},
archivePrefix = {arXiv},
       eprint = {2110.13820},
 primaryClass = {astro-ph.GA},
       adsurl = {https://ui.adsabs.harvard.edu/abs/2022A&A...658A.104G}
}

@ARTICLE{Greaves_2010,
       author = {{Greaves}, J.~S. and {Rice}, W.~K.~M.},
        title = "{Have protoplanetary discs formed planets?}",
      journal = {\mnras},
         year = 2010,
        month = sep,
       volume = {407},
       number = {3},
        pages = {1981-1988},
          doi = {10.1111/j.1365-2966.2010.17043.x},
archivePrefix = {arXiv},
       eprint = {1006.1220},
 primaryClass = {astro-ph.GA},
       adsurl = {https://ui.adsabs.harvard.edu/abs/2010MNRAS.407.1981G}
}

@ARTICLE{Williams_2012,
       author = {{Williams}, Jonathan P.},
        title = "{Astronomical evidence for the rapid growth of millimeter-sized particles in protoplanetary disks}",
      journal = {\maps},
         year = 2012,
        month = dec,
       volume = {47},
       number = {12},
        pages = {1915-1921},
          doi = {10.1111/maps.12028},
archivePrefix = {arXiv},
       eprint = {1205.2461},
 primaryClass = {astro-ph.EP},
       adsurl = {https://ui.adsabs.harvard.edu/abs/2012M&PS...47.1915W}
}

@ARTICLE{Najita_2014,
       author = {{Najita}, J.~R. and {Kenyon}, S.~J.},
        title = "{The mass budget of planet-forming discs: isolating the epoch of planetesimal formation}",
      journal = {\mnras},
         year = 2014,
        month = dec,
       volume = {445},
       number = {3},
        pages = {3315-3329},
          doi = {10.1093/mnras/stu1994},
archivePrefix = {arXiv},
       eprint = {1409.7021},
 primaryClass = {astro-ph.SR},
       adsurl = {https://ui.adsabs.harvard.edu/abs/2014MNRAS.445.3315N}
}

@ARTICLE{Hamer_2019,
       author = {{Hamer}, Jacob H. and {Schlaufman}, Kevin C.},
        title = "{Hot Jupiters Are Destroyed by Tides While Their Host Stars Are on the Main Sequence}",
      journal = {\aj},
         year = 2019,
        month = nov,
       volume = {158},
       number = {5},
          eid = {190},
        pages = {190},
          doi = {10.3847/1538-3881/ab3c56},
archivePrefix = {arXiv},
       eprint = {1908.06998},
 primaryClass = {astro-ph.EP},
       adsurl = {https://ui.adsabs.harvard.edu/abs/2019AJ....158..190H}
}

@ARTICLE{Winter_2020,
       author = {{Winter}, Andrew J. and {Kruijssen}, J.~M. Diederik and {Longmore}, Steven N. and {Chevance}, M{\'e}lanie},
        title = "{Stellar clustering shapes the architecture of planetary systems}",
      journal = {\nat},
         year = 2020,
        month = oct,
       volume = {586},
       number = {7830},
        pages = {528-532},
          doi = {10.1038/s41586-020-2800-0},
archivePrefix = {arXiv},
       eprint = {2010.10531},
 primaryClass = {astro-ph.EP},
       adsurl = {https://ui.adsabs.harvard.edu/abs/2020Natur.586..528W}
}

@ARTICLE{Guerra-Alvarado_2025,
       author = {{Guerra-Alvarado}, Osmar M. and {van der Marel}, Nienke and {Williams}, Jonathan P. and {Pinilla}, Paola and {Mulders}, Gijs D. and {Lambrechts}, Michiel and {Sanchez}, Mariana},
        title = "{A high-resolution survey of protoplanetary disks in Lupus and the nature of compact disks}",
      journal = {\aap},
         year = 2025,
        month = apr,
       volume = {696},
          eid = {A232},
        pages = {A232},
          doi = {10.1051/0004-6361/202453338},
archivePrefix = {arXiv},
       eprint = {2503.19504},
 primaryClass = {astro-ph.EP},
       adsurl = {https://ui.adsabs.harvard.edu/abs/2025A&A...696A.232G}
}

@ARTICLE{Liffman_2003,
       author = {{Liffman}, Kurt},
        title = "{The Gravitational Radius of an Irradiated Disk}",
      journal = {\pasa},
         year = 2003,
        month = jan,
       volume = {20},
       number = {4},
        pages = {337-339},
          doi = {10.1071/AS03019},
       adsurl = {https://ui.adsabs.harvard.edu/abs/2003PASA...20..337L}
}

@ARTICLE{Mori_2025,
       author = {{Mori}, Shoji and {Kunitomo}, Masanobu and {Ogihara}, Masahiro},
        title = "{Long-term evolution of the temperature structure in magnetized protoplanetary disks and its implication for the dichotomy of planetary composition}",
      journal = {\aap},
         year = 2025,
        month = may,
       volume = {697},
          eid = {A192},
        pages = {A192},
          doi = {10.1051/0004-6361/202453362},
archivePrefix = {arXiv},
       eprint = {2504.08042},
 primaryClass = {astro-ph.EP},
       adsurl = {https://ui.adsabs.harvard.edu/abs/2025A&A...697A.192M}
}

@ARTICLE{Zhao_2025,
       author = {{Zhao}, Haichen and {Lau}, Tommy Chi Ho and {Birnstiel}, Tilman and {Stammler}, Sebastian M. and {Dr{\k{a}}{\.z}kowska}, Joanna},
        title = "{Planetesimal formation in a pressure bump induced by infall}",
      journal = {\aap},
         year = 2025,
        month = feb,
       volume = {694},
          eid = {A205},
        pages = {A205},
          doi = {10.1051/0004-6361/202452941},
archivePrefix = {arXiv},
       eprint = {2501.17857},
 primaryClass = {astro-ph.EP},
       adsurl = {https://ui.adsabs.harvard.edu/abs/2025A&A...694A.205Z}
}

@ARTICLE{Vorobyov_2015,
       author = {{Vorobyov}, Eduard I. and {Lin}, D.~N.~C. and {Guedel}, Manuel},
        title = "{The effect of external environment on the evolution of protostellar disks}",
      journal = {\aap},
         year = 2015,
        month = jan,
       volume = {573},
          eid = {A5},
        pages = {A5},
          doi = {10.1051/0004-6361/201424583},
archivePrefix = {arXiv},
       eprint = {1410.1743},
 primaryClass = {astro-ph.SR},
       adsurl = {https://ui.adsabs.harvard.edu/abs/2015A&A...573A...5V}
}

@ARTICLE{Grinin_2024,
       author = {{Grinin}, V.~P. and {Demidova}, T.~V.},
        title = "{Clumpy Accretion As a Possible Cause of Prolonged Eclipses in UX Ori Stars}",
      journal = {Astronomy Letters},
         year = 2024,
        month = mar,
       volume = {50},
       number = {3},
        pages = {194-202},
          doi = {10.1134/S1063773724700075},
archivePrefix = {arXiv},
       eprint = {2403.20065},
 primaryClass = {astro-ph.SR},
       adsurl = {https://ui.adsabs.harvard.edu/abs/2024AstL...50..194G}
}

@ARTICLE{Cassen+1981,
       author = {{Cassen}, P. and {Moosman}, A.},
        title = "{On the formation of protostellar disks}",
      journal = {\icarus},
         year = 1981,
        month = dec,
       volume = {48},
       number = {3},
        pages = {353-376},
          doi = {10.1016/0019-1035(81)90051-8},
       adsurl = {https://ui.adsabs.harvard.edu/abs/1981Icar...48..353C}
}

@ARTICLE{Cassen+1983,
       author = {{Cassen}, P. and {Summers}, A.},
        title = "{Models of the formation of the solar nebula}",
      journal = {\icarus},
         year = 1983,
        month = jan,
       volume = {53},
       number = {1},
        pages = {26-40},
          doi = {10.1016/0019-1035(83)90018-0},
       adsurl = {https://ui.adsabs.harvard.edu/abs/1983Icar...53...26C}
}

@ARTICLE{Jin+2010,
       author = {{Jin}, Liping and {Sui}, Ning},
        title = "{The Evolution of the Solar Nebula I. Evolution of the Global Properties and Planet Masses}",
      journal = {\apj},
         year = 2010,
        month = feb,
       volume = {710},
       number = {2},
        pages = {1179-1194},
          doi = {10.1088/0004-637X/710/2/1179},
       adsurl = {https://ui.adsabs.harvard.edu/abs/2010ApJ...710.1179J}
}

@ARTICLE{Liu+2017,
       author = {{Liu}, Chunjian and {Li}, Min and {Yao}, Zhen and {Mao}, Xiaodong},
        title = "{The ability of a protostellar disc to fragment and the properties of molecular cloud cores}",
      journal = {\apss},
         year = 2017,
        month = jan,
       volume = {362},
       number = {1},
          eid = {5},
        pages = {5},
          doi = {10.1007/s10509-016-2986-7},
       adsurl = {https://ui.adsabs.harvard.edu/abs/2017Ap&SS.362....5L}
}

@ARTICLE{Moeckel_2009,
       author = {{Moeckel}, Nickolas and {Throop}, Henry B.},
        title = "{Bondi-Hoyle-Lyttleton Accretion Onto a Protoplanetary Disk}",
      journal = {\apj},
         year = 2009,
        month = dec,
       volume = {707},
       number = {1},
        pages = {268-277},
          doi = {10.1088/0004-637X/707/1/268},
archivePrefix = {arXiv},
       eprint = {0910.3539},
 primaryClass = {astro-ph.SR},
       adsurl = {https://ui.adsabs.harvard.edu/abs/2009ApJ...707..268M}
}

@ARTICLE{Joos_2012,
       author = {{Joos}, M. and {Hennebelle}, P. and {Ciardi}, A.},
        title = "{Protostellar disk formation and transport of angular momentum during magnetized core collapse}",
      journal = {\aap},
         year = 2012,
        month = jul,
       volume = {543},
          eid = {A128},
        pages = {A128},
          doi = {10.1051/0004-6361/201118730},
archivePrefix = {arXiv},
       eprint = {1203.1193},
 primaryClass = {astro-ph.SR},
       adsurl = {https://ui.adsabs.harvard.edu/abs/2012A&A...543A.128J}
}

@ARTICLE{Huhn+2025,
       author = {{H{\"u}hn}, L.-A. and {Dullemond}, C.~P. and {Lebreuilly}, U. and {Klessen}, R.~S. and {Maury}, A. and {Rosotti}, G.~P. and {Hennebelle}, P. and {Pacetti}, E. and {Testi}, L. and {Molinari}, S.},
        title = "{Planetesimal formation via the streaming instability in simulations of infall-dominated young disks}",
      journal = {\aap},
         year = 2025,
        month = apr,
       volume = {696},
          eid = {A162},
        pages = {A162},
          doi = {10.1051/0004-6361/202452689},
archivePrefix = {arXiv},
       eprint = {2503.13606},
 primaryClass = {astro-ph.EP},
       adsurl = {https://ui.adsabs.harvard.edu/abs/2025A&A...696A.162H}
}

@ARTICLE{Huhn+2026,
       author = {{H{\"u}hn}, L.-A. and {Kimmig}, C.~N. and {Dullemond}, C.~P.},
        title = "{Spiral formation caused by late infall onto protoplanetary disks}",
      journal = {\aap},
         year = 2026,
        month = apr,
       volume = {708},
          eid = {A93},
        pages = {A93},
          doi = {10.1051/0004-6361/202558773},
archivePrefix = {arXiv},
       eprint = {2603.03442},
 primaryClass = {astro-ph.EP},
       adsurl = {https://ui.adsabs.harvard.edu/abs/2026A&A...708A..93H}
}

@ARTICLE{Huhn+2025b,
       author = {{H{\"u}hn}, L.-A. and {Dullemond}, C.~P.},
        title = "{Emergence of streamers in simulations of late infall}",
      journal = {\aap},
         year = 2025,
        month = dec,
       volume = {704},
          eid = {A222},
        pages = {A222},
          doi = {10.1051/0004-6361/202556203},
archivePrefix = {arXiv},
       eprint = {2510.24269},
 primaryClass = {astro-ph.EP},
       adsurl = {https://ui.adsabs.harvard.edu/abs/2025A&A...704A.222H}
}

@INPROCEEDINGS{Manara_2023,
       author = {{Manara}, C.~F. and {Ansdell}, M. and {Rosotti}, G.~P. and {Hughes}, A.~M. and {Armitage}, P.~J. and {Lodato}, G. and {Williams}, J.~P.},
        title = "{Demographics of Young Stars and their Protoplanetary Disks: Lessons Learned on Disk Evolution and its Connection to Planet Formation}",
    booktitle = {Protostars and Planets VII},
         year = 2023,
       editor = {{Inutsuka}, S. and {Aikawa}, Y. and {Muto}, T. and {Tomida}, K. and {Tamura}, M.},
       series = {Astronomical Society of the Pacific Conference Series},
       volume = {534},
        month = jul,
        pages = {539},
          doi = {10.48550/arXiv.2203.09930},
archivePrefix = {arXiv},
       eprint = {2203.09930},
 primaryClass = {astro-ph.SR},
       adsurl = {https://ui.adsabs.harvard.edu/abs/2023ASPC..534..539M}
}

@ARTICLE{Sanchis_2021,
       author = {{Sanchis}, E. and {Testi}, L. and {Natta}, A. and {Facchini}, S. and {Manara}, C.~F. and {Miotello}, A. and {Ercolano}, B. and {Henning}, Th. and {Preibisch}, T. and {Carpenter}, J.~M. and {de Gregorio-Monsalvo}, I. and {Jayawardhana}, R. and {Lopez}, C. and {Mu{\v{z}}i{\'c}}, K. and {Pascucci}, I. and {Santamar{\'\i}a-Miranda}, A. and {van Terwisga}, S. and {Williams}, J.~P.},
        title = "{Measuring the ratio of the gas and dust emission radii of protoplanetary disks in the Lupus star-forming region}",
      journal = {\aap},
         year = 2021,
        month = may,
       volume = {649},
          eid = {A19},
        pages = {A19},
          doi = {10.1051/0004-6361/202039733},
archivePrefix = {arXiv},
       eprint = {2101.11307},
 primaryClass = {astro-ph.EP},
       adsurl = {https://ui.adsabs.harvard.edu/abs/2021A&A...649A..19S}
}

@ARTICLE{Miotello_2021,
       author = {{Miotello}, A. and {Rosotti}, G. and {Ansdell}, M. and {Facchini}, S. and {Manara}, C.~F. and {Williams}, J.~P. and {Bruderer}, S.},
        title = "{Compact disks. An explanation to faint CO emission in Lupus disks}",
      journal = {\aap},
         year = 2021,
        month = jul,
       volume = {651},
          eid = {A48},
        pages = {A48},
          doi = {10.1051/0004-6361/202140550},
archivePrefix = {arXiv},
       eprint = {2104.09109},
 primaryClass = {astro-ph.SR},
       adsurl = {https://ui.adsabs.harvard.edu/abs/2021A&A...651A..48M}
}

@ARTICLE{Visser_2009,
       author = {{Visser}, R. and {van Dishoeck}, E.~F. and {Doty}, S.~D. and {Dullemond}, C.~P.},
        title = "{The chemical history of molecules in circumstellar disks. I. Ices}",
      journal = {\aap},
         year = 2009,
        month = mar,
       volume = {495},
       number = {3},
        pages = {881-897},
          doi = {10.1051/0004-6361/200810846},
archivePrefix = {arXiv},
       eprint = {0901.1313},
 primaryClass = {astro-ph.SR},
       adsurl = {https://ui.adsabs.harvard.edu/abs/2009A&A...495..881V}
}

@ARTICLE{Suzuki_2016_tum,
       author = {{Suzuki}, Takeru K. and {Ogihara}, Masahiro and {Morbidelli}, Alessandro and {Crida}, Aur{\'e}lien and {Guillot}, Tristan},
        title = "{Evolution of protoplanetary discs with magnetically driven disc winds (Corrigendum)}",
      journal = {\aap},
         year = 2022,
        month = dec,
       volume = {668},
          eid = {C1},
        pages = {C1},
          doi = {10.1051/0004-6361/201628955e},
       adsurl = {https://ui.adsabs.harvard.edu/abs/2022A&A...668C...1S}
}

\clearpage
\begin{appendix}

\section{Parameter dependencies}
\label{sec:para_dep}

As a complement to Section~\ref{Result}, we present the parameter dependencies of the results in more detail, with a particular focus on the models without the effective torque.
\label{parameter study in fiducial}

\subsection{Reducing infall radius}
\label{ssec:apdx_rinfall}

\begin{figure}
    \centering
    \includegraphics[width=8cm]{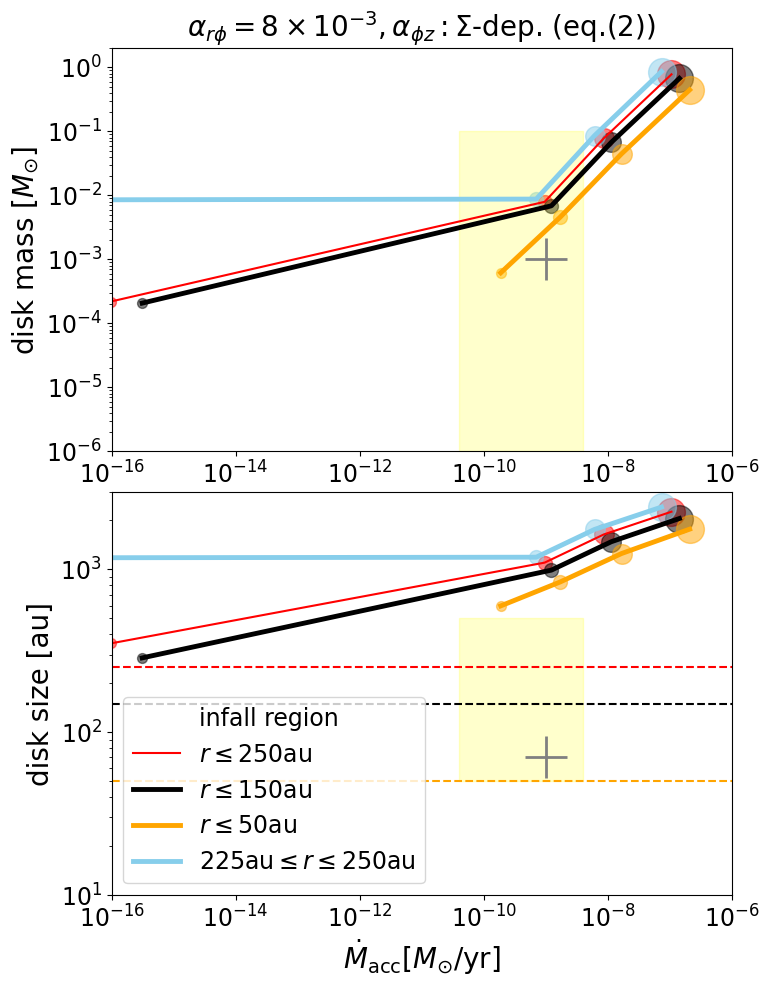}
\caption{
Same as Figure~\ref{fig:Acc_Diskmass_size_MRI}, but for different spatial distributions of infall. The models adopt $\alpha_{r \phi} = 8 \times 10^{-3}$ and the $\Sigma$-dependent $\alpha_{\phi z}$ (Eq.~\eqref{alpha_pz_fiducial}). The red, black, orange, and sky blue lines correspond to cases with infall occurring within $r \leq 250\ \mathrm{au}$, $r \leq 150\ \mathrm{au}$, $r \leq 50\ \mathrm{au}$, and $225\ \mathrm{au} \leq r \leq 250\ \mathrm{au}$, respectively. The dashed lines show the outer boundary of the infall region. The red lines are identical to those in Figure~\ref{fig:Acc_Diskmass_size_MRI}. The circles represent different infall rates of $\mdoti = 5\times10^{-10}$, $5\times10^{-9}$, $5\times10^{-8}$, and $5\times10^{-7}\ \msunyr$, respectively, in order of increasing circle size.
    }
    \label{fig:Acc_Infall_Rinfall}
\end{figure}

\begin{figure}
    \centering
    \includegraphics[width=8cm]{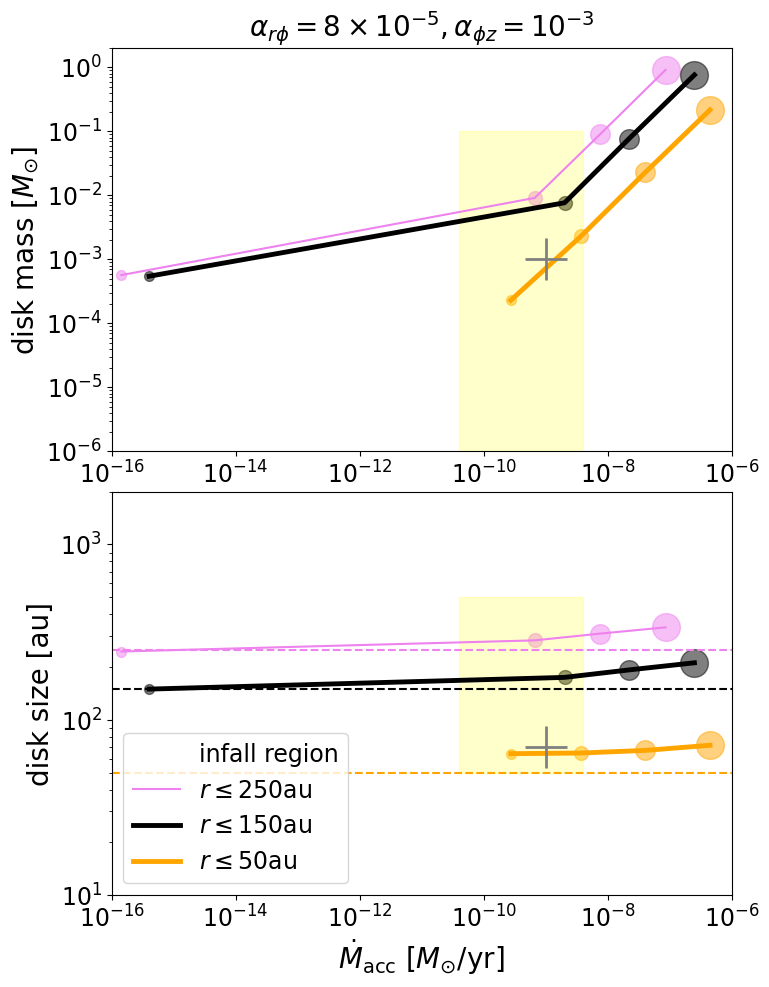}
\caption{
Same as Figure~\ref{fig:Acc_Infall_Rinfall} but for the models 
with the strong MHD disk wind torque of $\alpha_{\phi z} = 10^{-3}$ and weak turbulent viscosity of $\alpha_{r \phi}=8 \times 10^{-5}$. 
The fiducial model with $r_{\rm infall}=250$~au is shown by the magenta line, unlike in Figure~\ref{fig:Acc_Infall_Rinfall}.
}
    \label{fig:Acc_Infall_Rinfall_Alpha1e3}
\end{figure}

We examine the effects of concentrating late infall closer to the central star by reducing the characteristic infall radius, $r_{\rm infall}$, from the fiducial value of 250~au to 150~au and 50~au. Figures~\ref{fig:Acc_Infall_Rinfall} and \ref{fig:Acc_Infall_Rinfall_Alpha1e3} present the resulting trends for models with $\Sigma$-dependent and strong MHD disk wind torque, respectively.

Figure~\ref{fig:Acc_Infall_Rinfall} shows that for models with the $\Sigma$-dependent wind torque, decreasing $r_{\rm infall}$ (from red to black to orange) shifts the model tracks toward the lower right in each panel, corresponding to higher stellar accretion rate, $\dot{M}_*$, and smaller disk masses and sizes. The trend is relatively weak, however, because strong viscous torques efficiently spread the disk outward. As a result, the disk size depends only weakly on $r_{\rm infall}$, and reducing $r_{\rm infall}$ within this range does not resolve the disk-expansion problem.

Figure~\ref{fig:Acc_Infall_Rinfall_Alpha1e3} shows a similar trend for different values of $\alpha_{r \phi}$ and $\alpha_{\phi z}$, but with a stronger dependence on $r_{\rm infall}$. This reflects the fact that gas deposited closer to the star is accreted more efficiently. For example, with $r_{\rm infall} = 50$~au (orange line), even the lowest infall rate considered ($\dot{M}_{\rm infall} = 5 \times 10^{-10}~M_\odot~{\rm yr}^{-1}$; orange circle) yields $\dot{M}_* \gtrsim 10^{-10}~M_\odot~{\rm yr}^{-1}$, since photoevaporation fails to open a gap. Consequently, models with $r_{\rm infall} = 50$~au can reproduce both the observed accretion rates and disk properties, consistent with the gray crosses representing \citet{Deng_2025}. These results further suggest that the observed spread in disk sizes in Lupus \citep{Ansdell+2018,Guerra-Alvarado_2025} may be explained by variations in $r_{\rm infall}$.

\subsection{Reducing infall duration}
\label{ssec:apdx_tinfall}

Since the duration of late infall is highly uncertain, we explore its effect by considering three representative cases: continuous infall over the entire simulation duration (hereafter "infinite" duration), and finite durations of initial 1~Myr and 0.1~Myr.

\begin{figure}
    \centering
    \includegraphics[width=8cm]{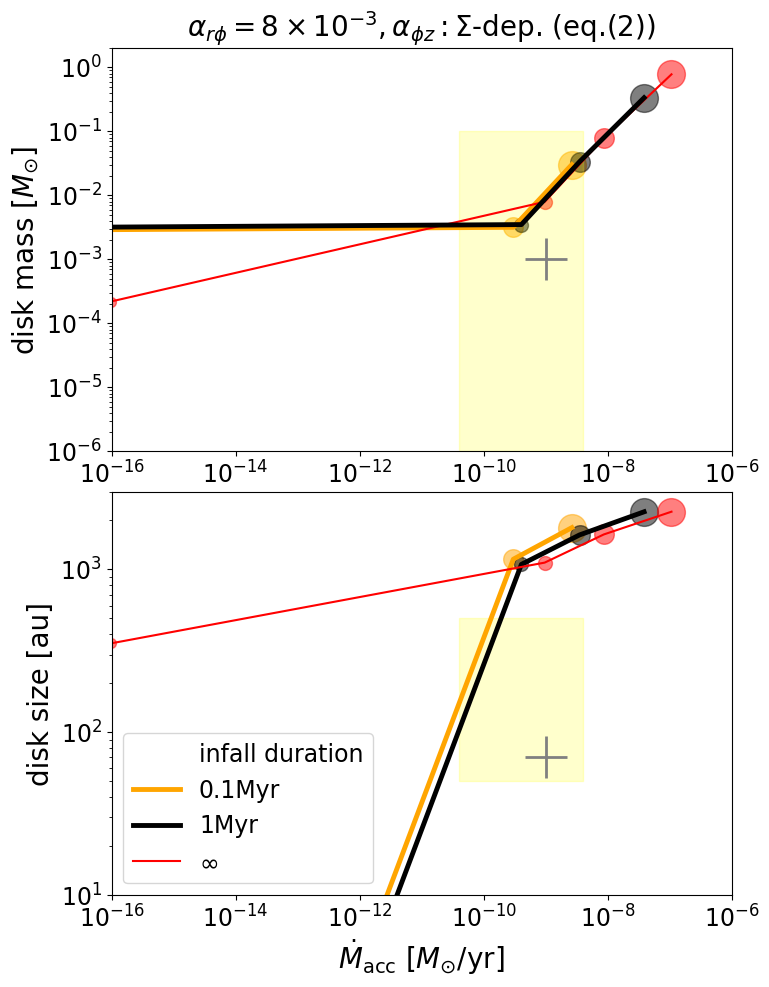}
\caption{
Same as Figure~\ref{fig:Acc_Infall_Rinfall}, but for different durations of infall. The red, black, and orange lines correspond to infall durations of $\infty$, 1~Myr, and 0.1~Myr, respectively, with the red lines identical to those in Figure~\ref{fig:Acc_Diskmass_size_MRI}.
}
    \label{fig:Acc_Infall_Tinfall}
\end{figure}

Figure~\ref{fig:Acc_Infall_Tinfall} presents the resulting trends for models with the $\Sigma$-dependent MHD disk wind torque. The lines in each panel (red, black, and orange) represent decreasing infall durations. As shown in the upper panel, shorter infall durations lead to lower $\dot{M}_*$ and disk masses, shifting the model tracks toward the lower left. This trend reflects the limited mass supply to the inner disk when infall persists only for a short time. In contrast, the lower panel shows that the disk size remains nearly unchanged across different infall durations, as it is primarily determined by viscous diffusion. Therefore, with only a few exceptions, the disk-size constraint makes it difficult to reproduce the scenario proposed by \citet{Winter+2024a}, even for short infall durations. In models with durations shorter than 1~Myr, the marked decrease in disk size at the lowest $\dot{M}_{\rm infall}$ reflects the depletion of disk gas, at which point the disk size is no longer well defined.

\begin{figure}
    \centering
    \includegraphics[width=8cm]{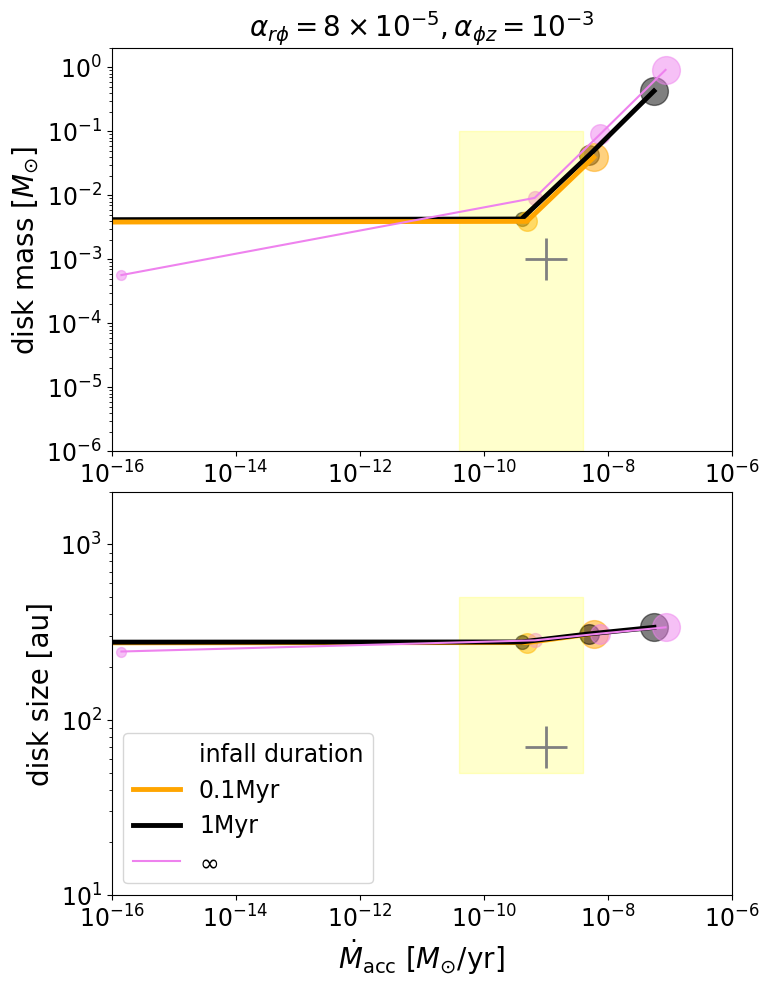}
\caption{
Same as Figure~\ref{fig:Acc_Infall_Tinfall} but for the models 
with the strong MHD disk wind torque of $\alpha_{\phi z} = 10^{-3}$ and weak turbulent viscosity of $\alpha_{r \phi}=8 \times 10^{-5}$. 
The fiducial model with the infinite duration is shown by the magenta line, unlike in Figure~\ref{fig:Acc_Infall_Tinfall}.
}
    \label{fig:Acc_Infall_Tinfall_alpha}
\end{figure}

Figure~\ref{fig:Acc_Infall_Tinfall_alpha} shows a similar trend for models with strong MHD disk wind torque. The main difference is that the disk size is primarily set by $r_{\rm infall}$. Nevertheless, the overall conclusion remains unchanged: the disk size does not vary significantly with infall duration.

Overall, shorter infall durations require higher values of the infall rate, $\dot{M}_{\rm infall}$, to reproduce the observed stellar accretion rates. However, for durations of $\sim$1~Myr, which are typically associated with late infall \citep{Winter+2024b}, the resulting variation in stellar accretion rates remains within a factor of a few at fixed $\dot{M}_{\rm infall}$. Moreover, since the disk size is largely insensitive to the infall duration, reducing it does not significantly alter our overall conclusions.

\subsection{Varying photoevaporation rates}
\label{ssec:apdx_pevrates}

The photoevaporation rate is subject to uncertainties arising from factors such as accretion-driven FUV emission and stellar evolution. In light of this uncertainty, we specifically assess the robustness of our conclusion that, within the range consistent with the Lupus observations, late infall can enhance stellar accretion only if supported by a strong MHD disk wind torque.

\begin{figure}
    \centering
    \includegraphics[width=8cm]{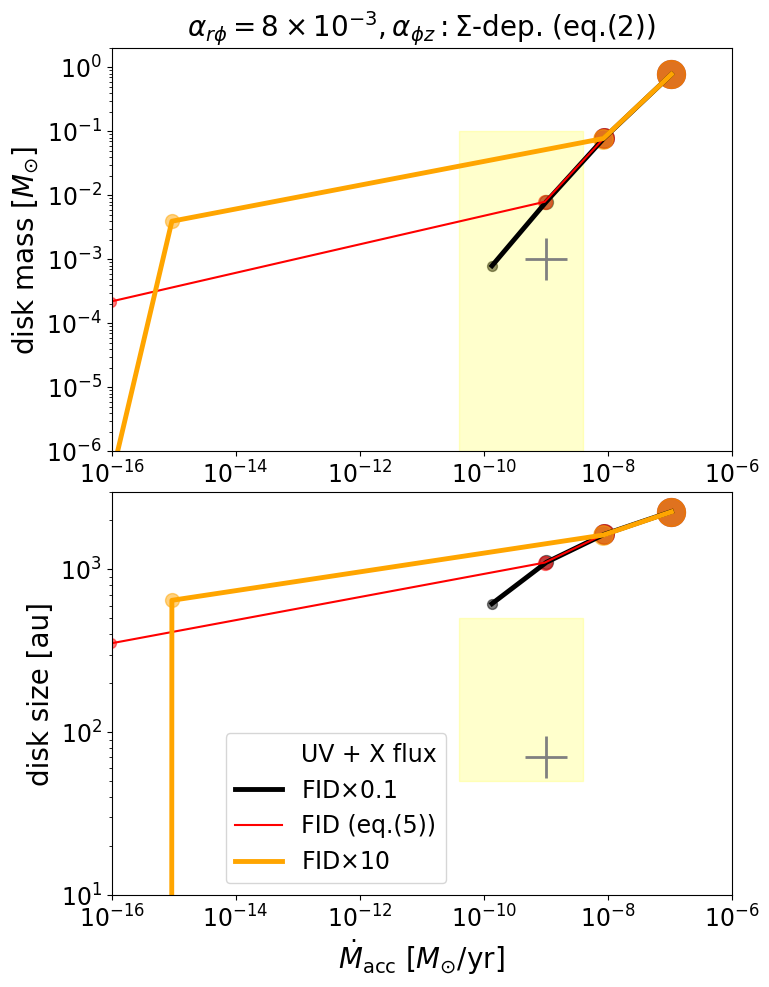}
\caption{
Same as Figure~\ref{fig:Acc_Infall_Rinfall}, but for different strengths of photoevaporation. The black, red, and orange lines correspond to stellar irradiation at $0.1\times$, $1\times$ (fiducial), and $10\times$ the fiducial values of FUV luminosity, EUV photon flux, and X-ray luminosity, respectively, with the red lines identical to those in Figure~\ref{fig:Acc_Diskmass_size_MRI}. 
}
\label{fig:Acc_Infall_Photo}
\end{figure}

Figure~\ref{fig:Acc_Infall_Photo} shows the effects of varying stellar irradiation for models with the $\Sigma$-dependent MHD disk wind torque. In the upper panel, the gap-opening transition manifests as a sharp drop in $\dot{M}_*$. For example, for fiducial UV and X-ray luminosities (red line), the transition occurs between $\dot{M}_{\rm infall} = 5 \times 10^{-10}$ and $5 \times 10^{-9}~\msunyr$. When the irradiation is ten times stronger (orange line), the transition shifts to $5 \times 10^{-9}$–$5 \times 10^{-8}~\msunyr$. In contrast, no transition is observed over $5 \times 10^{-10}$–$5 \times 10^{-7}~\msunyr$ when the irradiation is ten times weaker (black line). Thus, the critical infall rate for gap opening depends on the irradiation strength. The lower panel shows that the disk size is largely insensitive to irradiation strength. Consequently, varying the photoevaporation rate does not resolve the disk-size constraint, making the scenario of \citet{Winter+2024a} difficult to reproduce.

\begin{figure}
    \centering
    \includegraphics[width=8cm]{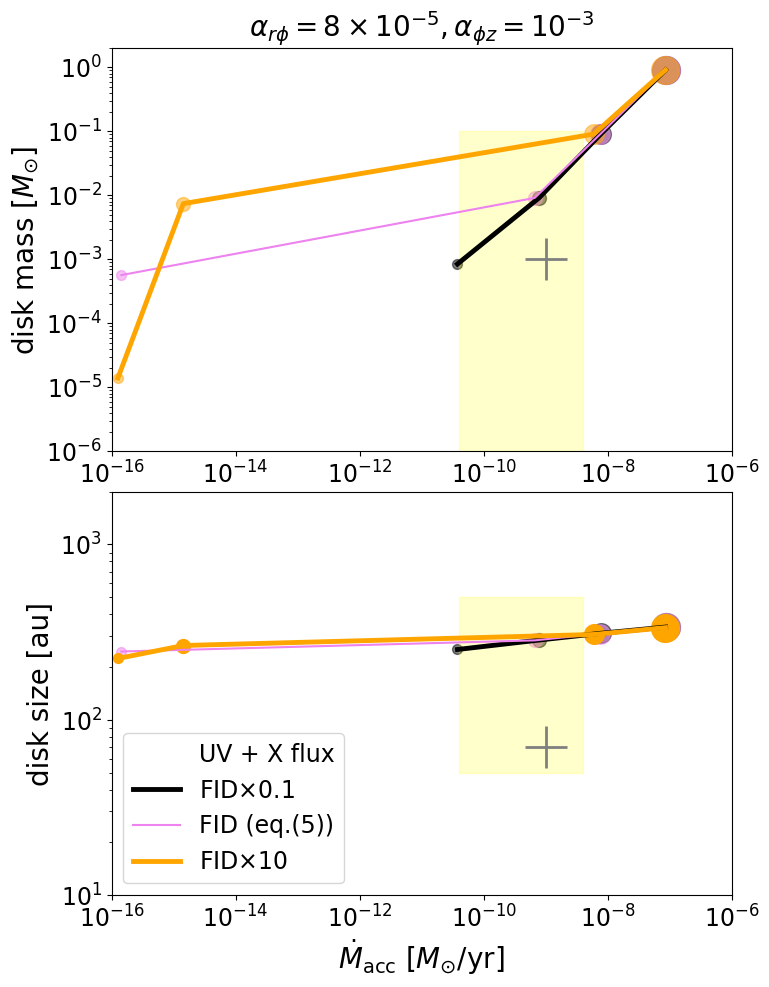}
\caption{
Same as Figure~\ref{fig:Acc_Infall_Photo}, but for models with the strong MHD disk wind torque of $\alpha_{\phi z} = 10^{-3}$ and weak turbulent viscosity of $\alpha_{r \phi}=8 \times 10^{-5}$. The fiducial model with the photoevaporation rate given by Eq.~\eqref{Photo} is shown by the magenta line, unlike in Figure~\ref{fig:Acc_Infall_Photo}.
}
\label{fig:Acc_Infall_Photo_alpha}
\end{figure}

Figure~\ref{fig:Acc_Infall_Photo_alpha} shows a similar trend for models with strong MHD disk wind torque. Comparison with the yellow shaded region indicates that weaker photoevaporation is more favorable for reproducing the observed disk masses and stellar accretion rates. The upper panel further shows that, if the photoevaporation rate is an order of magnitude higher than the fiducial value, it becomes difficult to simultaneously reproduce the observed disk masses and stellar accretion rates, even in the presence of strong MHD disk wind torque.

\section{Extension to $1M_{\odot}$ central star}

\begin{figure}
    \centering
    \includegraphics[width=8cm]{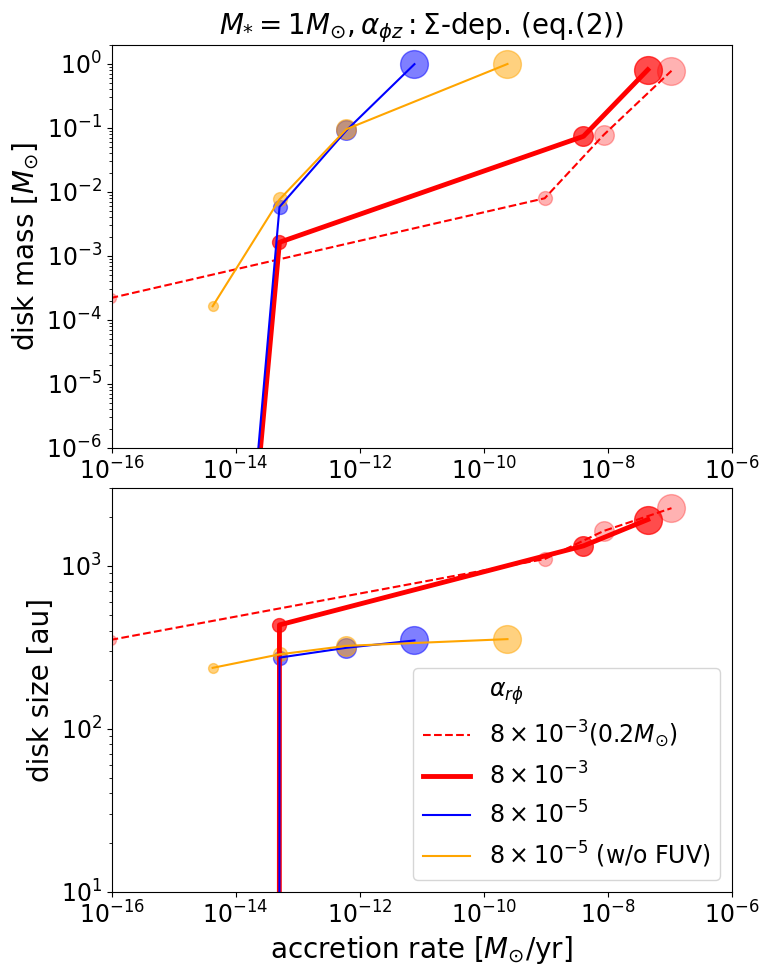}
\caption{Same as Figure~\ref{fig:Acc_Diskmass_size_MRI}, but for models with a $1~\Msun$ central star. The red, blue, and orange lines represent models with $\alpha_{r\phi}=8\times10^{-3}$, $\alpha_{r\phi}=8\times10^{-5}$ with FUV photoevaporation, and $\alpha_{r\phi}=8\times10^{-5}$ without FUV photoevaporation, respectively. The circles represent different infall rates of $\dot{M}_{\rm infall} = 5\times10^{-10}$, $5\times10^{-9}$, $5\times10^{-8}$, and $5\times10^{-7}\ \msunyr$, respectively, in order of increasing circle size.
}
    \label{fig:Acc_Diskmass_size_1Msun}
\end{figure}

Figure~\ref{fig:Acc_Diskmass_size_1Msun} shows the correlations between stellar $\mdota$, disk mass, and disk radius in models with a $1~\Msun$ central star. Compared to the $0.2~\Msun$ cases, $\mdota$ is generally lower, and the threshold infall rate at which a gap significantly affects $\mdota$ becomes larger. This is because a $1~\Msun$ star emits stronger UV and X-ray radiation, so late infall does not sustain sufficient stellar accretion. Even without FUV photoevaporation, an infall rate of $\mdoti > 5 \times 10^{-8} \msunyr$ is required to fill the gap. On the other hand, the typical $\mdoti$ values may exceed those in the $0.2 \Msun$ cases when the BHL accretion rate is considered. Thus, late infall can still influence disks around $1~\Msun$ stars, warranting further investigation.

\end{appendix}

\end{document}